\documentclass[a4paper,11pt]{article}
\usepackage{jheppub} 
\usepackage{amsmath,amssymb,amsfonts,amsthm,slashed}
\usepackage{empheq}
\usepackage[utf8]{inputenc}
\usepackage{graphicx}
\usepackage{hyperref}
\usepackage{bm,bbold}
\usepackage{accents}
\usepackage{ulem}
\usepackage{slashed}
\usepackage{color}
\usepackage{orcidlink}
\newcommand{\bra}[1]{\langle #1|}
\newcommand{\ket}[1]{|#1\rangle}

\newcommand{\wrho}{\widehat{\rho}}
\newcommand{\wW}{\widehat{W}}
\newcommand{\wP}{\widehat{P}}
\newcommand{\wJ}{\widehat{J}}
\newcommand{\wj}{\widehat{j}}
\newcommand{\wT}{\widehat{T}}
\newcommand{\wS}{\widehat{S}}
\newcommand{\SP}{\mathfrak{S}}

\newcommand{\wadf}[2]{{\widehat a}^\dagger_#1 (#2)}
\newcommand{\waf}[2]{{\widehat a}_#1 (#2)}
\newcommand{\wbdf}[2]{\widehat b^\dagger_{#1}(#2)}
\newcommand{\wbf}[2]{\widehat b_{#1}(#2)}
\newcommand{\di}{{\rm d}}
\newcommand{\ii}{{\rm i}}
\newcommand{\e}{{\rm e}}
\newcommand{\p}{{\rm p}}

\newcommand{\tr}{{\rm tr}}
\newcommand{\Tr}{{\rm Tr}}
\def\codev{{\stackrel{\leftrightarrow}{\partial}}}
\def\codevR{{\stackrel{\rightarrow}{\partial}}}
\def\codevL{{\stackrel{\leftarrow}{\partial}}}

\usepackage{xcolor}
\usepackage{ulem}
\definecolor{purple}{rgb}{0.8,0,0.6}

\arxivnumber{2610.37382}
\title{Covariant Wigner function and spin polarization of the Rarita-Schwinger field}

\author[1]{Matteo Buzzegoli\,\orcidlink{0000-0002-2114-5431}}
\emailAdd{matteo.buzzegoli@ifj.edu.pl}
\affiliation[1]{Institute of Nuclear Physics Polish Academy of Sciences, PL-31-342 Krakow, Poland}
\author[2]{Aleksandar Geci\'{c}\,\orcidlink{0009-0002-3032-9360}}
\emailAdd{aleksandar.gecic10@e-uvt.ro}
\affiliation[2]{Department of Physics, West University of Timisoara, Bulevardul Vasile P\^arvan 4, Timisoara 300223, Romania}
\author[3]{Rajeev Singh\,\orcidlink{0000-0001-5855-4039}}
\emailAdd{rsingh@itp.uni-frankfurt.de}
\affiliation[3]{Institut für Theoretische Physik, Goethe University, Max-von-Laue-Straße 1, D-60438 Frankfurt am Main, Germany}

\abstract{We study the covariant Wigner function for the spin-$3/2$ Rarita-Schwinger field and its relation to the spin density matrix and the mean spin polarization vector. We obtain a formula relating the covariant Wigner function to the mean spin vector for a weakly interacting field. We then calculate the mean spin vector induced by thermal vorticity, thermal shear, spin potential and spin Hall effect at local thermal equilibrium in the Belinfante, canonical and GLW pseudo-gauges. The resulting formula can be used to predict the spin polarization of spin-$3/2$ hadrons, such as the $\Omega^-$, in heavy-ion collisions. Differences in the thermal-shear induced polarization in the canonical and Belinfante pseudo-gauge make it possible to use local spin polarization measurements of the $\Omega^-$ to discriminate between the two pseudo-gauges. Lastly, we obtain the axial vortical effect conductivity for the free massive spin-$3/2$ Rarita-Schwinger field.}

\begin{document}
\maketitle
\flushbottom

\section{Introduction}
\label{sec:Intro}
The discovery of global spin alignment and polarization of hyperons in relativistic heavy-ion collisions \cite{STAR:2017ckg,STAR:2018gyt,STAR:2019erd,ALICE:2019onw,STAR:2022fan,Liang:2004ph,Liang:2004xn,Becattini:2024uha} opened a new avenue for investigating the quantum properties of strongly interacting matter under extreme conditions. Experimental measurements, which initially focused on the $\Lambda$ hyperon, have now expanded to include other light-flavor and strange hadrons \cite{STAR:2020xbm}, such as the multistrange $\Xi$ and $\Omega$ baryons. Furthermore, upcoming high-precision measurements from experiments such as ALICE are expected to provide dedicated results for the spin polarization of $\Omega$ hyperons~\cite{Alice:2026Moriond}.

To interpret these measurements, theoretical models rely fundamentally on quantum statistical mechanics and field theory, establishing a connection between the mean spin polarization vector $S^\mu(p)$ and the covariant Wigner function $W(x,p)$ \cite{Becattini:2020sww,Becattini:2024uha}. For spin-$1/2$ particles, this relation is rigorously derived from quantum field theory (QFT) as
\begin{equation}\label{eq:SFormulaDirac}
	S_{(1/2)}^\mu(p) = \frac{1}{2} \frac{\int_\Sigma \di\Sigma\cdot p\,\tr\left[\gamma^\mu\gamma^5 W(x,p)\right]}{\int_\Sigma \di\Sigma\cdot p\,\tr\left[ W(x,p)\right]}\;.
\end{equation}
For spin-$3/2$ particles, however, a full field-theoretic derivation has not yet been provided. Instead, one relies on the single-particle limit for massive spin-$S$ particles in global thermal equilibrium, where the spin vector induced by the thermal vorticity $\varpi_{\rho\nu}$ takes the form~\cite{Becattini:2020sww,Becattini:2024uha}:
\begin{equation}\label{eq:SVectorAnySpin}
	S_{(S)}^\mu(p)_{\rm GE} = -\frac{1}{2m}\frac{S(S+1)}{3}\epsilon^{\mu\rho\nu\tau} p_\tau \varpi_{\rho\nu}\;.
\end{equation}
Earlier phenomenological studies \cite{Xie:2017upb,Fu:2020oxj,Li:2026udm} modeled higher-spin hadron polarization, including spin-$3/2$ hyperons like the $\Omega^-$, by extending Eq.~\eqref{eq:SVectorAnySpin} directly to local thermal equilibrium (LTE):
\begin{equation}\label{eq:SVectorLTEExtension}
	S_{(3/2)}^\mu(p)_{\rm LE} = -\frac{5}{8m}\epsilon^{\mu\rho\nu\tau} p_\tau
    \frac{\int_\Sigma \di\Sigma\cdot p\ \varpi_{\rho\nu}(x)n_F(x,p)(1-n_F(x,p))}{\int_\Sigma \di\Sigma\cdot p\ n_F(x,p)}.
\end{equation}
While this is a reasonable ansatz for global spin polarization, understanding local spin polarization measurements requires the inclusion of out-of-equilibrium effects, such as contributions from thermal shear~\cite{Becattini:2021suc,Liu:2021uhn,Becattini:2021iol,Fu:2021pok} and the spin Hall effect~\cite{Liu:2020dxg,Buzzegoli:2022qrr,Fu:2022myl}. Because these effects vanish in equilibrium, they cannot be captured by the global equilibrium spin polarization~\eqref{eq:SVectorAnySpin} and therefore require a rigorous derivation.

In this work, following the established methodology adopted for the Dirac field~\cite{Becattini:2020sww,Becattini:2024uha}, we provide a \textit{complete} quantum field-theoretic derivation connecting the mean spin polarization vector to the covariant Wigner function for the massive spin-$3/2$ Rarita-Schwinger field (see~\cite{Rarita:1941mf,Fierz:1939ix} and the textbooks~\cite[Sec.~1.7]{Lurie:1968} and \cite[p.~232]{Weinberg:1995mt} and the review~\cite{Pascalutsa:2006up}). Furthermore, using the recently developed improved methods for integrating over the decoupling hypersurface~\cite{Sheng:2025cjk}, we also calculate the resulting spin polarization for a relativistic fluid in local thermal equilibrium, taking into account the linear response to hydrodynamic gradients and the spin potential. Lastly, as a theoretical application of the developed Wigner function framework, we compute the axial vortical conductivity for the massive spin-3/2 field and discuss its relation with quantum anomalies.

For the reader's convenience, the final general expression of the mean spin polarization vector for the spin-$3/2$ Rarita-Schwinger field is given in Eq.~\eqref{eq:SpinPol3/2}, which, in local thermal equilibrium and at first-order expansion in gradients, reduces to Eq.~\eqref{eq:FinalLTEPolGLW} for the de Groot--van Leeuwen--van Weert (GLW) pseudo-gauge, to Eq.~\eqref{eq:FinalLTEPolBel} for the Belinfante pseudo-gauge, and to Eq.~\eqref{eq:FinalLTEPolCan} for the canonical one.
\subsection*{Notations and conventions}

We use natural units with $\hbar=c=k_{\rm B}=1$ and the relativistic notation with repeated indices assumed to be saturated.
Contractions of indices are sometimes denoted with e.g., $\beta_\mu p^\mu = \beta \cdot p$ and $\varpi:J = \varpi_{\mu\nu}J^{\mu\nu}$.
The Minkowskian metric tensor is ${\rm diag}(1,-1,-1,-1)$ and the Levi-Civita symbol follows the convention $\epsilon^{0123}=1$.
Unit vectors are denoted with a small upper hat ($\hat u$), while operators in Hilbert space are denoted by a large upper hat $(\widehat{T})$, with the exception of the Rarita-Schwinger field, which is expressed by $\Psi_\mu$ without an upper hat.
The symbol $\Tr$ with a capital T stands for the trace over all states in the Hilbert space, whereas the symbol $\tr$ stands for a trace over polarization states or traces of finite dimensional matrices.

\section{The Rarita-Schwinger field}
\label{sec:RSField}
In this section, we review the spin-$3/2$ Rarita-Schwinger field and its underlying field equations. Using Noether's theorem, we derive the canonical energy-momentum tensor $\wT_C^{\mu\nu}$ and the canonical spin tensor $\wS_C^{\lambda,\mu\nu}$ for the free field. We then discuss pseudo-gauge transformations that map these canonical quantities into the symmetrized Belinfante energy-momentum tensor $\wT_{\rm B}^{\mu\nu}$ with a vanishing spin tensor. Finally, we formulate the symmetrized de Groot--van Leeuwen--van Weert (GLW) energy-momentum tensor $\wT_{\rm GLW}^{\mu\nu}$ alongside its non-vanishing, conserved GLW spin tensor $\wS_{\rm GLW}^{\lambda,\mu\nu}$.

A free, massive spin-$3/2$ field is represented by a vector-spinor field operator $\Psi_\mu(x)$ ($\mu \in \{0,1,2,3\}$), where each spacetime component transforms as a four-component Dirac spinor~\cite{Rarita:1941mf,Fierz:1939ix}. The field satisfies the Rarita-Schwinger equation of motion:
\begin{equation}
\label{rseq}
(\ii \gamma^\alpha \partial_\alpha - m) \Psi_\mu(x) = 0,
\end{equation}
subject to two auxiliary physical state constraints that eliminate lower-spin (spin-$1/2$) degrees of freedom:
\begin{equation}
\label{rsconstraints}
\gamma^\mu \Psi_\mu(x) = 0, \qquad \partial^\mu \Psi_\mu(x) = 0.
\end{equation}
The dynamics of the spin-$3/2$ vector-spinor field $\Psi_\mu(x)$ are described by the Hermitian Rarita-Schwinger Lagrangian density~\cite{Pascalutsa:2006up}:
\begin{equation}
\label{rs_lagrangian}
\mathcal{L} = \frac{\ii}{2} \left[ \bar{\Psi}_\mu \gamma^{\mu\nu\alpha} \partial_\alpha \Psi_\nu - (\partial_\alpha \bar{\Psi}_\mu) \gamma^{\mu\nu\alpha} \Psi_\nu \right] - m \bar{\Psi}_\mu \gamma^{\mu\nu} \Psi_\nu,
\end{equation}
where $\gamma^{\mu\nu\alpha}$ denotes the totally antisymmetric product of three Dirac gamma matrices:
\begin{subequations}
\begin{align}
\gamma^{\mu\nu} \equiv& \frac{1}{2}[\gamma^\mu, \gamma^\nu] = -2\ii \Sigma^{\mu\nu},\\
\gamma^{\mu\nu\alpha} \equiv& \frac{1}{2}\left\{\gamma^{\mu\nu},\,\gamma^\alpha\right\}
    = -\ii\epsilon^{\mu\nu\alpha\rho}\gamma_\rho\gamma^5
    = \gamma^\mu\gamma^\nu\gamma^\alpha - g^{\nu\alpha}\gamma^\mu - g^{\mu\nu}\gamma^\alpha
        + g^{\mu\alpha}\gamma^\nu. \label{eq:cantigamma}
\end{align}
\end{subequations}
Evaluating the Euler-Lagrange equations,
\begin{equation}
\frac{\partial \mathcal{L}}{\partial \bar{\Psi}_\mu} - \partial_\alpha \frac{\partial \mathcal{L}}{\partial(\partial_\alpha \bar{\Psi}_\mu)} = 0    
\end{equation}
yields the field equation:
\begin{equation}\label{eq:ELrw}
\ii \gamma^{\mu\nu\alpha} \partial_\alpha \Psi_\nu - m \gamma^{\mu\nu} \Psi_\nu = 0.
\end{equation}
Contracting Eq.~\eqref{eq:ELrw} with $\partial_\mu$ leads to the physical state constraint
\begin{equation}\label{eq:con1}
\partial_\mu \left(\ii \gamma^{\mu\nu\alpha}\partial_\alpha - m \gamma^{\mu\nu}\right) \Psi_\nu = 0
    =-m\gamma^{\mu\nu}\partial_\mu \Psi_\nu=0,
\end{equation}
where we have used $\gamma^{\mu\nu\alpha}\partial_\mu\partial_\alpha=0$ due to the antisymmetry of $\gamma^{\mu\nu\alpha}$.
Contracting Eq.~\eqref{eq:ELrw} with $\gamma_\mu$ and employing the gamma-matrix identities $\gamma_\mu\gamma^{\mu\nu\alpha}=2\gamma^{\nu\alpha}$ and $\gamma_\mu\gamma^{\mu\nu}=3\gamma^\nu$ gives the further constraint
\begin{equation}\label{eq:con2}
\gamma_\mu \left(\ii \gamma^{\mu\nu\alpha}\partial_\alpha - m \gamma^{\mu\nu}\right) \Psi_\nu = 0
    =(2\ii\gamma^{\mu\nu}\partial_\mu -3m\gamma^\nu)\Psi_\nu= -3m\gamma^\nu\Psi_\nu=0,
\end{equation}
where Eq.~\eqref{eq:con1} has been applied.
Using the constraint equations~\eqref{eq:con1} and \eqref{eq:con2}, the Euler-Lagrange equation simplifies to $(\ii \slashed{\partial} - m) \Psi_\mu(x) = 0$.
Furthermore, contracting the Euler-Lagrange equation with $\gamma_\mu$ and expanding $\gamma^{\mu\nu\alpha}$ yields
\begin{equation*}
\begin{split}
0= &\gamma_\mu \left( \ii \gamma^{\mu\nu\alpha} \partial_\alpha \Psi_\nu - m \gamma^{\mu\nu} \Psi_\nu\right) \\
= & \ii \gamma_\mu \left(\gamma^\mu\gamma^\nu\gamma^\alpha - g^{\nu\alpha}\gamma^\mu - g^{\mu\nu}\gamma^\alpha + g^{\mu\alpha}\gamma^\nu\right)\partial_\alpha \Psi_\nu
    - m \gamma_\mu\gamma^{\mu\nu} \Psi_\nu\\
= & 4\gamma^\nu (\ii\slashed\partial\Psi_\nu) -\ii 4(\partial\cdot\Psi) 
    - \ii\gamma^\nu\slashed\partial\Psi_\nu + \ii \gamma^\alpha\gamma^\nu\partial_\alpha\Psi_\nu
    - 3m\gamma^\nu\Psi_\nu.
\end{split}
\end{equation*}
Imposing $\gamma^\nu\Psi_\nu=0$ together with the equation of motion $\ii\slashed\partial\Psi_\nu=m\Psi_\nu$, we obtain
\begin{equation}
\begin{split}
4m\gamma^\nu \Psi_\nu -\ii 4(\partial\cdot\Psi) 
    - 4 m\gamma^\nu\Psi_\nu + \ii \gamma^\alpha\partial_\alpha\gamma^\nu\Psi_\nu
= -\ii 4(\partial\cdot\Psi)= 0,
\end{split}
\end{equation}
where we again used $\gamma^\nu\Psi_\nu=0$, confirming that $(\partial\cdot\Psi)= 0$.
In summary, the complete set of equations and constraints governing the Rarita-Schwinger field reads:
\begin{subequations}
\label{eq:EoMAll}
\begin{align}
(\ii \slashed{\partial} - m) \Psi_\mu(x) =& 0,&
    \gamma^\nu\Psi_\nu=& 0,\\
\label{eq:con3}
\gamma^{\mu\nu}\partial_\mu \Psi_\nu = & 0,&
    \partial\cdot\Psi=& 0.
\end{align}
\end{subequations}
In free space, the field operator $\Psi_\mu(x)$ can be expanded in plane-wave modes as:
\begin{equation}
\label{rsfieldexp}
\Psi_\mu(x) = \sum_{r=-3/2}^{+3/2} \int \frac{\di^3 p}{(2\pi)^{3/2} \, 2 E_p} \left[ \waf{r}{p} u_{\mu,r}(p) \e^{-\ii p \cdot x} + \wbdf{r}{p} v_{\mu,r}(p) \e^{\ii p \cdot x} \right],
\end{equation}
where $E_p = \sqrt{|\mathbf{p}|^2 + m^2}$, $\waf{r}{p}$, and $\wbdf{r}{p}$ are the annihilation operators for particles and creation operators for antiparticles of spin projection $r \in \{-3/2, -1/2, +1/2, +3/2\}$, respectively. They satisfy standard canonical anticommutation relations:
\begin{equation}
\{ \waf{r}{p}, \wadf{s}{p'} \} = \{ \wbf{r}{p}, \wbdf{s}{p'} \} = 2 E_p \delta_{rs} \delta^3(\mathbf{p} - \mathbf{p}').
\end{equation}
The vector-spinors $u_{\mu,r}(p)$ and $v_{\mu,r}(p)$ satisfy the momentum-space field equation and constraints:
\begin{subequations}
\begin{align}
\label{eq:EoMu}
(\slashed{p} - m) u_{\mu,r}(p) =& 0,& \gamma^\mu u_{\mu,r}(p) =& 0,& p^\mu u_{\mu,r}(p) =& 0,\\
\label{eq:EoMv}
(\slashed{p} + m) v_{\mu,r}(p) =& 0,& \gamma^\mu v_{\mu,r}(p) =& 0,& p^\mu v_{\mu,r}(p) =& 0.
\end{align}    
\end{subequations}
Under the standard relativistic normalization for Dirac spinors ($\bar{u} u = 2m$) and spatial polarization vectors ($\epsilon_\mu \epsilon^{*\mu} = -1$), the normalization conditions for the spin-$3/2$ vector-spinors are chosen as:
\begin{equation}\label{eq:SpinorNorm}
\bar{u}_{\mu,r}(p) u_s^\mu(p) = -2m \delta_{rs}, \qquad \bar{v}_{\mu,r}(p) v_s^\mu(p) = 2m \delta_{rs},
\end{equation}
where $\bar{u}_{\mu,r}(p) \equiv u_{\mu,r}^\dagger(p) \gamma^0$, and the spatial metric contraction accounts for the spacelike nature of the polarization vectors. The completeness relation (spin sum) for positive-energy spin-$3/2$ vector-spinors is given by:
\begin{equation}
\label{spinsum_rs}
\sum_{r=-3/2}^{+3/2} u_{\mu,r}(p) \bar{u}_{\nu,r}(p) = \Pi_{\mu\nu}(p)
    = - (\slashed{p} + m) P_{\mu\nu}(p),
\end{equation}
where $P_{\mu\nu}(p)$ is the spin-$3/2$ transverse projection tensor:
\begin{equation}
\label{projection_def}
P_{\mu\nu}(p) = g_{\mu\nu} - \frac{1}{3}\gamma_\mu\gamma_\nu - \frac{1}{3m}(\gamma_\mu p_\nu - \gamma_\nu p_\mu) - \frac{2}{3m^2}p_\mu p_\nu.
\end{equation}
From the spinor identities~\eqref{eq:EoMu}, the projector inherits the following properties:
\begin{subequations}
\begin{align}
\label{eq:pPiIdentity}
p^\mu \Pi_{\mu\nu}(p) = & 0, & \Pi_{\mu\nu}(p) p^\nu = & 0,\\
\label{eq:GammaPiIdentity}
\gamma^\mu \Pi_{\mu\nu}(p) = & 0, & \Pi_{\mu\nu}(p) \gamma^\nu = & 0,
\end{align}
and
\begin{equation}
\Pi^\mu_{\ \mu}(p) = \Pi_{\mu\nu}(p) g^{\mu\nu} = -2\left(\slashed{p} + m\right).
\end{equation}
\end{subequations}
%

\subsection{The canonical energy-momentum and spin tensors}
\label{subsec:canonical_emt}
By Noether's theorem, spacetime translation invariance $x^\mu \to x^\mu + a^\mu$ defines the canonical energy-momentum tensor:
\begin{equation}
\label{canonical_emt_def}
\widehat{T}_C^{\mu\nu} = -\frac{\partial \mathcal{L}}{\partial(\partial_\mu \Psi_\beta)} \partial^\nu \Psi_\beta - (\partial^\nu \bar{\Psi}_\alpha) \frac{\partial \mathcal{L}}{\partial(\partial_\mu \bar{\Psi}_\alpha)} - g^{\mu\nu} \mathcal{L}.
\end{equation}
Evaluating the functional derivatives of the Lagrangian~\eqref{rs_lagrangian} with respect to the field gradients gives:
\begin{equation}
\frac{\partial \mathcal{L}}{\partial(\partial_\mu \Psi_\beta)} = \frac{\ii}{2} \bar{\Psi}_\alpha \gamma^{\alpha\mu\beta}, \qquad \frac{\partial \mathcal{L}}{\partial(\partial_\mu \bar{\Psi}_\alpha)} = -\frac{\ii}{2} \gamma^{\alpha\mu\beta} \Psi_\beta.
\end{equation}
Substituting these expressions into Eq.~\eqref{canonical_emt_def} and noting that $\mathcal{L} = 0$ is on-shell, we find:
\begin{equation}
\label{canonical_emt_rs}
\widehat{T}_C^{\mu\nu} = -\frac{\ii}{2} \left[ \bar{\Psi}_\alpha \gamma^{\alpha\mu\beta} \partial^\nu \Psi_\beta - (\partial^\nu \bar{\Psi}_\alpha) \gamma^{\alpha\mu\beta} \Psi_\beta \right].
\end{equation}
Under the constraint conditions ($\gamma^\alpha \Psi_\alpha = 0$ and $\bar{\Psi}_\beta \gamma^\beta = 0$), this simplifies to:
\begin{equation}\label{eq:set_c}
\widehat{T}_C^{\mu\nu}(x) = -\frac{\ii}{2} \bar{\Psi}_\alpha(x) \gamma^\mu \codev\,^\nu \Psi^\alpha(x),
\end{equation}
where $A \codev\,^\nu B \equiv A (\partial^\nu B) - (\partial^\nu A) B$.
Analogously to the Dirac field, $\widehat{T}_C^{\mu\nu}$ is conserved ($\partial_\mu \widehat{T}_C^{\mu\nu} = 0$), but asymmetric ($\widehat{T}_C^{\mu\nu} \neq \widehat{T}_C^{\nu\mu}$). Moreover, plugging the plane-wave expansion \eqref{rsfieldexp} and using the identity $\bar{u}^\alpha_{s}(p) \gamma^0 u_{\alpha,r}(p)=-2E_p\delta_{r,s}$, we obtain the Hamiltonian
\begin{equation}
\widehat{H} = \int \di^3 x\ \wT^{00}_C(x) = \int\frac{\di^3 p}{2 E_p}\, E_p
    \left[ \wadf{r}{p} \waf{r}{p} + \wbdf{r}{p} \wbf{r}{p} \right].
\end{equation}
Under an infinitesimal Lorentz transformation $x^\mu \to x^\mu + \omega^{\mu\nu} x_\nu$ ($\omega^{\mu\nu} = -\omega^{\nu\mu}$), the vector-spinor field transforms as~\cite[Sec.~1.7]{Lurie:1968}:
\begin{equation}\label{eq:16LorentzTransField}
\delta \Psi_{r\mu}(x) = \frac{1}{2} \omega_{\rho\sigma}
    \left( -\ii\mathcal{J}^{\rho\sigma}_{rs} \right)_\mu^{\ \nu} \Psi_{s\nu}(x)
= \frac{1}{2} \omega_{\rho\sigma} \left[
    \frac{1}{2}\gamma^{\rho\sigma}_{rs} \delta_\mu^\nu
    + (M^{\rho\sigma})_\mu^{\ \nu}\mathbb{1}_{rs} \right] \Psi_{s\nu}(x),
\end{equation}
where $\gamma^{\rho\sigma}$ acts on Dirac spinor indices and $(M^{\rho\sigma})_\mu^{\ \nu} = \delta^\rho_\mu g^{\sigma\nu} - \delta^\sigma_\mu g^{\rho\nu}$ acts on vector indices.
Noether's theorem yields the conserved total angular momentum current $\widehat{J}_C^{\lambda,\mu\nu} = x^\mu \widehat{T}_C^{\lambda\nu} - x^\nu \widehat{T}_C^{\lambda\mu} + \widehat{S}_C^{\lambda,\mu\nu}$, where the canonical spin tensor is given by
\begin{equation}
\widehat{S}_C^{\lambda,\mu\nu} =
-\frac{\partial \mathcal{L}}{\partial(\partial_\lambda \Psi_\alpha)} \left(-\ii\mathcal{J}^{\mu\nu}\right)_\alpha^{\ \beta} \Psi_\beta
- \bar{\Psi}_\alpha \left(-\ii\mathcal{J}^{\mu\nu}\right)^\alpha_{\ \beta} \frac{\partial \mathcal{L}}{\partial(\partial_\lambda \bar{\Psi}_\beta)},
\end{equation}
which decomposes into Dirac spinor and spacetime vector contributions~\cite[Sec.~1.7]{Lurie:1968}:
\begin{equation}
\widehat{S}_C^{\lambda,\mu\nu} = \widehat{S}_{(1/2)}^{\lambda,\mu\nu}
    + \widehat{S}_{(1)}^{\lambda,\mu\nu}.
\end{equation}
Computing the field variations explicitly from the Lagrangian~\eqref{rs_lagrangian} yields:
\begin{subequations}
\begin{align}
\widehat{S}_{(1/2)}^{\lambda,\mu\nu} & = -\frac{\ii}{4} \bar{\Psi}_\alpha \gamma^{\alpha\lambda\beta} \gamma^{\mu\nu} \Psi_\beta + \frac{\ii}{4} \bar{\Psi}_\alpha  \gamma^{\mu\nu} \gamma^{\alpha\lambda\beta} \Psi_\beta, \label{s_dirac} \\
\widehat{S}_{(1)}^{\lambda,\mu\nu} &= -\frac{\ii}{2} \left[ \bar{\Psi}_\alpha \gamma^{\alpha\lambda\mu} \Psi^\nu - \bar{\Psi}_\alpha \gamma^{\alpha\lambda\nu} \Psi^\mu \right] + \frac{\ii}{2} \left[ \bar{\Psi}^\mu \gamma^{\nu\lambda\beta} \Psi_\beta - \bar{\Psi}^\nu \gamma^{\mu\lambda\beta} \Psi_\beta \right]. \label{s_vector}
\end{align}    
\end{subequations}
Substituting the definition~\eqref{eq:cantigamma} and imposing the on-shell constraints, $\gamma^\alpha \Psi_\alpha = 0$ and $\bar{\Psi}_\alpha \gamma^\alpha = 0$, the canonical spin tensor reduces to:
\begin{equation}
\label{canonical_spin_rs}
\widehat{S}_C^{\lambda,\mu\nu} = -\frac{1}{2} \bar{\Psi}_\alpha \left\{\gamma^\lambda,\, \Sigma^{\mu\nu}\right\} \Psi^\alpha -\ii\left(\bar{\Psi}^\mu\gamma^\lambda\Psi^\nu - \bar{\Psi}^\nu\gamma^\lambda\Psi^\mu\right).
\end{equation}
This can again be partitioned into spinor (Dirac) and vector components:
\begin{subequations}
\begin{align}
\widehat{S}_D^{\lambda,\mu\nu} =& -\frac{1}{2} \bar{\Psi}_\alpha \left\{\gamma^\lambda,\, \Sigma^{\mu\nu}\right\} \Psi^\alpha,\\
\label{eq:SpinTenCanV}
\widehat{S}_V^{\lambda,\mu\nu} =& -\ii\left(\bar{\Psi}^\mu\gamma^\lambda\Psi^\nu - \bar{\Psi}^\nu\gamma^\lambda\Psi^\mu\right).
\end{align}
\end{subequations}
Direct calculation shows that the vector contribution is separately conserved, $\partial_\lambda \widehat{S}_V^{\lambda,\mu\nu}=0$, and that the divergence of the canonical spin tensor corresponds to the antisymmetric part of the canonical energy-momentum tensor:
\begin{equation}\label{eq:div_can_spin_tensor}
\partial_\lambda \widehat{S}_C^{\lambda,\mu\nu} = \widehat{T}_C^{\nu\mu} - \widehat{T}_C^{\mu\nu},
\end{equation}
which guarantees the conservation of total angular momentum, $\partial_\lambda \widehat{J}_C^{\lambda,\mu\nu}=0$.

\subsection{Pseudo-gauge transformation and the Belinfante form}
\label{subsec:pseudo_gauge_Belinfante}
A pseudo-gauge transformation~\cite{Hehl:1976vr,Buzzegoli:2021wlg,Becattini:2024uha,Singh:2024qvg} redefines the energy-momentum and spin tensors while preserving the integrated energy, momentum, and angular-momentum charges:
\begin{subequations}\label{eq:pgt}
\begin{align}
\widehat{T}_{\text{pgt}}^{\mu\nu} &= \widehat{T}^{\mu\nu} + \frac{1}{2} \partial_\lambda \left( \widehat{\Phi}^{\lambda,\mu\nu} + \widehat{\Phi}^{\nu,\mu\lambda} + \widehat{\Phi}^{\mu,\nu\lambda} \right), \label{pgt_emt} \\
\widehat{S}_{\text{pgt}}^{\lambda,\mu\nu} &= \widehat{S}^{\lambda,\mu\nu} - \widehat{\Phi}^{\lambda,\mu\nu}, \label{pgt_spin}
\end{align}
\end{subequations}
where $\widehat{\Phi}^{\lambda,\mu\nu} = -\widehat{\Phi}^{\lambda,\nu\mu}$ is a superpotential tensor.

In the Belinfante approach, starting from the canonical form, the superpotential is identified with the canonical spin tensor: $\widehat{\Phi}_{\rm B}^{\lambda,\mu\nu} = \widehat{S}_C^{\lambda,\mu\nu}$.
This choice identically eliminates the resulting spin tensor: $\widehat{S}_{\rm B}^{\lambda,\mu\nu} = 0$.

Substituting $\widehat{\Phi}_{\rm B}^{\lambda,\mu\nu} = \widehat{S}_C^{\lambda,\mu\nu}$ into Eq.~\eqref{pgt_emt}, utilizing Eq.~\eqref{eq:div_can_spin_tensor}, and noting that the spinor spin tensor $\widehat{S}_D^{\nu,\mu\lambda}$ is completely antisymmetric, we have:
\begin{equation}
\begin{split}
\widehat{T}_{\rm B}^{\mu\nu}  = & \widehat{T}_C^{\mu\nu}
    +\frac{1}{2}\partial_\lambda \widehat{S}_C^{\lambda,\mu\nu}
    +\frac{1}{2}\partial_\lambda \left(\widehat{S}_C^{\nu,\mu\lambda} +\widehat{S}_C^{\mu,\nu\lambda}\right)\\
=& \widehat{T}_C^{\mu\nu}
    +\frac{1}{2}\left( \widehat{T}_C^{\nu\mu} - \widehat{T}_C^{\mu\nu}\right)
    +\frac{1}{2}\partial_\lambda \left(\widehat{S}_D^{\nu,\mu\lambda} -\widehat{S}_D^{\nu,\mu\lambda} + \widehat{S}_V^{\nu,\mu\lambda} +\widehat{S}_V^{\mu,\nu\lambda}\right)\\
=& \frac{1}{2}\left( \widehat{T}_C^{\mu\nu} + \widehat{T}_C^{\nu\mu} \right)
    +\frac{1}{2}\partial_\lambda \left(\widehat{S}_V^{\nu,\mu\lambda} +\widehat{S}_V^{\mu,\nu\lambda}\right).
\end{split}
\end{equation}
Substituting the explicit form of $\widehat{S}_V^{\nu,\mu\lambda}$ and employing Eq.~\eqref{eq:con3}, we obtain
\begin{equation*}
\begin{split}
\frac{1}{2}\partial_\lambda \left(\widehat{S}_V^{\nu,\mu\lambda} +\widehat{S}_V^{\mu,\nu\lambda}\right)
= & -\frac{\ii}{2}\partial_\lambda\left(\bar{\Psi}^\mu\gamma^\nu\Psi^\lambda
    + \bar{\Psi}^\nu\gamma^\mu\Psi^\lambda
    - \bar{\Psi}^\lambda\gamma^\nu\Psi^\mu
    - \bar{\Psi}^\lambda\gamma^\mu\Psi^\nu \right)\\
= & -\frac{\ii}{2}\left[(\partial_\lambda\bar{\Psi}^\mu)\gamma^\nu\Psi^\lambda
    + (\partial_\lambda\bar{\Psi}^\nu)\gamma^\mu\Psi^\lambda
    - \bar{\Psi}^\lambda\gamma^\nu\partial_\lambda\Psi^\mu
    - \bar{\Psi}^\lambda\gamma^\mu\partial_\lambda\Psi^\nu \right].
\end{split}
\end{equation*}
Combining these results yields
\begin{equation}
\begin{split}
\widehat{T}_{\rm B}^{\mu\nu}  = &
    \frac{1}{2}\left( \widehat{T}_C^{\mu\nu} + \widehat{T}_C^{\nu\mu} \right) + \frac{1}{2}\partial_\lambda \left(\widehat{S}_V^{\nu,\mu\lambda} +\widehat{S}_V^{\mu,\nu\lambda}\right)\\
=& -\frac{\ii}{4} \bar{\Psi}_\alpha \gamma^\mu \codev\,^\nu \Psi^\alpha
    -\frac{\ii}{4} \bar{\Psi}_\alpha \gamma^\nu \codev\,^\mu \Psi^\alpha\\
    &-\frac{\ii}{2}\left[(\partial_\lambda\bar{\Psi}^\mu)\gamma^\nu\Psi^\lambda
    + (\partial_\lambda\bar{\Psi}^\nu)\gamma^\mu\Psi^\lambda
    - \bar{\Psi}^\lambda\gamma^\nu\partial_\lambda\Psi^\mu
    - \bar{\Psi}^\lambda\gamma^\mu\partial_\lambda\Psi^\nu \right],
\end{split}
\end{equation}
and taking advantage of $\partial\cdot\Psi=0$, we finally arrive at
\begin{equation}
\label{bel_emt_rs}
\begin{split}
\widehat{T}_{\rm B}^{\mu\nu}
=& -\frac{\ii}{4} \bar{\Psi}_\alpha \gamma^\mu \codev\,^\nu \Psi^\alpha
    -\frac{\ii}{4} \bar{\Psi}_\alpha \gamma^\nu \codev\,^\mu \Psi^\alpha\\
    &+\frac{\ii}{2}\bar{\Psi}_\iota\left[ g^{\iota\mu}\codev\,^\zeta\gamma^\nu
    +g^{\iota\nu}\codev\,^\zeta\gamma^\mu + g^{\zeta\mu}\codev\,^\iota\gamma^\nu
    + g^{\zeta\nu}\codev\,^\iota\gamma^\mu\right]\Psi_{\zeta}.
\end{split}
\end{equation}
The Belinfante energy-momentum tensor $T_{\rm B}^{\mu\nu}$ is manifestly symmetric, $T_{\rm B}^{\mu\nu} = T_{\rm B}^{\nu\mu}$, and conserved, $\partial_\mu T_{\rm B}^{\mu\nu} = 0$.

\subsection{The GLW energy-momentum and spin tensors}
The GLW pseudo-gauge framework for the Dirac field~\cite{DeGroot:1980dk,Speranza:2020ilk,Buzzegoli:2021wlg} translates directly to the Rarita-Schwinger field, and it is obtained from the canonical tensors by applying the pseudo-gauge transformation~\eqref{eq:pgt} with the superpotential
\begin{equation}
\widehat{\Phi}^{\lambda,\mu\nu}_{\rm GLW} =
    \frac{1}{4m}\bar{\Psi}_\alpha\left(
    \gamma^{\lambda\mu}\codev\,^\nu
    - \gamma^{\lambda\nu}\codev\,^\mu
    \right)\Psi^\alpha\, .
\end{equation}
The resulting spin tensor is given by
\begin{equation}\label{eq:GLW_SpinTensor}
\widehat{S}^{\lambda,\mu\nu}_{\rm GLW} = \widehat{S}_C^{\lambda,\mu\nu}
    -\frac{1}{4m}\bar{\Psi}_\alpha\left(
    \gamma^{\lambda\mu}\codev\,^\nu
    - \gamma^{\lambda\nu}\codev\,^\mu
    \right)\Psi^\alpha\, .
\end{equation}
By employing the field equations together with Dirac algebra identities, one can readily verify that this operator is conserved:
\begin{equation}
\partial_\lambda \widehat{S}^{\lambda,\mu\nu}_{\rm GLW} =0.
\end{equation}
The GLW energy-momentum tensor is obtained via Eq.~\eqref{eq:pgt}:
\begin{equation}
\begin{split}
\widehat{T}_{\text{GLW}}^{\mu\nu} =& \widehat{T}_C^{\mu\nu} + \frac{1}{2} \partial_\lambda \left( \widehat{\Phi}_{\rm GLW}^{\lambda,\mu\nu} + \widehat{\Phi}_{\rm GLW}^{\nu,\mu\lambda} + \widehat{\Phi}_{\rm GLW}^{\mu,\nu\lambda} \right),
\end{split}
\end{equation}
yielding:
\begin{equation}
\label{glw_emt_rs}
\widehat{T}_{\text{GLW}}^{\mu\nu} = \frac{1}{4m} \bar{\Psi}_\alpha \codev\,^\mu \codev\,^\nu \Psi^\alpha = \frac{1}{4m} \left[ \bar{\Psi}_\alpha \left( \partial^\mu \partial^\nu \Psi^\alpha \right) - 2 (\partial^\mu \bar{\Psi}_\alpha) (\partial^\nu \Psi^\alpha) + \left( \partial^\mu \partial^\nu \bar{\Psi}_\alpha \right) \Psi^\alpha \right].
\end{equation}
%

\section{The covariant Wigner function}
\label{sec:sec:WignerFunc}
The covariant Wigner operator~\cite{Wigner:1932eb} is an important quantity for studying the properties of a macroscopic fluid starting from a quantum field theory, as it allows one to obtain, for instance, the particle density in momentum space, the spin density matrix, as well as the expectation values of numerous physical observables. In this section, we define the Wigner operator for the Rarita-Schwinger field and list its most important properties. 

Similarly to the Dirac field, the covariant Wigner operator for the Rarita-Schwinger field is defined as a Fourier transform of the two-point function of the quantum field; however, in this case, the Wigner operator carries two Lorentz vector indices $(\mu, \nu)$ alongside its $4\times 4$ Dirac matrix structure:
\begin{equation}
	\label{rswignerop}
	\widehat{W}_{\mu\nu,AB}(x,k) = \int \frac{\di^4 y}{(2\pi)^4} \e^{-\ii k \cdot y} : \bar{\Psi}_{\nu,B}\left(x + \frac{y}{2}\right) \Psi_{\mu,A}\left(x - \frac{y}{2}\right) :,
\end{equation}
where $::$ denotes normal ordering and $A$, $B$ are the four-spinor indices. For a quantum statistical ensemble described by the density operator $\widehat{\rho}$, the covariant Wigner function is defined as the ensemble average:
\begin{equation}\label{eq:rsWignerfunction}
W_{\mu\nu}(x,k) = \Tr\left[ \widehat{\rho} \, \widehat{W}_{\mu\nu}(x,k) \right].
\end{equation}
From the Rarita-Schwinger equations of motion \eqref{rseq} and constraints \eqref{rsconstraints}, the free Wigner operator satisfies the Wigner equation:
\begin{equation}
\label{rswignereq}
\left( m - \slashed{k} - \frac{\ii}{2}\slashed{\partial} \right) \widehat{W}_{\mu\nu}(x,k) = 0,
\end{equation}
together with the operator constraints:
\begin{equation}
\label{rswignerconstraints}
\gamma^\mu \widehat{W}_{\mu\nu}(x,k) = 0 = \widehat{W}_{\mu\nu}(x,k) \gamma^\nu.
\end{equation}
Denoting by $\Gamma^I$ a generic combination of gamma matrices with a set of indices $I$, for example $\Gamma^I=\mathbb{1}$, $\Gamma^I=\gamma^\alpha$, or $\Gamma^I=\gamma^{\alpha\beta}$, all local operators bilinear in the fields can be obtained by integrating the Wigner operator over momentum, taking advantage of the identities
\begin{subequations}\label{eq:WignBilin}
\begin{align}
\sum_{A,B}\int\di^4 k\, \Gamma^I_{BA} \widehat{W}_{\mu\nu,AB}(x,k) = &
	\bar{\Psi}_\nu(x) \Gamma^I \Psi_\mu(x),\\
	\int\di^4 k\, k^\alpha \widehat{W}_{\mu\nu}(x,k) = &
	\frac{\ii}{2}\bar{\Psi}_\nu(x) \codev\!\,^\alpha \Psi_\mu(x)
	= \frac{\ii}{2}\bar{\Psi}_\nu(x) \left(\stackrel{\rightarrow}{\partial}\!\,^\alpha - \stackrel{\leftarrow}{\partial}\!\,^\alpha \right)\Psi_\mu(x).
\end{align}
\end{subequations}
Similarly, the thermal expectation values of these operators are obtained from the previous relations by replacing the Wigner operator with the Wigner function.

\paragraph{Hermiticity:}
The covariant Wigner operator is Dirac-Hermitian ($\gamma^0$-Hermitian) under the combined action of Dirac conjugation and vector index transposition:
\begin{equation}
\widehat{W}_{\mu\nu}^\dagger(x, k) = \gamma^0 \widehat{W}_{\nu\mu}(x, k) \gamma^0, \quad
	\widehat{\bar{W}}_{\mu\nu}(x, k) \equiv \gamma^0 \widehat{W}_{\nu\mu}^\dagger(x, k) \gamma^0 = \widehat{W}_{\mu\nu}(x, k).
\end{equation}
This can be proven by taking the Dirac-matrix Hermitian conjugate $^\dagger$ of $\widehat{W}_{\mu\nu}(x, k)$:
\begin{equation}
(\widehat{W}_{\mu\nu}(x, k))^\dagger = \int \frac{d^4 y}{(2\pi)^4} \e^{+i k \cdot y}
	\left[ \psi_\mu\left(x - \frac{y}{2}\right) \bar{\psi}_\nu\left(x + \frac{y}{2}\right) \right]^\dagger.
\end{equation}
We evaluate the operator conjugate product inside the expectation value using the relation $(AB)^\dagger = B^\dagger A^\dagger$:
\begin{equation}
\left[ \psi_\mu\left(x - \frac{y}{2}\right) \bar{\psi}_\nu\left(x + \frac{y}{2}\right) \right]^\dagger = \left[ \bar{\psi}_\nu\left(x + \frac{y}{2}\right) \right]^\dagger \left[ \psi_\mu\left(x - \frac{y}{2}\right) \right]^\dagger.
\end{equation}
Using $\bar{\psi}_\nu^\dagger = (\psi_\nu^\dagger \gamma^0)^\dagger = \gamma^0 \psi_\nu$ and $\psi_\mu^\dagger = \bar{\psi}_\mu \gamma^0$:
\begin{equation}
\left[ \psi_\mu\left(x - \frac{y}{2}\right) \bar{\psi}_\nu\left(x + \frac{y}{2}\right) \right]^\dagger = \gamma^0 \psi_\nu\left(x + \frac{y}{2}\right) \bar{\psi}_\mu\left(x - \frac{y}{2}\right) \gamma^0.
\end{equation}
Substituting this back into the integral yields:
\begin{equation}
(\widehat{W}_{\mu\nu}(x, k))^\dagger = \int \frac{d^4 y}{(2\pi)^4} \e^{+i k \cdot y}
	\gamma^0 \psi_\nu\left(x + \frac{y}{2}\right) \bar{\psi}_\mu\left(x - \frac{y}{2}\right) \gamma^0.
\end{equation}
We then perform the change of integration variable $y \to -y$, under which the measure $d^4 y \to d^4 y$, the exponent $\e^{+i k \cdot y} \to \e^{-i k \cdot y}$, and the field arguments flip: $x + \frac{y}{2} \to x - \frac{y}{2}$ and $x - \frac{y}{2} \to x + \frac{y}{2}$.
Since $\gamma^0$ is a constant matrix independent of $y$, it can be factored out of the integral:
\begin{equation}
(\widehat{W}_{\mu\nu}(x, k))^\dagger = \gamma^0 \left[ \int \frac{d^4 y}{(2\pi)^4} \e^{-i k \cdot y} \left\langle \psi_\nu\left(x - \frac{y}{2}\right) \bar{\psi}_\mu\left(x + \frac{y}{2}\right) \right\rangle \right] \gamma^0.
\end{equation}
Recognizing that the integral inside the brackets is precisely $\widehat{W}_{\nu\mu}(x, k)$, we obtain:
\begin{equation}
\label{eq:WignerDagger}
(\widehat{W}_{\mu\nu}(x, k))^\dagger = \gamma^0 \widehat{W}_{\nu\mu}(x, k) \gamma^0.
\end{equation}
Multiplying both sides by $\gamma^0$ from the left and right (using $(\gamma^0)^2 = \mathbb{1}$) yields the equivalent Dirac self-adjointness condition:
\begin{equation}
\gamma^0 (\widehat{W}_{\mu\nu}(x, k))^\dagger \gamma^0 = \widehat{W}_{\nu\mu}(x, k).
\end{equation}
This property guarantees that physical observables extracted from traces over Clifford algebra structures (such as scalar density, vector current, and spin vector $S^\mu$) evaluate to purely real numbers.

\paragraph{Decomposition into particle, antiparticle, and spacelike parts:}
Substituting the plane-wave expansion \eqref{rsfieldexp} into the Wigner operator definition \eqref{rswignerop}, we obtain:
\begin{subequations}
\label{rswignerdop}
\begin{align}
\label{eq:wignerPart}
\widehat{W}_{\mu\nu}(x,k) = & \sum_{r,s} \frac{1}{(2\pi)^3} \int \frac{\di^3 p}{2E_p} \frac{\di^3 p'}{2E_{p'}}
	\left\{ \e^{-\ii(p-p')\cdot x} \, \delta^4\left(k - \frac{p+p'}{2}\right) \wadf{s}{p'} \waf{r}{p} u_{\mu,r}(p) \bar{u}_{\nu,s}(p') \right.\\
\label{eq:wignerAntiPart}
&- \e^{-\ii(p-p')\cdot x} \, \delta^4\left(k + \frac{p+p'}{2}\right) \wbdf{r}{p'} \wbf{s}{p} v_{\mu,r}(p') \bar{v}_{\nu,s}(p) \\
\label{eq:wignerSpace1}
& - \delta^4\left(k - \frac{p-p'}{2}\right) \left[ \e^{-\ii(p+p')\cdot x} \, \waf{r}{p} \wbf{s}{p'} u_{\mu,r}(p) \bar{v}_{\nu,s}(p') \right. \\
\label{eq:wignerSpace2}
	&\left.\left. + \, \e^{\ii(p+p')\cdot x} \, \wbdf{r}{p'} \wadf{s}{p} v_{\mu,r}(p') \bar{u}_{\nu,s}(p) \right] \right\}.
\end{align}
\end{subequations}
Analogously to the spin-$1/2$ case, the four-momentum $k$ in $\widehat{W}_{\mu\nu}(x,k)$ is not restricted to the mass shell, i.e., $k^2\neq m^2$, even for the free field. However, the Wigner operator can be decomposed into future timelike (particle, Eq.~\eqref{eq:wignerPart}), past timelike (antiparticle, Eq.~\eqref{eq:wignerAntiPart}), and spacelike parts (Eqs.~\eqref{eq:wignerSpace1} and~\eqref{eq:wignerSpace2}):
\begin{equation}
\widehat{W}_{\mu\nu}(x,k) = \widehat{W}_{\mu\nu,+}(x,k) + \widehat{W}_{\mu\nu,-}(x,k) + \widehat{W}_{\mu\nu,S}(x,k),
\end{equation}
where:
\begin{subequations}
\begin{align}
\widehat{W}_{\mu\nu,+}(x,k) &\equiv \widehat{W}_{\mu\nu}(x,k) \, \theta(k^2)\theta(k^0), \\
	\widehat{W}_{\mu\nu,-}(x,k) &\equiv \widehat{W}_{\mu\nu}(x,k) \, \theta(k^2)\theta(-k^0), \\
	\widehat{W}_{\mu\nu,S}(x,k) &\equiv \widehat{W}_{\mu\nu}(x,k) \, \theta(-k^2).
\end{align}
\end{subequations}
Similarly to the Dirac field, and unlike in the scalar case, despite this decomposition, one cannot identify a phase-space distribution function simply by integrating the covariant Wigner function over $k$; however, it is possible to establish an exact relation between the particle density in momentum space and the covariant Wigner function.

\paragraph{Particle density in momentum space:}
The conserved vector current for the Rarita-Schwinger field is defined as
\begin{equation}\label{eq:Vcurrent}
\widehat{j}^\lambda(x) = -\bar{\Psi}_\alpha(x) \gamma^\lambda \Psi^\alpha(x).
\end{equation}
Using the equations of motion~\eqref{eq:EoMAll}, it is straightforward to show that the current is conserved: $\partial_\lambda \widehat{j}^\lambda(x)=0$.
The expectation value of the current is directly related to the Wigner function and is obtained via the trace:
\begin{equation}
j^\lambda(x) = -\langle :  \bar{\Psi}_\alpha(x) \gamma^\lambda \Psi^\alpha(x) :\rangle
	= -\int\di^4 k\, \tr\left[ \gamma^\lambda g^{\alpha\beta} W_{\alpha\beta}(x,k)\right].
\end{equation}
Considering only the particle contribution to the current and employing the expansion \eqref{rswignerdop}, we obtain the mean value
\begin{equation}\label{eq:meanj+}
\begin{split}
j_+^\lambda(x) =& -\int\di^4 k\, \tr\left[ \gamma^\lambda g^{\alpha\beta} W_{\alpha\beta,+}(x,k)\right]\\
	=& \sum_{r,s} \frac{(-1)}{(2\pi)^3} \int \frac{\di^3 p}{2E_p} \frac{\di^3 p'}{2E_{p'}}
	\e^{-\ii(p-p')\cdot x} \langle \wadf{s}{p'} \waf{r}{p} \rangle
	\bar{u}^\alpha_{s}(p') \gamma^\lambda u_{\alpha,r}(p).    
\end{split}
\end{equation}
As mentioned, this expression demonstrates that one cannot factorize the momentum integration because the spinors $u_\alpha(p)$ and $\bar{u}^\alpha(p')$ depend on different momenta, which precludes defining a sensible spin-dependent particle distribution function $f(x, k)_{rs}$. Nevertheless, the covariant Wigner function can still be used to determine the particle density in momentum space.

Notice that, using the spinor equation~\eqref{eq:EoMu}, the particle part of the current is independently conserved, i.e., $\partial_\lambda j_+^\lambda(x)=0$. Since the divergence of $j_+$ vanishes, integrating the particle current~\eqref{eq:meanj+} over an arbitrary 3D spacelike hypersurface yields a constant particle number, provided that boundary fluxes vanish. Consequently, the total number of particles can be obtained by integrating $j^0_+$ over a constant-time hyperplane $t = \text{const.}$, yielding
\begin{equation}
\begin{split}
N =& \int\di^3 x\, j^0_+(x) =
	-\int\di^3 x\int\di^4 k\, \tr\left[ \gamma^0 g^{\alpha\beta} W_{\alpha\beta,+}(x,k)\right]\\
=& \sum_{r,s} \frac{(-1)}{(2\pi)^3} \int \frac{\di^3 p}{2E_p} \frac{\di^3 p'}{2E_{p'}}
	\delta^3(\bm{p}-\bm{p}') \langle \wadf{s}{p'} \waf{r}{p} \rangle
	\bar{u}^\alpha_{s}(p') \gamma^0 u_{\alpha,r}(p)\\
=& -\sum_{r,s} \int \frac{\di^3 p}{4E_p^2} \langle \wadf{s}{p} \waf{r}{p} \rangle
	\bar{u}^\alpha_{s}(p) \gamma^0 u_{\alpha,r}(p)
	= \sum_{r,s} \int \frac{\di^3 p}{4E_p^2} \langle \wadf{s}{p} \waf{r}{p} \rangle
	2E_p\delta_{r,s}\\
=& \int \di^3 p\,\sum_r \frac{\langle \wadf{r}{p} \waf{r}{p} \rangle}{2E_p} ,
\end{split}
\end{equation}
where we used the spinor identity $\bar{u}^\alpha_{s}(p) \gamma^0 u_{\alpha,r}(p)=-2E_p\delta_{r,s}$. As expected, we recover the particle density in momentum space:
\begin{equation}
\frac{\di N}{\di^3 p} = \sum_r \frac{\langle \wadf{r}{p} \waf{r}{p} \rangle}{2E_p}.
\end{equation}
From this relation and the Wigner function result in Eq.~\eqref{eq:wmunuplus}, we obtain the connection between the momentum spectrum and the covariant Wigner function:
\begin{equation}
E_k \frac{\di N}{\di^3 k} = -\int\di k^0 \frac{E_k}{m}\int\di\Sigma\cdot k\,
	\tr\left[g^{\alpha\beta} W_{\alpha\beta,+}(x,k) \right].
\end{equation}
This is easily verified by using Eq.~\eqref{eq:wmunuplus}, which we will prove later,  along with the spinor normalization identity~\eqref{eq:SpinorNorm}, that is $\bar{u}_{\alpha,s}(k)u^\alpha_r(k)=-2m\delta_{r,s}$:
\begin{equation}
\begin{split}
E_k \frac{\di N}{\di^3 k} =& -\int\di k^0 \frac{E_k}{m}\frac{\delta(k^0-E_k)}{4 E_k} \sum_{r,s}
	\tr\left[\bar{u}_{\alpha,s}(k)u^\alpha_r(k)\right] \langle \wadf{s}{k} \waf{r}{k} \rangle\\
= & -\frac{1}{4m} \sum_{r,s} \tr\left[\bar{u}_{\alpha,s}(k)u^\alpha_r(k)\right]
	\langle \wadf{s}{k} \waf{r}{k} \rangle
	=\frac{1}{2} \sum_r  \langle \wadf{r}{k} \waf{r}{k} \rangle.
\end{split}
\end{equation}
Anticipating that the trace of the Wigner function in local thermal equilibrium is given by
\begin{equation}
\tr\left[g^{\alpha\beta} W_{\alpha\beta,+}(x,k)_{\rm LTE} \right]
	=-8m\frac{\delta\left(k^2 - m^2\right)\theta(p_0)}{(2\pi)^3} n_F\left(\beta(x)\cdot k -\zeta(x)\right),
\end{equation}
we obtain the Cooper-Frye formula for a spin-$3/2$ field:
\begin{equation}
E_k \frac{\di N}{\di^3 k} = \frac{4}{(2\pi)^3}\int\di\Sigma\cdot k\, n_F(\beta(x)\cdot k - \zeta(x)),
\end{equation}
where $n_F$ is the Fermi-Dirac distribution function and $\zeta=\beta\mu$, with $\mu$ denoting the chemical potential.

\paragraph{On-shell condition for the integrated Wigner operator:}
An important property of the free-field Wigner operator is that integrating it over a 3D hypersurface places the momentum vector $k$ strictly on-shell. To prove this, take the spacetime directional derivative $k \cdot \partial$ of each component of $\widehat{W}_{\mu\nu}(x,k)$ in Eq.~\eqref{rswignerdop}. For the particle component $\widehat{W}_{\mu\nu,+}(x,k)$, the spatial derivative produces a factor of $-\ii (p - p')_\lambda$:
\begin{equation}
k^\lambda \partial_\lambda \e^{-\ii(p-p')\cdot x} = -\ii k \cdot (p - p')\, \e^{-\ii(p-p')\cdot x}.
\end{equation}
Due to the four-dimensional delta function $\delta^4\left(k - \frac{p+p'}{2}\right)$, the momentum $k$ is constrained to $\frac{p+p'}{2}$. Since both $p$ and $p'$ are on-shell ($p^2 = p'^2 = m^2$), we have:
\begin{equation}
k \cdot (p - p') = \frac{1}{2}(p + p') \cdot (p - p') = \frac{1}{2}(p^2 - p'^2) = \frac{1}{2}(m^2 - m^2) = 0.
\end{equation}
Thus, we obtain:
\begin{equation}
\label{div_free}
k^\lambda \partial_\lambda \widehat{W}_{\mu\nu,+}(x,k) = 0, \qquad k^\lambda \partial_\lambda \widehat{W}_{\mu\nu,-}(x,k) = 0, \qquad k^\lambda \partial_\lambda \widehat{W}_{\mu\nu,S}(x,k) = 0.
\end{equation}
Applying Gauss's divergence theorem in 4D spacetime, it follows that, under suitable boundary conditions (fields vanishing sufficiently rapidly at spatial infinity), the integral of the Wigner operator $\wW_{\mu\nu}(x,k)$ over a spacelike 3D hypersurface $\Sigma$,
\begin{equation}
\widehat{I}_{\mu\nu}(k) = \int_\Sigma \di \Sigma\cdot k \,  \wW_{\mu\nu}(x,k),
\end{equation}
is independent of the choice of hypersurface $\Sigma$.
Therefore, without loss of generality, we may perform the integration over the constant-time hyperplane $t = 0$, where $\di \Sigma_\lambda = (1, 0, 0, 0) \di^3 \mathbf{x}$. The integral simplifies to:
\begin{equation}
\widehat{I}_{\mu\nu}(k) = \int_{t=0} \di \Sigma\cdot k \,  \widehat{W}_{\mu\nu}(x,k) = k^0 \int \di^3 \mathbf{x} \, \widehat{W}_{\mu\nu}(x,k).
\end{equation}
Substituting the explicit plane-wave expansion \eqref{rswignerdop}, the spatial integration yields a three-dimensional Dirac delta function:
\begin{equation}
\int \di^3 \mathbf{x} \, \e^{-\ii (\mathbf{p} - \mathbf{p}') \cdot \mathbf{x}} = (2\pi)^3 \delta^3(\mathbf{p} - \mathbf{p}').
\end{equation}
Enforcing $\mathbf{p}' = \mathbf{p}$ implies $E_{p'} = E_p = \sqrt{|\mathbf{p}|^2 + m^2}$.
For the particle term $\widehat{W}_{\mu\nu,+}$, the delta function becomes $\delta^4(k - p) = \delta(k^0 - E_p) \delta^3(\mathbf{k} - \mathbf{p})$.
Using the relativistic delta identity $2 k^0 \delta(k^2 - m^2) \theta(k^0) = \delta(k^0 - E_k)$, integrating over $\mathbf{p}$ gives:
\begin{align}
\widehat{I}_{\mu\nu}(k) = 
	\int_{t=0} \di \Sigma\cdot k \,  \widehat{W}_{\mu\nu}(x,k) = & \frac{1}{2} \delta(k^2 - m^2) \sum_{r,s} \left[ \theta(k^0) \wadf{s}{k} \waf{r}{k} u_{\mu,r}(k) \bar{u}_{\nu,s}(k) \right. \nonumber \\
& \left. + \theta(-k^0) \wbdf{r}{-k} \wbf{s}{-k} v_{\mu,r}(-k) \bar{v}_{\nu,s}(-k) \right],
\label{rswignerint_onshell}
\end{align}
where cross-terms vanish identically due to the factor $k^0 \delta(k^0)$. This establishes that the integral of the Wigner operator over any spacelike 3D hypersurface $\Sigma$ is proportional to $\delta(k^2 - m^2)$, placing $k$ strictly on the mass shell after integration.
From Eq.~\eqref{rswignerint_onshell}, the integrated particle and antiparticle Wigner operators, respectively, reduce to:
\begin{subequations}
\begin{align}\label{eq:wmunuplus}
\int \di k^0 \int \di \Sigma\cdot k \,  \widehat{W}_{\mu\nu,+}(x,k) &= \frac{1}{2 E_k} \int \di k^0\delta(k^0 - E_k) \widehat{w}_{\mu\nu,+}(k)=\frac{1}{2 E_k} \widehat{w}_{\mu\nu,+}(k), \\
	\int \di k^0\int \di \Sigma\cdot k \,  \widehat{W}_{\mu\nu,-}(x,k) &= \frac{1}{2 E_k} \int \di k^0\delta(k^0 + E_k) \widehat{w}_{\mu\nu,-}(k)=\frac{1}{2 E_k} \widehat{w}_{\mu\nu,-}(k),
\end{align}    
\end{subequations}
where the Wigner operators reduced on-shell are defined as:
\begin{subequations}
\label{reduced_rs_wigner}
\begin{align}
\widehat{w}_{\mu\nu,+}(k) =& \frac{1}{2} \sum_{r,s} \wadf{s}{k} \waf{r}{k} u_{\mu,r}(k) \bar{u}_{\nu,s}(k),\\
\widehat{w}_{\mu\nu,-}(k) =& \frac{1}{2} \sum_{r,s} \wbdf{r}{-k} \wbf{s}{-k} v_{\mu,r}(-k) \bar{v}_{\nu,s}(-k).    
\end{align}
\end{subequations}
It is convenient, at this point, to derive several identities that will be used below.
From the equation of motion for the spinors~\eqref{eq:EoMu}, it immediately follows that
\begin{equation}
(\slashed{k}-m)\widehat{w}_{\mu\nu,+}(k) = \widehat{w}_{\mu\nu,+}(k)(\slashed{k}-m) = 0,
\end{equation}
or, equivalently,
\begin{equation}
\int \di \Sigma\cdot k \, \, \slashed{k}\widehat{W}_{\mu\nu,+}(x,k)
	= \int \di \Sigma\cdot k \, \, \widehat{W}_{\mu\nu,+}(x,k)\slashed{k}
	= \int \di \Sigma\cdot k \, \, m\,\widehat{W}_{\mu\nu,+}(x,k).
\label{eq:pslash_int}
\end{equation}
Similarly, from the orthogonality of the spinors~\eqref{eq:EoMu}, it follows that, after the momentum $k$ is placed on-shell by the integration, the free-field Wigner operator~\eqref{rswignerdop} satisfies:
\begin{equation}
\label{eq:WPerpMom}
\int \di \Sigma\cdot k \, k^\mu \widehat{W}_{\mu\nu,+}(x,k) =
    \int \di \Sigma\cdot k \, \widehat{W}_{\mu\nu,+}(x,k) k^\nu = 0.
\end{equation}
Lastly, taking advantage of the previous identities \eqref{rswignerconstraints}, \eqref{eq:pslash_int} and \eqref{eq:WPerpMom}, the explicit projector expression \eqref{projection_def} and the well known Clifford anticommutator $\{\gamma_\mu,\gamma_\nu\}=2 g_{\mu\nu}$, it is straightforward to prove that
\begin{subequations}
\label{eq:WProjector}
\begin{align}
\int \di \Sigma\cdot k \, (\slashed{k} + m)P^{\lambda\mu}(k)\widehat{W}_{\mu\nu,+}(x,k)
	=& 2m \int \di \Sigma\cdot k \, \, \widehat{W}^\lambda_{\ \nu,+}(x,k),\\
\int \di \Sigma\cdot k \, \widehat{W}_{\mu\nu,+}(x,k)(\slashed{k} + m)P^{\nu\rho}(k)
	=& 2m\int \di \Sigma\cdot k \, \widehat{W}_\mu^{\ \rho,+}(x,k).
\end{align}
\end{subequations}
%

\paragraph{Clifford decomposition:}
For the Rarita-Schwinger field, as in the Dirac case, the $4 \times 4$ matrix structure of the Wigner function $W_{\mu\nu}(x,p)$ (and likewise for the corresponding operator) can be decomposed in terms of the 16 generators of the Clifford algebra $C\ell_{1,3}$:
\begin{equation}
\label{rs_clifford}
W_{\mu\nu}(x,p) = \frac{1}{4} \left[ \mathcal{F}_{\mu\nu} + \ii \gamma^5 \mathcal{P}_{\mu\nu} + \gamma^\alpha \mathcal{V}_{\alpha, \mu\nu} + \gamma^5 \gamma^\alpha \mathcal{A}_{\alpha, \mu\nu} + \Sigma^{\alpha\beta} \mathcal{S}_{\alpha\beta, \mu\nu} \right],
\end{equation}
where $\Sigma^{\alpha\beta} = \frac{\ii}{4}[\gamma^\alpha, \gamma^\beta]$ and all Wigner function components are evaluated at $(x, p)$. The above decomposition applies separately to the particle, antiparticle, and spacelike parts of the Wigner function as well.
To extract the corresponding components, it suffices to take appropriate traces of the Wigner function:
\begin{subequations}
\begin{align}
\mathcal{F}_{\mu\nu}(x,p) =& \tr_4\left[ W_{\mu\nu}(x,p)\right],&
    \mathcal{P}_{\mu\nu}(x,p) =& -\ii\tr_4\left[\gamma^5 W_{\mu\nu}(x,p)\right],\\
\mathcal{V}^\alpha_{\mu\nu}(x,p) =& \tr_4\left[ \gamma^\alpha W_{\mu\nu}(x,p)\right],&
	\mathcal{A}^\alpha_{\mu\nu}(x,p) =& \tr_4\left[ \gamma^\alpha  \gamma^5 W_{\mu\nu}(x,p)\right],\\
\mathcal{S}^{\alpha\beta}_{\mu\nu}(x,p) =& 2\tr_4\left[ \Sigma^{\alpha\beta} W_{\mu\nu}(x,p)\right].
\end{align}
\end{subequations}
Following the standard Dirac case~\cite{Das:2022azr}, it is straightforward to show that the Wigner equation~\eqref{rswignereq} decomposes into the set of equations
\begin{subequations}
\begin{align}
p_\alpha \mathcal{V}^\alpha_{\mu\nu} =& m \mathcal{F}_{\mu\nu},&
	\partial_\alpha \mathcal{V}^\alpha_{\mu\nu} =&0,\\
\label{eq:WEqpdotA}
p_{\alpha }\mathcal{A}^{\alpha}_{\mu \nu} =& 0,&
	\frac{1}{2}\partial^\alpha \mathcal{P}_{\mu \nu}
	-\frac{1}{2}\epsilon^{\alpha\beta\gamma\lambda}\mathcal{S}_{\beta\gamma,\mu \nu} p_\lambda
	=& m\,\mathcal{A}^\alpha_{\mu\nu},\\
\frac{1}{2}\partial^{[\alpha}\mathcal{V}^{\beta]}_{\mu\nu}-m\mathcal{S}^{\alpha\beta}_{\mu\nu} - \epsilon^{\lambda\rho\alpha\beta}p_\lambda\mathcal{A}_{\rho,\mu\nu} =&0,&
	\partial_\alpha \mathcal{A}^\alpha_{\mu\nu} + 2m\mathcal{P}_{\mu\nu} = & 0,\\
\label{eq:WEqDFandS}
\frac{1}{2}\partial_\beta\mathcal{S}^{\alpha\beta}_{\mu\nu} + p^\alpha \mathcal{F}_{\mu\nu} -m\mathcal{V}^\alpha_{\mu\nu}=&0, &
	\partial^\alpha \mathcal{F}_{\mu\nu} + 2 p_\lambda \mathcal{S}^{\lambda\alpha}_{\mu\nu} =&0,\\
\frac{1}{2}\epsilon^{\lambda\rho\alpha\beta}\partial_\lambda \mathcal{A}_{\rho,\mu\nu}+p^{[\alpha}\mathcal{V}^{\beta]}_{\mu\nu}=&0,&
	\frac{1}{2}\epsilon^{\lambda\rho\beta\alpha}\partial_\lambda\mathcal{S}_{\rho\beta,\mu\nu}+2p^\alpha\mathcal{P}_{\mu\nu} =&0.
\end{align}    
\end{subequations}
Owing to Eq.~\eqref{eq:WPerpMom}, all Clifford components of the Wigner function are orthogonal to the momentum $k$.
Similarly, the constraint equations~\eqref{rswignerconstraints} can be decomposed into their respective Clifford components.

Furthermore, we prove several properties that will be useful in subsequent calculations.
Multiplying identity~\eqref{eq:pslash_int} from the left by $\gamma^5$ and taking the trace, the left-hand side yields $k_\alpha \mathcal{A}^\alpha_{\mu\nu}(x,k)$, which vanishes according to Eq.~\eqref{eq:WEqpdotA}. On the right-hand side, the trace selects the pseudoscalar component of the Wigner function; hence, we conclude that
\begin{equation}\label{eq:WPIntegrated}
\int \di \Sigma\cdot k\, \mathcal{P}_{\mu \nu,+}(x,k) = 0.
\end{equation}
Similarly, multiplying Eq.~\eqref{eq:pslash_int} by $\gamma^\alpha \gamma^5$, tracing, and using Eq.~\eqref{eq:WPIntegrated}, we obtain
\begin{equation}
\int \di \Sigma\cdot k\, \mathcal{A}^{\alpha}_{\mu\nu, +}(x, k)=
	-\frac{1}{2m}\epsilon^{\alpha\tau\beta\rho} k_{\rho}
    \int \di \Sigma\cdot k\,  \mathcal{S}_{\tau\beta;\mu \nu}(x,k) .
\end{equation}
It is also useful to determine the symmetrization and antisymmetrization of the scalar component of the Wigner function. Consider the complex conjugate of the scalar part:
\begin{equation}
\mathcal{F}^*_{\mu\nu,+}(x,p) = \tr_4\left[ W_{\mu\nu,+}(x,p)\right]^*
	= \tr_4\left[ W_{\mu\nu,+}^\dagger(x,p)\right]
	= \tr_4\left[ \gamma^ 0 W_{\nu\mu,+}(x,p) \gamma^0\right],
\end{equation}
where we used Eq.~\eqref{eq:WignerDagger}. Exploiting the cyclicity of the trace and $(\gamma^0)^2=\mathbb{1}$, we obtain
\begin{equation}
\mathcal{F}^*_{\mu\nu,+}(x,p) = \mathcal{F}_{\nu\mu,+}(x,p).
\end{equation}
Hence, the symmetric and antisymmetric parts of the scalar component correspond, respectively, to its real and imaginary parts:
\begin{subequations}
\begin{align}
\mathcal{F}_{(\mu\nu),+}(x,p) =&
	\frac{1}{2}\left( \mathcal{F}_{\mu\nu,+}(x,p) + \mathcal{F}_{\nu\mu,+}(x,p)\right) = 
	\text{Re}\left[ \mathcal{F}_{\mu\nu,+}(x,p)\right],\\
\mathcal{F}_{[\mu\nu],+}(x,p) =&
	\frac{1}{2}\left( \mathcal{F}_{\mu\nu,+}(x,p) - \mathcal{F}_{\nu\mu,+}(x,p)\right) =
	\ii\text{Im}\left[ \mathcal{F}_{\mu\nu,+}(x,p)\right].
\label{eq:ImF}
\end{align}    
\end{subequations}
%

\section{The Rarita-Schwinger spin polarization}
\label{sec:RSSpin}
In this section, we derive an \textit{exact} relation connecting the Wigner function and the mean spin vector for the Rarita-Schwinger spin-$3/2$ fermion. First, we give the general definitions of the spin density matrix and the mean spin vector in a relativistic quantum framework. Then, we express the spin density matrix in terms of the Wigner function defined in the previous section and derive the spin polarization formula.

\subsection{The mean spin vector and the spin density matrix}
\label{subsec:SpinIntro}
In relativistic quantum mechanics, the spin vector of a single massive particle is defined via the Pauli-Lubanski operator:
\begin{equation}\label{luba}
 \widehat S^\mu = -\frac{1}{2m} \epsilon^{\mu \nu \rho \sigma} \wJ_{\nu \rho} 
 \widehat P_\sigma ,
\end{equation} 
where $\wJ_{\nu \rho}$ and $\wP_\sigma$ denote the angular momentum-boost operator and the energy-momentum operator, respectively. This operator is orthogonal to the energy-momentum four-vector, $\widehat S \cdot \widehat P = 0$, and obeys the angular momentum algebra:
\begin{equation}
\left[\widehat S_\mu, \widehat P_\nu\right] = 0,\quad
\left[\widehat S_\mu, \widehat S_\nu\right] = -i \epsilon_{\mu \nu \rho \sigma} \widehat S^\rho \widehat P^\sigma .
\end{equation}
Consequently, an eigenstate $\ket{p}$ of $\widehat P$, representing a momentum state observed by a detector, is likewise an eigenvector of $\widehat S^\mu$. For any given momentum $p$, let $\widehat S^\mu(p)$ represent the restriction of $\widehat S^\mu$ to the eigenspace spanned by $\ket{p}$. Owing to the relation $\widehat S(p) \cdot p = 0$, there are three independent operators $\widehat S_i(p)$ that satisfy an $\text{SU}(2)$ algebra and serve as generators of the \emph{little group} for massive particles. 
Choosing one such generator, say $\widehat S_3(p)$, allows us to construct a set of mutually commuting operators $\{\wP,\, \widehat S_3(p),\, \widehat S^2\}$. A state $\ket{p,s}$ corresponds to a joint eigenvector with respective eigenvalues $\{p,\, S_z,\, S(S+1)\}$, where $S_z$ is the spin projection along the $z$-axis (the eigenvalue of $\widehat S_3(p)$) and $S$ denotes the particle's intrinsic spin. 
To streamline notation, we hereafter abbreviate $S_z$ as $r$, $s$, or $t$.

Now, consider a single relativistic quantum particle belonging to a statistical ensemble described by the density operator $\wrho$. The mean spin vector, denoted as $S^\mu$, is given by the expectation value of operator~\eqref{luba}:
\begin{equation}\label{spinvector}
  S^\mu = \Tr (\, \wrho \, \widehat S^\mu \,) .
\end{equation} 
Because experimental measurements select a definite momentum state, we project operator~\eqref{luba} onto the four-momentum $p$ subspace, yielding the average spin vector for a particle with momentum $p$:
\begin{equation}\label{means}
  S^\mu(p) = \Tr[ \wrho \widehat S^\mu(p)] = \Tr [ \widehat \Theta(p) \widehat S^\mu(p)],
\end{equation}
where we have introduced the \emph{spin density operator} $\widehat \Theta(p)$. Evaluating matrix elements in the eigenbasis of $\widehat S_3(p)$ defines the spin density matrix:
\begin{equation}\label{sdm}
 \Theta(p)_{rs} \equiv \bra{p,r} \wrho \, \ket{p,s}
    = \bra{p,r} \widehat \Theta(p)  \ket{p,s} .
\end{equation} 
The spin density matrix completely characterizes the particle's spin state, inheriting the properties of being Hermitian, positive definite, and normalized to unit trace from $\wrho$. To express the  expectation value of the spin vector strictly in terms of spin matrix elements, we first expand  the trace in~\eqref{means}:
\begin{equation}\label{meansp2}
S^\mu(p) = \sum_r \bra{p,r} \widehat S^\mu(p) \widehat \Theta(p) \ket{p,r}
  = \sum_{r,s} \bra{p,r} \widehat S^\mu(p) \ket{p,s}\bra{p,s}  \widehat \Theta(p) \ket{p,r}.
\end{equation} 
The matrix elements of the Pauli-Lubanski vector can be derived by exploiting the fact that $\widehat S^\mu(p)$ generates the little group $\text{SU}(2)$ algebra within the subspace spanned by $\ket{p}$, alongside the fundamental properties of the Pauli-Lubanski vector. Explicitly, one finds~\cite{Becattini:2020sww, Becattini:2024uha}:
\begin{equation}
\bra{p,r} \widehat S^\mu(p) \ket{p,s} = -\frac{1}{2m} \epsilon^{\mu\nu\rho\tau} p_\tau
 \left[D^S([p])^{-1} D^S( J_{\nu\rho}) D^S([p])\right]_{rs} ,
\end{equation} 
where $D^S(J_{\lambda\nu})$ are the angular momentum generator matrices in the spin-$S$ representation, $[p]$ denotes the \emph{standard Lorentz transformation} mapping the rest-frame vector $p_0 = (m,0,0,0)$ to  the four-momentum $p$, and $D^S([p])$ is its corresponding matrix representation. The mean spin vector is, thus, expressed as:
\begin{equation}\label{eq:SpinVectorSRepresentation}
S^\mu(p) = -\frac{1}{2m} \epsilon^{\mu\nu\rho\tau} p_\tau
      \tr \left(D^S([p])^{-1} D^S( J_{\nu\rho}) D^S([p]) \Theta(p)\right) .
\end{equation} 
Starting from this expression and considering a global thermal equilibrium with constant thermal vorticity, one obtains~\cite{Becattini:2020sww, Becattini:2024uha} that the mean spin polarization is given by Eq.~\eqref{eq:SVectorAnySpin}.

Within quantum field theory, the spin operator~\eqref{luba} and the density operator $\wrho$ remain well-defined notions, though the concept of a single-particle state with sharp momentum requires refinement. A rigorous definition of a single particle applies primarily to free or weakly interacting theories, such as QCD in the perturbative regime. In these contexts, a single-particle state with momentum $p$ and spin $s$ is generated by acting on the vacuum with the creation operator $\wadf{s}{p}$, i.e., $\ket{p,s}=\wadf{s}{p}\ket{0}$. 
In quantum field theory, the mean spin vector for a particle with momentum $p$ is still given by Eq.~\eqref{means}, but with the spin density matrix redefined as:
\begin{equation}\label{spindens}
\Theta(p)_{r s} = \frac{\Tr [\wrho \, \wadf{s}{p} \waf{r}{p} ]}
{\sum_{t} \Tr [\wrho \, \wadf{t}{p} \waf{t}{p} ]}.
\end{equation}
For the spin-$1/2$ Dirac field, the spin density matrix \eqref{spindens} can be expressed in terms of the covariant Dirac Wigner function; then, using Eq.~\eqref{eq:SpinVectorSRepresentation} one eventually obtains~\cite{Becattini:2020sww, Becattini:2024uha} the Dirac field mean spin polarization formula~\eqref{eq:SFormulaDirac}. Following the same procedure, we obtain the Rarita-Schwinger field mean spin polarization formula.

First, we notice that the identity in Eq.~\eqref{reduced_rs_wigner} already establishes a relation between the Wigner operator and the creation and annihilation operators appearing in the density matrix~\eqref{spindens}. To properly isolate the needed operator from~\eqref{reduced_rs_wigner}, we take advantage of the spinor identities by multiplying  $\widehat{w}_{\iota\zeta,+}(p)$ expressed as in Eq.~\eqref{reduced_rs_wigner} with $\bar{u}_r^\iota(p)$ on the left and $u_s^\zeta(p)$ on the right, obtaining:
\begin{equation}
\bar{u}_r^\iota(p) \widehat{w}_{\iota\zeta,+}(p) u_s^\zeta(p) = \frac{1}{2} \sum_{r',s'} \wadf{{s'}}{p} \waf{{r'}}{p} \left[ \bar{u}_r^\iota(p) u_{\iota,r'}(p) \right] \left[ \bar{u}_{\zeta,s'}(p) u_s^\zeta(p) \right].
\end{equation}
Using the spinor orthogonality condition $\bar{u}_r^\iota(p) u_{\iota,r'}(p) = -2m \delta_{rr'}$, we get:
\begin{equation}
\bar{u}_r^\iota(p) \widehat{w}_{\iota\zeta,+}(p) u_s^\zeta(p) = \frac{1}{2} \sum_{r',s'} \wadf{{s'}}{p} \waf{{r'}}{p} (-2m \delta_{rr'}) (-2m \delta_{ss'}) = 2 m^2 \wadf{s}{p} \waf{r}{p}.
\end{equation}
Finally, taking the ensemble average $w_{\iota\zeta,+}(p) = \Tr[\widehat{\rho} \, \widehat{w}_{\iota\zeta,+}(p)]$ and substituting into~\eqref{spindens}, we arrive at the relation:
\begin{equation}
\label{rs_theta_wigner}
\Theta(p)_{rs} = \frac{\int\di p^0\int \di \Sigma \cdot p \; \bar{u}_r^\iota(p) W_{\iota\zeta,+}(x,p) u_s^\zeta(p)}{\sum_{t} \int\di p^0\int \di \Sigma \cdot p \; \bar{u}_t^\iota(p) W_{\iota\zeta,+}(x,p) u_t^\zeta(p)}.
\end{equation}
%

\subsection{The Rarita-Schwinger spin polarization formula}
\label{subsec:RSSpin}
We are now in a position to obtain the spin polarization formula for the spin $3/2$ fermion: simply plug Eq.~\eqref{rs_theta_wigner} in Eq.~\eqref{eq:SpinVectorSRepresentation} for the spin $3/2$, that reads:
\begin{equation}\label{eq:SpinVector3/2Representation}
S^\mu(p) = -\frac{1}{2m} \epsilon^{\mu\alpha\beta\sigma} p_\sigma
      \tr \left(D^{3/2}([p])^{-1} D^{3/2}( J_{\alpha\beta}) D^{3/2}([p]) \Theta(p)\right) .
\end{equation} 
However, while $D^{3/2}$ belongs to the $(0,3/2)$ representation, that is the 4-dimensional irreducible unitary representation of the proper orthochronous Lorentz group SO$(1,3)$, the Wigner function is given in terms of the Rarita-Schwinger vector-spinor field operator $\psi^\mu_A(x)$. The field carrying composite Lorentz vector ($\mu \in \{0,1,2,3\}$) and Dirac spinor indices ($A \in \{1,2,3,4\}$) lives in a 16-dimensional representation space:
\begin{equation}
V_{16} = \left(\frac{1}{2}, \frac{1}{2}\right) \otimes \left[\left(\frac{1}{2},0\right) \oplus \left(0, \frac{1}{2}\right)\right]
= \left(1, \frac{1}{2}\right) \oplus \left(0, \frac{1}{2}\right) = \frac{3}{2} \oplus \frac{1}{2} \oplus \frac{1}{2}\, .
\end{equation}
The calculation of the trace in Eq.~\eqref{eq:SpinVector3/2Representation} would be greatly simplified if we could establish the relations between the spin-$3/2$ irreducible representations of Lorentz transformations and their generators and the 16-dimensional vector-spinor field representations used for the Rarita-Schwinger field. The vector-spinor wavefunctions $u_s^\mu(p)$ act as embedding or intertwiner matrices connecting the $(0,3/2)$ and $V_{16}$ representations~\cite{Weinberg:1995mt}. Taking advantage of the 16D finite-dimensional representation of Lorentz transformations for the Rarita-Schwinger field, it is possible to show (see Appendix~\ref{app:AppIntertwiner}) that the $(0,3/2)$ representation matrices in Eq.~\eqref{eq:SpinVector3/2Representation} are related to the 16-dimensional representation of Lorentz transformation
\begin{equation}\label{eq:16DLorentzJ}
\mathcal{J}_{\alpha\beta, \ \ \delta}^{\ \ \gamma} = \frac{1}{2}\delta^\gamma_\delta \sigma_{\alpha\beta} 
    + \ii\left(\delta^\gamma_\alpha g_{\beta\delta} - \delta^\gamma_\beta g_{\alpha\delta}\right)\mathbb{1},
\end{equation}
where $\sigma_{\alpha\beta}=\frac{\ii}{2}[\gamma_\alpha,\,\gamma_\beta]$, see also Eq.~\eqref{eq:16LorentzTransField}, through the identity
\begin{equation}\label{eq:intertwinerLubansky}
\left[ D^{3/2}([p])^{-1} D^{3/2}(J_{\alpha\beta}) D^{3/2}([p]) \right]_{sr} = -\frac{1}{2m} \bar{u}_s^\gamma(p) \mathcal{J}_{\alpha\beta, \gamma \delta} u_r^\delta(p).
\end{equation}
Now, using Eqs. \eqref{rs_theta_wigner} and \eqref{eq:intertwinerLubansky} in the spin polarization \eqref{eq:SpinVector3/2Representation}, we obtain
\begin{equation}
S^{\mu}(p) =  \frac{1}{4m^2} \epsilon^{\mu\alpha\beta\sigma} p_\sigma\frac{\int\di p^0\int \di \Sigma \cdot p \; \tr_4 \left[ \bar{u}_\gamma(p) \mathcal{J}_{\alpha\beta, \ \delta}^{\ \ \gamma} \, u^\delta(p) \bar{u}^\iota(p) W_{\iota\zeta,+}(x,p) u^\zeta(p) \right]}{\sum_{t} \int\di p^0\int \di \Sigma \cdot p \; \bar{u_t}^\iota(p) W_{\iota\zeta,+}(x,p) u_t^\zeta(p)},
\end{equation}
which is in a suitable form to be further simplified.
Exploiting the cyclic property of a trace in both the numerator and denominator, and subsequently employing the completeness relation \eqref{spinsum_rs}, we arrive at
\begin{align*}
S^\mu(p) &=  \frac{1}{4m^2} \epsilon^{\mu\alpha\beta\sigma} p_\sigma\frac{\int\di p^0\int \di \Sigma \cdot p \; \tr_4 \left[  \mathcal{J}_{\alpha\beta, \ \delta}^{\ \ \gamma}\, u^\delta(p) \bar{u}^\iota(p) W_{\iota\zeta,+}(x,p)u^\zeta(p)\bar{u}_\gamma(p) \right]}{\int\di p^0\int \di \Sigma \cdot p \; \tr\left[u^\zeta(p)\bar{u}^\iota(p) W_{\iota\zeta,+}(x,p)\right] } \nonumber \\
&=-\frac{1}{4m^2} \epsilon^{\mu\alpha\beta\sigma} p_\sigma\frac{\int\di p^0\int \di \Sigma \cdot p \; \tr_4 \left[\mathcal{J}_{\alpha\beta, \  \delta}^{\ \ \gamma} \, (\slashed{p}+m) P^{\delta \iota}(p) W_{\iota\zeta,+}(x,p)(\slashed{p}+m)P^{\zeta}_{\ \ \gamma}(p)  \right]}{ \int\di p^0\int \di \Sigma \cdot p \; \tr \left[(\slashed{p}+m) P^{\iota\zeta}(p) W_{\iota\zeta,+}(x,p)\right] } .
\end{align*}
Then, using the previously derived identities in Eq. \eqref{eq:WProjector} in both the numerator and the denominator the expression simplifies into
\begin{equation}
S^\mu(p) = -\frac{1}{2m} \epsilon^{\mu\alpha\beta\sigma} p_\sigma\frac{\int\di p^0\int \di \Sigma \cdot p \; \tr_4 \left[\mathcal{J}_{\alpha\beta, \  \delta}^{\ \ \gamma} \, W^\delta_{\ \gamma,+}(x,p)\right]}{ \int\di p^0\int \di \Sigma \cdot p \; \tr \left[ W^\iota_{\ \iota,+}(x,p)\right] } .
\end{equation}
Plugging in the Lorentz transformation representation \eqref{eq:16DLorentzJ}, taking advantage of the identity $\frac{1}{2}\epsilon^{\mu\alpha\beta\sigma} \sigma_{\alpha \beta}=-\ii \gamma^5 \sigma^{\mu\sigma}$ and reminding the notation $A_{[\mu,\nu]}=(A_{\mu\nu}-A_{\nu\mu})/2$, we can reduce the numerator to the trace of Wigner function and gamma matrices as
\begin{subequations}
\begin{align}
S^\mu(p) &= -\frac{1}{2m} \epsilon^{\mu\alpha\beta\sigma} p_\sigma\frac{\int\di p^0\int \di \Sigma \cdot p \; \tr_4 \left[
    \left(\frac{1}{2}\sigma_{\alpha\beta}\, \delta^\gamma_\delta
        + \ii \, \delta^\gamma_\alpha \,  g_{\beta\delta} - \ii \, \delta^\gamma_\beta \, g_{\alpha\delta}\right)
    W^\delta_{\ \gamma,+}(x,p)\right]}{\int\di p^0\int \di \Sigma \cdot p \; \tr \left[ W^\iota_{\ \iota,+}(x,p)\right] }\\
&= \frac{1}{2m}  p_\sigma\frac{\int\di p^0\int \di \Sigma \cdot p \; \tr_4 \left[
    \ii\gamma^5\sigma^{\mu\sigma} W^\delta_{\ \delta,+}(x,p) + 2\ii\epsilon^{\mu\alpha\beta\sigma} W_{[\alpha\beta],+}(x,p)
    \right]}{ \int\di p^0\int \di \Sigma \cdot p \; \tr \left[ W^\iota_{\ \iota,+}(x,p)\right] }\\
\label{eq:SPolAlmostFinal}
&= \frac{1}{2m}  p_\sigma\frac{\int\di p^0\int \di \Sigma \cdot p \; \tr_4 \left[
    \ii\gamma^5\sigma^{\mu\sigma} W^\delta_{\ \delta,+}(x,p)\right] + 2\ii\epsilon^{\mu\alpha\beta\sigma} \mathcal{F}_{\alpha\beta,+}(x,p)
    }{ \int\di p^0\int \di \Sigma \cdot p \; \mathcal{F}^\iota_{\ \iota,+}(x,p)},
\end{align}
\end{subequations}
where we recognized the scalar part of the Wigner function in the Clifford decomposition, see Eq. \eqref{rs_clifford}.
The trace in the numerator can be simplified by expanding the definition of the Pauli matrix and then using $\slashed{p}\gamma^\mu=2p^\mu-\gamma^\mu\slashed{p}$ and Eqs. \eqref{eq:pslash_int}, \eqref{rs_clifford} and \eqref{eq:WPIntegrated}:
\begin{equation}
\begin{split}
\ p_\sigma&\, \int \di \Sigma \cdot p\,  \tr \left[ \ii \gamma^5\sigma^{\mu\sigma} W^{\delta}_{\ \delta,+}(x,p)\right] = -\frac{1}{2}\int \di \Sigma \cdot p \  \tr \left[ \gamma^5 (\gamma^{\mu} \slashed{p}-\slashed{p}\gamma^{\mu})W^{\delta}_{\ \delta,+}(x,p)\right] \\
&= -\int \di \Sigma \cdot p \  \tr \left[ \gamma^5 (\gamma^{\mu} \slashed{p}-p^\mu)W^{\delta}_{\ \delta,+}(x,p)\right] \\
&= -m\int \di \Sigma \cdot p \  \tr \left[ \gamma^5 \gamma^{\mu}W^{\delta}_{\delta,+}(x,p) \right]
    +\int \di \Sigma \cdot p\ p^{\mu} \ \tr \left[ \gamma^5 W^{\delta}_{\delta,+}(x,p) \right]\\
&= m\int \di \Sigma \cdot p \  \tr \left[ \gamma^{\mu} \gamma^5 W^{\delta}_{\delta,+}(x,p) \right]
    + \ii p^\mu\int \di \Sigma \cdot p  \  \mathcal{P}^{\delta}_{\delta,+}(x,p)\\
&= m\int \di \Sigma \cdot p \ \mathcal{A}^{\mu,\delta}_{\hphantom{\mu,\delta} \delta,+}(x,p)
    = m\int \di \Sigma \cdot p \ g^{\alpha\beta}\mathcal{A}^\mu_{\alpha\beta,+}(x,p).
\end{split}
\end{equation}
Finally, replacing this expression back in Eq. \eqref{eq:SPolAlmostFinal}, we obtain the result
\begin{equation}
\label{eq:SpinPol3/2}
S^\mu(p) = \frac{1}{2}\frac{\int\di p^0\int\di\Sigma\cdot p\,\left[g^{\alpha\beta}\mathcal{A}^\mu_{\alpha\beta,+}(x,p)+2\ii\epsilon^{\mu\alpha\beta\sigma}\frac{p_\sigma}{m}\mathcal{F}_{\alpha\beta,+}(x,p)\right]}{\int\di p^0\int\di\Sigma\cdot p\, \mathcal{F}^\alpha_{\ \alpha,+}(x,p)}.
\end{equation}
Compared with the Dirac field formula \eqref{eq:SFormulaDirac}, there is an extra contribution related to the scalar part of the Clifford decomposition, obtained by taking the dual of the vector indices of the Wigner function, which is obviously not possible for the Dirac field. Given the explicit appearance of the imaginary unit, it might seem that the extra contribution could be imaginary. On the contrary, due to the property \eqref{eq:ImF}, the spin polarization vector is guaranteed to be a real valued quantity.
For future convenience, we refer to $\mathcal{A}^{\mu\,\alpha}_{\alpha,+}$ as the \textit{direct} axial part of the Wigner function and to the axial vector
\begin{equation}
\label{eq:DefAxialF5}
\mathcal{F}^\mu_{5,+}(x,p)= 2\ii\epsilon^{\mu\alpha\beta\sigma}\frac{p_\sigma}{m}\mathcal{F}_{\alpha\beta,+}(x,p) ,
\end{equation}
as the \textit{dual} axial part of Wigner function, such that the mean spin polarization vector can be written as
\begin{equation}
\label{eq:SpinPol3/2Short}
S^\mu(p) = \frac{1}{2}\frac{\int\di p^0\int\di\Sigma\cdot p\,\left[g^{\alpha\beta}\mathcal{A}^\mu_{\alpha\beta,+}(x,p)+\mathcal{F}^\mu_{5,+}(x,p)\right]}{\int\di p^0\int\di\Sigma\cdot p\, \mathcal{F}^\alpha_{\ \alpha,+}(x,p)}.
\end{equation}

\section{Spin polarization at local thermal equilibrium}
\label{sec:SpinLTE}
In the previous sections, we established the relation between the spin density matrix, the mean spin vector, and the covariant Wigner function. Evaluating the spin polarization, therefore, requires determining the Wigner function. The latter depends on the statistical operator $\wrho$ [see, e.g., Eq.~\eqref{eq:rsWignerfunction}], which must be specified to calculate the spin polarization. In this section, we formulate the statistical operator describing heavy-ion collisions and derive the expression relating the spin polarization of spin-$3/2$ particles measured in heavy-ion collisions to the hydrodynamic fields of the fluid, thereby providing the key theoretical input to extract spin polarization predictions from relativistic hydrodynamics or transport models.

In relativistic heavy-ion collisions, strongly interacting matter under extremely high temperatures—known as the quark-gluon plasma (QGP)—is produced \citep{Collins:1974ky,Shuryak:1977ut,Busza:2018rrf}. 
After a brief pre-equilibrium stage, the system is assumed to attain local thermodynamic equilibrium (LTE), and its subsequent evolution is effectively described by relativistic hydrodynamics \citep{Stoecker:1986ci,Rischke:1995ir,Rischke:1995mt,Jaiswal:2020hvk,Arslandok:2023utm}. 
Physical quantities are then characterized by slowly varying macroscopic fields, such as flow velocity, temperature, and chemical potential. The non-equilibrium statistical operator approach~\citep{Zubarev:1966,Zubarev:1979afm,Zubarev:1989su,vanWeert1982} in the modern Becattini formulation \citep{Becattini:2014yxa,Becattini:2019dxo,Buzzegoli:2020ycf} provides a rigorous framework to determine physical observables in a relativistic fluid within quantum statistical field theory at local equilibrium. This framework is particularly well-suited for systems with spin degrees of freedom~\citep{Becattini:2013fla,Becattini:2019dxo,Becattini:2020sww,Becattini:2021iol,Becattini:2021suc,Buzzegoli:2025zud}, which is inherently a quantum observable \citep{STAR:2017ckg,STAR:2018gyt,STAR:2019erd,ALICE:2019onw,STAR:2022fan,Liang:2004ph,Liang:2004xn,Becattini:2024uha}.

If the QGP achieves LTE on an initial hypersurface $\Sigma_0$, the density operator is obtained by maximizing the entropy $S=-\Tr(\wrho\,\log\wrho)$ subject to constraints on the energy-momentum, angular momentum, and charge densities. Representing these charges via the stress-energy operator $\wT^{\lambda\rho}$, the spin tensor operator $\wS^{\lambda,\rho\nu}$, and the current operator $\wj^\lambda$, the density operator takes the form \cite{Zubarev:1966,vanWeert1982,Becattini:2012tc,Becattini:2019dxo,Becattini:2018duy,Buzzegoli:2021wlg}:
\begin{equation}\label{eq:NEDO}
	\widehat{\rho}=\frac{1}{Z}\exp\left[-\int_{\Sigma_0} \!\!\!\di\Sigma_\lambda(y)\left(
	\wT^{\lambda\rho}(y)\beta_{\rho}(y)
	\!-\!\frac{1}{2}{\wS^{\lambda,\rho\nu}(y)\SP_{\rho\nu}(y)}
	\!-\!{\wj^\lambda(y)\zeta(y)}\right)\right]\;,
\end{equation}
where $\beta$ is the four-temperature vector, $\SP$ is the spin potential, and $\zeta = \mu/T$ is the reduced chemical potential. By applying Gauss's theorem and the conservation laws $\partial_\lambda \wT^{\lambda\rho}=0$, $\partial_\lambda \wS^{\lambda,\rho\nu}=-2\wT^{[\rho\nu]}$, and $\partial_\lambda \wj^\lambda=0$, the exponent in Eq.~\eqref{eq:NEDO} can be transformed into the sum of a 3D integral over the decoupling hypersurface $\Sigma_{\rm D}$ (when the fluid breaks up)
\footnote{In this work, the stage at which the fluid description ceases to be valid is called decoupling. Freeze-out refers to the stage when interactions between hadrons cease, which is conceptually distinct from decoupling.}
and a 4D integral over the spacetime region $\Omega$ enclosed by $\Sigma_{\rm D}$ and $\Sigma_0$, representing the volume occupied by the QGP \cite{Becattini:2019dxo,Buzzegoli:2025zud}:
\begin{equation}\label{gauss}
	\begin{split}
		\widehat{\rho}=&\frac{1}{Z}\exp\left\{
		-\int_{\Sigma_{\rm D}} \!\! \di \Sigma_{\lambda}(y) \; \left[\wT^{\lambda\rho}\beta_{\rho} \!-\!\frac{1}{2}{\wS^{\lambda,\rho\nu}\SP_{\rho\nu}}
		\!-\!\wj^\lambda\zeta\right]\right.\\
		&\left.+\int_\Omega \di^4 y \; \left[\wT^{(\lambda\rho)} \partial_\lambda \beta_\rho
		+\wT^{[\rho\nu]} \left(\SP_{\rho\nu} - \partial_\nu \beta_\rho\right)
		-\frac{1}{2}\wS^{\lambda,\rho\nu}\partial_\lambda\SP_{\rho\nu}
		-\wj^\lambda\partial_\lambda\zeta\right]\right\}\;,
	\end{split}
\end{equation}
where operators and hydrodynamic fields are intended to be evaluated in $y$.
While the first term represents the LTE integral at decoupling, the second term accounts for dissipative contributions to physical observables. In this work, we confine ourselves to calculating the Wigner function and spin polarization in LTE at decoupling. Dissipative terms can be addressed using the methods developed in Ref.~\cite{Buzzegoli:2025zud} and are assumed here to be small compared to the LTE contribution. We, therefore, neglect the second integral in Eq.~\eqref{gauss} and focus on the leading-order approximation, namely the LTE density operator:
\begin{equation}\label{LE-density-operator}
	\wrho_{\rm LE}=\frac{1}{Z}\exp\left[-\int_{\Sigma_{\rm D}} \!\!\!\di\Sigma_\lambda(y)\left(
	\wT^{\lambda\rho}(y)\beta_{\rho}(y)
	\!-\!\frac{1}{2}{\wS^{\lambda,\rho\nu}(y)\SP_{\rho\nu}(y)}
	\!-\!{\wj^\lambda(y)\zeta(y)}\right)\right]\;.
\end{equation}
Consequently, the spin polarization in heavy-ion collisions is well approximated by its LTE value, which can be readily evaluated using Eq.~\eqref{eq:SpinPol3/2Short} from the LTE Wigner function:
\begin{equation}\label{eq:rsWLTE}
	W_{\alpha\beta}(x,p)_{\rm LE} = \Tr\left[\wrho_{\rm LE} \wW_{\alpha\beta}(x,p)\right].
\end{equation}
Note that to compute the spin polarization, the Wigner function in Eq.~\eqref{eq:SpinPol3/2Short} must be integrated over the decoupling hypersurface $\Sigma_{\rm D}$. Evaluating the Wigner function at $x\in\Sigma_{\rm D}$ renders the use of the LTE statistical operator at decoupling particularly effective. However, an exact evaluation of the LTE Wigner function in Eq.~\eqref{eq:rsWLTE} using operator \eqref{LE-density-operator} is computationally intractable. Fortunately, we can exploit the hydrodynamic regime of the system to obtain a reliable approximation via linear response theory.

To calculate the LTE expectation value \eqref{eq:rsWLTE}, we re-express the four-temperature $\beta_\rho(y)$ and reduced chemical potential $\zeta(y)$ in Eq.~\eqref{LE-density-operator} as:
\begin{subequations}
	\begin{align}
		\beta_\rho(y) &= \beta_\rho(x) + \left[ \beta_\rho(y)-\beta_\rho(x) \right] \equiv \beta_\rho(x) + \Delta \beta_\rho(y,x)\;,\\
		\zeta(y) &= \zeta(x)+ [\zeta(y)-\zeta(x)] \equiv \zeta(x)+\Delta\zeta(y,x)\;,
	\end{align}    
\end{subequations}
allowing us to rewrite Eq.~\eqref{eq:rsWLTE} as:
\begin{multline}
\label{lteop}
W_{\alpha\beta}(x,p)_{\rm LE} =  \frac{1}{Z}  \Tr \left[ \exp\left\{- \beta(x) \cdot \widehat P +\zeta(x)\widehat{Q} \right.\right.\\
		\left.\left.-\int_{\Sigma_{\rm D}} \!\!\!\di\Sigma_\lambda(y)\left(
		\wT^{\lambda\rho}(y)\Delta\beta_{\rho}(y)
		\!-\!\frac{1}{2}{\wS^{\lambda,\rho\nu}(y)\SP_{\rho\nu}(y)}
		\!-\!{\wj^\lambda(y)\Delta\zeta(y)}\right)\right\}\, \wW_{\alpha\beta}(x,p)\right]\;,
\end{multline}
where $\beta_\rho(x)$ and $\zeta(x)$ have been factored out of the integral using the definitions of the four-momentum operator and the charge operator:
\begin{equation}\label{eq:ConservedCharges}
	\widehat{P}^\rho = \int_{\Sigma_{\rm D}} \!\! \di\Sigma_{\lambda}(y)\; \widehat{T}^{\lambda\rho}(y),\quad
	\widehat{Q}=\int_{\Sigma_{\rm D}} \!\! \di\Sigma_{\lambda}(y)\; \widehat{j}^{\lambda}(y).    
\end{equation}
In the hydrodynamic limit, where a clear separation exists between the microscopic correlation length of the operators and the macroscopic length scale over which $\beta$ and $\zeta$ vary, the first two terms in the exponent of Eq.~\eqref{lteop} dominate over the remaining integral term. The latter can be treated as a small perturbation, enabling linear response theory to yield an accurate approximation for the expectation value of $\wW_{\alpha\beta}(x,p)$ at LTE:
\begin{equation}\label{eq:WLEDeltas}
	W_{\alpha\beta}(x,p)_{\rm LE}=W_{\alpha\beta}(x,p)_{\rm GE}
	+\Delta_\beta W_{\alpha\beta}(x,p)_{\rm LE} +\Delta_\SP W_{\alpha\beta}(x,p)_{\rm LE}
	+\Delta_\zeta W_{\alpha\beta}(x,p)_{\rm LE}\;,
\end{equation}
where $W_{\alpha\beta}(x,p)_{\rm GE}$ denotes the global equilibrium (GE) expectation value evaluated with constant parameters $\beta(x)$ and $\zeta(x)$: 
\begin{equation}\label{eq:W-GE}
	W_{\alpha\beta}(x,p)_{\rm GE}\equiv\left\langle\wW_{\alpha\beta}(x,p)\right\rangle_{\rm GE}=\frac{\text{Tr}\left[\wW_{\alpha\beta}(x,p)\,
		\e^{-\widehat{A}(x)}\right]}{\text{Tr}\left[\e^{-\widehat{A}(x)}\right]}\;,
\end{equation}
with $\widehat{A}(x)\equiv\beta(x) \cdot \wP-\zeta(x)\widehat{Q}$. The terms $\Delta_\beta W_{\alpha\beta}(x,p)_{\rm LE}$, $\Delta_\SP W_{\alpha\beta}(x,p)_{\rm LE}$, and $\Delta_\zeta W_{\alpha\beta}(x,p)_{\rm LE}$ represent the leading linear response corrections to $\Delta\beta$, $\SP$, and $\Delta\zeta$, respectively:
\begin{subequations}\label{eq:DeltasW}
	\begin{align} \label{deltaO}
		\Delta_\beta W_{\alpha\beta}(x,p)_{\rm LE}=& -\!\!\int_{\Sigma_{\rm D}} \!\!\!\! \di \Sigma_\lambda(y) \!\!
		\int_{0}^{1} \!\!\di z \, \Delta\beta_{\rho}(y,x)\left\langle \wW_{\alpha\beta}(x,p),\,\e^{- z \widehat{A}(x) } \wT^{\lambda\rho}(y) 
		\, \e^{z\widehat{A}(x)} \right\rangle _{c,\text{GE}}\;,\\
		\Delta_\SP W_{\alpha\beta}(x,p)_{\rm LE}=& \frac{1}{2}\int_{\Sigma_{\rm D}} \!\!\!\! \di \Sigma_\lambda(y) \!\!
		\int_{0}^{1} \!\!\di z \, \SP_{\rho\nu}(y)\left\langle \wW_{\alpha\beta}(x,p),\,\e^{- z \widehat{A}(x) } \wS^{\lambda,\rho\nu}(y) 
		\, \e^{z\widehat{A}(x)} \right\rangle _{c,\text{GE}}\;,\\
		\Delta_\zeta W_{\alpha\beta}(x,p)_{\rm LE}=&\int_{\Sigma_{\rm D}} \!\!\!\! \di \Sigma_\lambda(y) \!\!
		\int_{0}^{1} \!\!\di z \, \Delta\zeta(y,x)\left\langle \wW_{\alpha\beta}(x,p),\,\e^{- z \widehat{A}(x) } \widehat{j}^\lambda(y) 
		\, \e^{z\widehat{A}(x)} \right\rangle _{c,\text{GE}}\;.
	\end{align}    
\end{subequations}
In the expressions above, $\left\langle\cdots\right\rangle_{c,\text{GE}}$ denotes the connected correlation function in global equilibrium, defined as:
\begin{equation}
	\left\langle\widehat{O}_1(x),\,\widehat{O}_2(y)\right\rangle_{c,\rm GE}=\left\langle\widehat{O}_1(x)\widehat{O}_2(y)\right\rangle_{\rm GE}-\left\langle\widehat{O}_1(x)\right\rangle_{\rm GE}\left\langle\,\widehat{O}_2(y)\right\rangle_{\rm GE}\;.
\end{equation}
%

\subsection{Wigner function in global equilibrium}
\label{subsec:WignerGE}
To evaluate the spin polarization, we are exclusively interested in the particle component of the Wigner function $W_{\alpha\beta}(x,\,p)_{\rm LE}$ [see Eq.~\eqref{eq:SpinPol3/2Short}]. Utilizing the plane-wave expansion of the Rarita-Schwinger field, we obtain the decomposition \eqref{rswignerdop}, whose particle component reads:
\begin{equation}
	\label{eq:rswignerdopPart}
	\widehat{W}_{\alpha\beta,+}(x,k) =  \sum_{r,s}  \frac{1}{(2\pi)^3}\int \frac{\di^3 p}{2E_p} \frac{\di^3 p'}{2E_{p'}} \e^{-\ii(p-p')\cdot x}
	 \, \delta^4\left(k - \frac{p+p'}{2}\right) \wadf{s}{p'} \waf{r}{p} u_{\alpha,r}(p) \bar{u}_{\beta,s}(p') .
\end{equation}
The leading-order contribution to the LTE Wigner function \eqref{eq:WLEDeltas} is the global equilibrium value evaluated at the local parameters $\beta(x)$ and $\zeta(x)$.
This global equilibrium Wigner function is obtained by taking the quantum statistical trace of Eq.~\eqref{eq:rswignerdopPart} with the GE statistical operator:
\begin{equation}
\label{eq:rswignerGE}
\begin{split}
W_{\alpha\beta,+}(x,k)_{\rm GE} =& \sum_{r,s} \frac{1}{(2\pi)^3}\int \frac{\di^3 p}{2E_p} \frac{\di^3 p'}{2E_{p'}}
	\e^{-\ii(p-p')\cdot x} \, \delta^4\left(k - \frac{p+p'}{2}\right)\\
    &\times u_{\alpha,r}(p) \bar{u}_{\beta,s}(p')
	\langle  \wadf{s}{p'} \waf{r}{p} \rangle_{\rm GE}\,.    
\end{split}
\end{equation}
The thermal expectation value of the creation and annihilation operators in global equilibrium \eqref{eq:W-GE} is given by:
\begin{equation}
	\langle  \wadf{s}{p'} \waf{r}{p} \rangle = 2E_p\delta_{sr}\delta^3(\bm{p-\bm{p'}})
	n_F(\beta(x)\cdot p -\zeta(x)),
\end{equation}
where $n_F$ denotes the Fermi-Dirac distribution function:
\begin{equation}
	n_F(x,p) =  n_F\left(\beta(x)\cdot p -\zeta(x)\right)
	= \frac{1}{\e^{\beta(x)\cdot p -\zeta(x)}+1}.
\end{equation}
Substituting this thermal expectation value into Eq.~\eqref{eq:rswignerGE} yields:
\begin{equation}
	\begin{split}
		W_{\alpha\beta,+}(x,k)_{\rm GE} =& \sum_{s} \frac{1}{(2\pi)^3} \int \frac{\di^3 p}{2E_p}\,
		\delta^4 \left(k - p\right) u_{\alpha,s}(p) \bar{u}_{\beta,s}(p)n_F(\beta\cdot p -\zeta)\\
		= & \frac{1}{(2\pi)^3} \frac{\delta\left(k_0 - E_k\right)}{2E_k}
		n_F(\beta\cdot k -\zeta)\sum_{s} u_{\alpha,s}(k) \bar{u}_{\beta,s}(k).
	\end{split}
\end{equation}
Applying the spin-sum completeness relation for vector-spinors \eqref{spinsum_rs} and relabeling the momentum $k$ as $p$ yields the result:
\begin{equation}
	W_{\alpha\beta,+}(x,p)_{\rm GE} = -\frac{\delta\left(p^2 - m^2\right)\theta(p_0)}{(2\pi)^3} 
	(\slashed{p} + m) P_{\alpha\beta}(p) n_F(\beta(x)\cdot p -\zeta(x)).
\end{equation}
The corresponding scalar and axial components are:
\begin{align}\label{eq:WFGE}
	\mathcal{F}_{\alpha\beta,+}(x,p)_{\rm GE} =& -\frac{\delta\left(p^2 - m^2\right)\theta(p_0)}{(2\pi)^3}
	\frac{8}{3}m\left(g_{\alpha\beta}-\frac{p_\alpha p_\beta}{m^2}\right)n_F(\beta(x)\cdot p -\zeta(x)),\\
	\mathcal{A}_{\mu, \alpha\beta,+}(x,p)_{\rm GE} =& \ii\frac{\delta\left(p^2 - m^2\right)\theta(p_0)}{(2\pi)^3}
	\frac{4}{3}\epsilon_{\mu\alpha\beta\lambda}p^\lambda n_F(\beta(x)\cdot p -\zeta(x)),
\end{align}
from which the direct and dual axial vectors [see Eq.~\eqref{eq:DefAxialF5}] follow immediately:
\begin{equation}
	\mathcal{F}^\mu_{5,+}(x,p)_{\rm GE} = 0,\quad
	\mathcal{A}^\mu_{\alpha\beta,+}(x,p)_{\rm GE} \, g^{\alpha\beta} = 0.
\end{equation}
Consequently, according to Eq.~\eqref{eq:SpinPol3/2Short}, there is no spin polarization at global equilibrium in the absence of vorticity. From these results and the spin polarization formula \eqref{eq:SpinPol3/2Short}, it also follows that for the first-order corrections, we only need to evaluate the first-order contributions in the numerator of Eq.~\eqref{eq:SpinPol3/2Short}—specifically, the Wigner function axial vectors $g^{\alpha\beta}\mathcal{A}^\mu_{\alpha\beta,+}$ and $\mathcal{F}^\mu_{5,+}$—while the denominator simplifies to:
\begin{equation}\label{eq:DenomSpin}
\begin{split}
\int\di p^0\int_{\Sigma_{\rm D}}\!\!\!\di\Sigma\cdot p\, \mathcal{F}^\alpha_{\ \alpha}(x,p) \simeq &
	\int\di p^0\int_{\Sigma_{\rm D}}\!\!\!\di\Sigma\cdot p\, \mathcal{F}^\alpha_{\ \alpha}(x,p)_{\rm GE}\\
	=&-\frac{8m}{(2\pi)^3}\int\di p^0\, \delta\left(p^2 - m^2\right) \theta(p^0) N_p ,    
\end{split}
\end{equation}
where
\begin{equation}
	N_p = \int_{\Sigma_D}\di\Sigma\cdot p \ n_F(x,p).
\end{equation}
We now recast the particle part of the Wigner operator \eqref{eq:rswignerdopPart} into a form more suitable for subsequent derivations.
We express the integration measure in Eq.~\eqref{eq:rswignerdopPart} as:
\begin{equation*}
	\int\frac{\di^3 \p}{2E_p}\rightarrow \int \di^4 p\; \delta(p^{2}-m^{2})\theta(p^{0})\;,
\end{equation*}
and perform the change of integration variables:
\begin{equation}\label{varchange}
	P \equiv \frac{p+p'}{2}\;, \qquad q \equiv p'-p\;, \qquad\implies\qquad p' = P + q/2\;, \qquad p = P - q/2\;,
\end{equation}
such that:
$$
\di^4 p' \; \di^4 p = \di^4 P \; \di^4 q\;,
$$
and 
$$
\delta(p^{\prime 2}-m^{2})\delta(p^{2}-m^{2})=\frac{1}{2}\delta\left(P^{2}+\frac{q^{2}}{4}-m^{2}\right)
\delta(P\cdot q)\;.
$$
Integrating over $\di^4 P$ is straightforward due to the factor $\delta^4 (p - P)$ in Eq.~\eqref{eq:rswignerdopPart}. Substituting all variables according to Eq.~\eqref{varchange}, one readily arrives at the integral representation:
\begin{equation}
	\label{eq:WignerParticle}
	\begin{split}
		\widehat{W}_{\alpha\beta}^+(x,\,p) =& \frac{1}{2(2\pi)^3}\sum_{r,s}\int\di^4 q\, \e^{i q\cdot x} \,
		\theta(p^0_+) \, \theta(p^0_-) \, \delta(p\cdot q) \,
		\delta\left( p^2 + \frac{q^2}{4} - m^2 \right)\\
		&\times\widehat{a}^\dagger_r (p_+) \widehat{a}_s (p_-) u_{\alpha,s}(p_-) \bar{u}_{\beta,r}(p_+),
	\end{split}
\end{equation}
where $p_\pm = p \pm q/2$ represents the on-shell four-momenta.

\subsection{Integration over the decoupling hypersurface} 
\label{subsec:hypersurface-int}
\begin{figure}
    \centering
	\includegraphics[width=8cm]{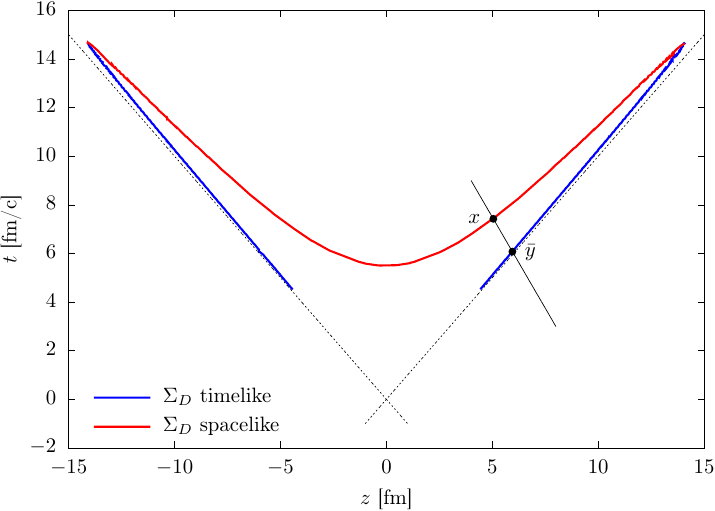}
	\caption{\label{fig:Freeze-out} A slice of the decoupling hypersurface $\Sigma_{\rm D}$
		(red and blue lines) on the transverse plane $x=y=0$, obtained by 3+1D viscous hydrodynamic
		simulations with the CLVisc model at $\sqrt{s_{\text{NN}}}=27\text{ GeV}$ \citep{Pang:2018zzo,Wu:2021fjf,Wu:2022mkr}, data taken from~\cite{Sheng:2026Data}.
		The black solid line represents the worldline of a particle, which intersects $\Sigma_{\rm D}$ at two points: one on the spacelike branch and the other on the timelike branch.}
\end{figure}
A delicate step in evaluating the contributions in Eqs.~\eqref{eq:DeltasW} is the integration over the decoupling hypersurface $\Sigma_{\rm D}$. While a three-dimensional hypersurface in relativistic hydrodynamics typically represents a regular time slice of a spacetime foliation, the decoupling hypersurface $\Sigma_{\rm D}$ in heavy-ion collisions can be topologically complex, containing both spacelike and timelike regions. In particular, for a given spatial position $\mathbf{x}$, there may exist multiple times $x^0$ belonging to $\Sigma_{\rm D}$ (see Fig.~\ref{fig:Freeze-out}). Nevertheless, the hypersurface can be partitioned into single-valued branches $x^0=f_k(\mathbf{x})$. To perform the integration over $\Sigma_{\rm D}$, we sum over these branches and define the point $\mathbf{y}_{i,k}$ as the $i$-th root of the equation
\begin{equation}\label{eq:Defybar}
	\bm{y} = \bm{x} + \frac{\bm{p}}{p^0}\left( y^0 - x^0\right)
	= \bm{x} + \frac{\bm{p}}{p^0}\left( f_k(\bm{y}) - x^0\right).
\end{equation}
A convenient formulation for integrating over $\Sigma_{\rm D}$ that avoids special geometric assumptions was recently established in Ref.~\cite{Sheng:2025cjk}. The procedure involves substituting the Wigner operator expression \eqref{eq:WignerParticle} into Eq.~\eqref{eq:DeltasW} and performing the integration over $q$ first.
Using Eq.~\eqref{eq:WignerParticle}, the contributions in Eq.~\eqref{eq:DeltasW} become
\begin{subequations}
	\begin{align}
		\Delta_\beta W_{\alpha\beta}(x,\,p)_{\rm LE} =& \frac{1}{(2\pi)^3}\int\di^4q\, \delta(p\cdot q)\,
		G^{\lambda\rho}_{\alpha\beta}F_{\lambda\rho}(q),\\
		\Delta_\SP W_{\alpha\beta}(x,\,p)_{\rm LE} =& \frac{1}{(2\pi)^3}\int\di^4q\, \delta(p\cdot q)\,
		G^{\lambda,\rho\nu}_{\alpha\beta}F_{\lambda,\rho\nu}(q),\\
		\Delta_\zeta W_{\alpha\beta}(x,\,p)_{\rm LE} =& \frac{1}{(2\pi)^3}\int\di^4q\, \delta(p\cdot q)\,
		G^{\lambda}_{\alpha\beta}F_{\lambda}(q),
	\end{align}
\end{subequations}
where
\begin{subequations}
	\label{eq:GFunctions}
	\begin{align}
		\label{eq:GEMT}
		G^{\lambda\rho}_{\alpha\beta}(q) =  & \frac{1}{2}\delta\left( p^2 + \frac{q^2}{4} - m^2 \right)
		\frac{1-\e^{\beta(x)\cdot q}}{\beta(x)\cdot q}\theta(p^0_+) \theta(p^0_-) \delta(p\cdot q)
		\\ & \times
		\sum_{r,s} u_{\alpha,s}(p_-) \bar{u}_{\beta,r}(p_+)\left\langle \widehat{a}^\dagger_r (p_+)
		\widehat{a}_s (p_-),\, \widehat{T}^{\lambda\rho}(0)\right\rangle_{c,{\rm GE}},\nonumber\\
		\label{eq:GSpin}
		G^{\lambda,\rho\nu}_{\alpha\beta}(q) =  & -\frac{1}{4}\delta\left( p^2 + \frac{q^2}{4} - m^2 \right)
		\frac{1-\e^{\beta(x)\cdot q}}{\beta(x)\cdot q}\theta(p^0_+) \theta(p^0_-) \delta(p\cdot q)
		\\ & \times
		\sum_{r,s} u_{\alpha, s}(p_-) \bar{u}_{\beta, r}(p_+)\left\langle \widehat{a}^\dagger_r (p_+)
		\widehat{a}_s (p_-),\, \widehat{S}^{\lambda,\rho\nu}(0)\right\rangle_{c,{\rm GE}},\nonumber\\
		\label{eq:GCurrent}
		G^{\lambda}_{\alpha\beta}(q) =  & -\frac{1}{2}\delta\left( p^2 + \frac{q^2}{4} - m^2 \right)
		\frac{1-\e^{\beta(x)\cdot q}}{\beta(x)\cdot q}\theta(p^0_+) \theta(p^0_-) \delta(p\cdot q)
		\\ & \times
		\sum_{r,s} u_{\alpha, s}(p_-) \bar{u}_{\beta, r}(p_+)\left\langle \widehat{a}^\dagger_r (p_+)
		\widehat{a}_s (p_-),\, \widehat{j}^{\lambda}(0)\right\rangle_{c,{\rm GE}},\nonumber
	\end{align}
\end{subequations}
and
\begin{subequations}
	\begin{align}
		F_{\lambda\rho}(q) =& \int_{\Sigma_D}\di\Sigma_\lambda(y) \e^{-i q\cdot (y-x)}
		\Delta\beta_\rho(y,x),\\
		F_{\lambda,\rho\nu}(q) =& \int_{\Sigma_D}\di\Sigma_\lambda(y) \e^{-i q\cdot (y-x)}
		\SP_{\rho\nu}(y),\\
		F_{\lambda}(q) =& \int_{\Sigma_D}\di\Sigma_\lambda(y) \e^{-i q\cdot (y-x)}
		\Delta\zeta(y,x).
	\end{align}
\end{subequations}
Exploiting the fact that the functions $F(q)$ are sharply peaked around $q=0$, due to the assumed hydrodynamic regime, and expanding $G(q)$ in a power series in $q$, the local equilibrium corrections to the Wigner function are obtained from Ref.~\cite{Sheng:2025cjk}:
\begin{subequations}\label{eq:DeltasW-int}
	\begin{align}
		\label{eq:DeltaOmegaWSeriesBeta}
		\Delta_\beta W^+_{\alpha\beta}(x,\,p)_{\rm LE} =& \sum_{N=0}^\infty \sum_{\bar{y}(x,p)}\frac{1}{N!} D^N_y(\bar{y})
		\left.\left[G^{\lambda\rho}_{\alpha\beta}(q)\frac{n_\lambda(y)}{|p\cdot n(y)|}\Delta\beta_\rho(y,x)\right]\right|_{q=0,y=\bar{y}},\\
		\label{eq:DeltaOmegaWSeries}
		\Delta_\SP W^+_{\alpha\beta}(x,\,p)_{\rm LE} =& \sum_{N=0}^\infty \sum_{\bar{y}(x,p)} \frac{1}{N!} D^N_y(\bar{y})
		\left.\left[G^{\lambda,\rho\nu}_{\alpha\beta}(q)\frac{n_\lambda(y)}{|p\cdot n(y)|}\SP_{\rho\nu}(y)\right]\right|_{q=0,y=\bar{y}},\\
		\label{eq:DeltaZetaWSeries}
		\Delta_\zeta W^+_{\alpha\beta}(x,\,p)_{\rm LE} =& \sum_{N=0}^\infty \sum_{\bar{y}(x,p)} \frac{1}{N!} D^N_y(\bar{y})
		\left.\left[G^{\lambda}_{\alpha\beta}(q)\frac{n_\lambda(y)}{|p\cdot n(y)|}\Delta\zeta(y,x)\right]\right|_{q=0,y=\bar{y}},
	\end{align}
\end{subequations}
where the sum over $\bar{y}$ runs over the roots \eqref{eq:Defybar}, $n$ is the unit vector orthogonal to the decoupling hypersurface $\Sigma_{\rm D}$, and
\begin{equation}
	D_y(\bar{y}) \equiv -i\Delta^{\kappa\sigma}(\bar{y})\partial_\kappa^q \partial_\sigma^y,
\end{equation}
with the projector operator $\Delta^{\kappa\sigma}$ defined as
\begin{equation}\label{eq:DeltaProjDef}
	\Delta^{\kappa\sigma}(\bar{y}) = g^{\kappa\sigma}- \frac{n^\kappa(\bar{y}) \, p^\sigma}{p\cdot n(\bar{y})}.
\end{equation}
%

\subsection{Spin polarization in the GLW pseudo-gauge}
\label{subsec:SpinLTEGLW}

Unlike the global equilibrium contribution~\eqref{eq:W-GE}, which depends solely on the globally conserved charges~\eqref{eq:ConservedCharges} and is therefore pseudo-gauge invariant, the local thermal equilibrium statistical operator~\eqref{LE-density-operator} is expressed in terms of local current operators ($\wT^{\lambda\rho}$ and $\wS^{\lambda,\rho\nu}$), making it inherently pseudo-gauge dependent~\cite{Becattini:2018duy,Buzzegoli:2021wlg,Becattini:2025oyi}.
Although the spin vector operator, defined via the Pauli-Lubanski vector~\eqref{luba}, is constructed from conserved quantities and is thus unaffected by pseudo-gauge transformations, evaluating its expectation value in LTE requires specifying a statistical operator, which renders the covariant Wigner function and the mean spin polarization pseudo-gauge dependent~\cite{Buzzegoli:2021wlg}. In this work, we derive the Wigner function and the spin polarization in both the GLW and Belinfante pseudo-gauges.

In the GLW pseudo-gauge, the LTE statistical operator~\eqref{LE-density-operator} takes the form
\begin{equation}\label{eq:StatOperGLW}
	\wrho_{\rm LE,GLW}=\frac{1}{Z}\exp\left[-\int_{\Sigma_{\rm D}} \!\!\!\di\Sigma_\lambda(y)\left(
	\wT^{\lambda\rho}_{\rm GLW}(y)\beta_{\rho}(y)
	\!-\!\frac{1}{2}{\wS_{\rm GLW}^{\lambda,\rho\nu}(y)\SP_{\rho\nu}(y)}
	\!-\!{\wj^\lambda(y)\zeta(y)}\right)\right]\;,
\end{equation}
where the free Rarita-Schwinger energy-momentum and spin tensors are given by Eqs.~\eqref{glw_emt_rs} and \eqref{eq:GLW_SpinTensor}. The linear response corrections to the Wigner function in the GLW framework—namely $\Delta_\beta W_{\alpha\beta}(x,p)_{\rm LE,GLW}$, $\Delta_\SP W_{\alpha\beta}(x,p)_{\rm LE,GLW}$, and $\Delta_\zeta W_{\alpha\beta}(x,p)_{\rm LE,GLW}$ as linear responses to $\Delta\beta$, $\SP$, and $\Delta\zeta$, respectively—are given by Eqs.~\eqref{eq:DeltasW-int} with the current operators evaluated in the GLW pseudo-gauge.

Note that the kernel functions $G$ in Eqs.~\eqref{eq:GFunctions} are obtained from the correlators of the currents with the particle creation and annihilation operators. For a free field, the only non-vanishing correlators involve the particle component of the currents. Applying the plane-wave expansion \eqref{rsfieldexp} to the current definitions \eqref{glw_emt_rs}, \eqref{eq:GLW_SpinTensor}, and \eqref{eq:Vcurrent}, and isolating the particle components, we obtain
\begin{subequations}\label{eq:CurrentsIn0}
	\begin{align}
		\wT^{\lambda\rho}_{+{\rm GLW}}(0)  =& -\sum_{s',r'}\int\frac{\di^3p'_-\di^3 p'_+}{4(2\pi)^3E'_+ E'_-} \frac{(p'_+ + p'_-)^\lambda(p'_+ + p'_-)^\rho}{4m}
		\bar{u}^\iota_{s'}(p'_-) u_{\iota,r'}(p'_+) \widehat{a}^\dagger_{s'}(p'_-) \widehat{a}_{r'} (p'_+),\\
		\wS^{\lambda,\rho\nu}_{+{\rm GLW}}(0) =& \sum_{s',r'}\int\frac{\di^3p'_-\di^3 p'_+}{4(2\pi)^3E'_+ E'_-}\bar{u}_{\iota,s'}(p'_-)\left(\mathcal{S}^{\lambda,\rho\nu}_{\rm GLW}\left(\frac{p'_+ + p'_-}{2}\right)\right)^{\iota\zeta}
		u_{\zeta,r'}(p'_+) \widehat{a}^\dagger_{s'}(p'_-) \widehat{a}_{r'} (p'_+),\\
		\wj_+^\lambda(0) =& -\sum_{s',r'}\int\frac{\di^3p'_-\di^3 p'_+}{4(2\pi)^3E'_+ E'_-}\bar{u}_{\iota,s'}(p'_-)\gamma^\lambda g^{\iota\zeta}
		u_{\zeta,r'}(p'_+) \widehat{a}^\dagger_{s'}(p'_-) \widehat{a}_{r'} (p'_+),
	\end{align}
\end{subequations}
where $E'_\pm=\sqrt{\bm{p}^{\prime 2}_\pm+m^2}$ and
\begin{subequations}\label{eq:SGLWDecomp}
	\begin{align}
		\left(\mathcal{S}^{\lambda,\rho\nu}_{\rm GLW}(p)\right)^{\iota\zeta} =&
		\mathcal{S}^{\lambda,\rho\nu}_{\rm GLW,D1}(p) g^{\iota\zeta}
		+ \mathcal{S}^{\lambda,\rho\nu}_{\rm GLW,D2}(p) g^{\iota\zeta}
		+ \left(\mathcal{S}^{\lambda,\rho\nu}_{\rm GLW,V}(p)\right)^{\iota\zeta},\\   
		\mathcal{S}^{\lambda,\rho\nu}_{\rm GLW,D1}(p) = & -\frac{\ii}{2}\gamma^{\lambda\rho\nu},\quad
		\mathcal{S}^{\lambda,\rho\nu}_{\rm GLW,D2}(p) = +\frac{\ii}{2m}
		\left( \gamma^{\lambda\rho}p^\nu - \gamma^{\lambda\nu}p^\rho \right),\\
		\left(\mathcal{S}^{\lambda,\rho\nu}_{\rm GLW,V}(p)\right)^{\iota\zeta} =& 
		-\ii \gamma^\lambda \left(g^{\rho\iota}g^{\nu\zeta} - g^{\rho\zeta}g^{\nu\iota}\right).
	\end{align}    
\end{subequations}
Substituting expressions \eqref{eq:CurrentsIn0} into the kernel functions $G$ in Eqs.~\eqref{eq:GFunctions}, the statistical averages reduce to the connected global equilibrium two-point expectation value of the creation and annihilation operators:
\begin{equation}
	\label{eq:4pointsGE}
	\begin{split}
		\left\langle \widehat{a}^\dagger_r(p_+) \widehat{a}_s(p_-),\,
		\widehat{a}^\dagger_{s'}(p'_-) \widehat{a}_{r'}(p'_+) \right\rangle_{c,{\rm GE}} &= 
		4E_+ E_-\delta^3(\bm{p}_- - \bm{p}'_-)\delta^3(\bm{p}_+ - \bm{p}'_+)\\
		&\times\delta_{rr'}\delta_{ss'} n_F(x,\,p_+)(1-n_F(x,\,p_-)),
	\end{split}
\end{equation}
where $E_\pm=\sqrt{m^2+\bm{p}_\pm^2}$ denotes the on-shell energies.
Using the Kronecker deltas $\delta_{rr'}\delta_{ss'}$ to perform the summation over $r'$ and $s'$, and the Dirac delta functions $\delta^3(\bm{p}_- - \bm{p}'_-)\delta^3(\bm{p}_+ - \bm{p}'_+)$ to integrate over $\bm{p}_-$ and $\bm{p}_+$, we obtain the linear response to $\beta$:
\begin{equation*}
\begin{split}
G^{\lambda\rho}_{\alpha\beta}(q)_{\rm GLW} =  & \frac{(-1)}{2(2\pi)^3}\delta\left( p^2 + \frac{q^2}{4} - m^2 \right)
	\theta(p^0_+) \theta(p^0_-) \delta(p\cdot q)\frac{1-\e^{\beta(x)\cdot q}}{\beta(x)\cdot q}\\
&\times n_F(x,\,p_+)(1-n_F(x,\,p_-))
	\sum_{r,s} u_{\alpha,s}(p_-) \bar{u}^\iota_{s}(p_-) \frac{p^\lambda p^\rho}{m} u_{\iota,r}(p_+) \bar{u}_{\beta,r}(p_+).
\end{split}
\end{equation*}
Noting that $\delta(p\cdot q)=\delta(p_+^2-p_-^2)=\delta(m^2-m^2)=1$, this factor can be omitted.
The spin sums are performed using the vector-spinor completeness relation \eqref{spinsum_rs}:
\begin{subequations}\begin{align}
		\sum_s u_{\alpha,s}(p_-) \bar{u}_{\iota,s}(p_-) =& -(\slashed{p}_- + m)P_{\alpha\iota}(p_-),\\
		\sum_r u_{\zeta,r}(p_+) \bar{u}_{\beta,r}(p_+) = & -(\slashed{p}_+ + m)P_{\zeta\beta}(p_+).
\end{align}\end{subequations}
Finally, employing the identity
\begin{equation}
	\left(1-\e^{\beta(x)\cdot q}\right)n_F(x,\,p_+)(1-n_F(x,\,p_-)) =
	n_F(x,\,p_+) - n_F(x,\,p_-),
\end{equation}
we arrive at
\begin{subequations}\label{eq:GFreeFieldGLW}
\begin{align}
	G^{\lambda\rho}_{\alpha\beta}(q)_{\rm GLW} =  & -G(q)
		\left( \slashed{p}_- + m\right) P_\alpha^{\ \iota}(p_-)\frac{p^\lambda p^\rho}{m}
		\left( \slashed{p}_+ + m\right)P_{\iota\beta}(p_+),\\
	G^{\lambda,\rho\nu}_{\alpha\beta}(q)_{\rm GLW} =  & -\frac{1}{2}G(q)
		\left( \slashed{p}_- + m\right) P_{\alpha\iota}(p_-)\left(\mathcal{S}^{\lambda,\rho\nu}_{\rm GLW}(p)\right)^{\iota\zeta}
		\left( \slashed{p}_+ + m\right)P_{\zeta\beta}(p_+), \\
	G^{\lambda}_{\alpha\beta}(q) =  & G(q)
		\left( \slashed{p}_- + m\right) P_\alpha^{\ \iota}(p_-)\gamma^\lambda
		\left( \slashed{p}_+ + m\right)P_{\iota\beta}(p_+),
\end{align}\end{subequations}
where
\begin{equation}
	\label{eq:SimpleG}
	G(q) =  \frac{n_F(x,\,p_+) - n_F(x,\,p_-)}{2(2\pi)^3(\beta(x)\cdot q)}\delta\left( p^2 + \frac{q^2}{4} - m^2 \right)    \theta(p^0_+) \theta(p^0_-).
\end{equation}
To evaluate the spin polarization, we are interested in the following axial vectors constructed from the Wigner function [see Eq.~\eqref{eq:SpinPol3/2Short}], which, using the series expansions \eqref{eq:DeltasW-int} and results \eqref{eq:GFreeFieldGLW}, read:
\begin{subequations}
	\begin{align}
		\label{eq:DeltaASeriesBeta}
		\Delta_\beta \mathcal{A}^{\mu,\alpha}_{\alpha,+}(x,\,p)_{\rm LE,GLW} =&
		\sum_{\stackrel{N=0}{\bar{y}(x,p)}}^\infty \frac{(-D^N_y(\bar{y}))}{N!} 
		\left.\left\{ G(q) A^{\mu,\lambda\rho}(q)_{\rm GLW}
		\frac{n_\lambda(y)}{|p\cdot n(y)|}\Delta\beta_\rho(y,x)\right\}\right|_{\stackrel{q=0}{y=\bar{y}}},\\
		\label{eq:DeltaF5SeriesBeta}
		\Delta_\beta \mathcal{F}_{5,+}^{\mu}(x,\,p)_{\rm LE,GLW} =&
		\sum_{\stackrel{N=0}{\bar{y}(x,p)}}^\infty \frac{(-D^N_y(\bar{y}))}{N!} 
		\left.\left\{ G(q) F_5^{\mu,\lambda\rho}(q)_{\rm GLW}
		\frac{n_\lambda(y)}{|p\cdot n(y)|}\Delta\beta_\rho(y,x)\right\}\right|_{\stackrel{q=0}{y=\bar{y}}},
	\end{align}
\end{subequations}
where we defined
\begin{subequations}
	\begin{align}
		\label{eq:ATraceBeta}
		A^{\mu,\lambda\rho}(q)_{\rm GLW} =&  \tr\left[ \gamma^\mu\gamma^5 g^{\alpha\beta} \left( \slashed{p}_- + m\right) P_\alpha^{\ \iota}(p_-)\frac{p^\lambda p^\rho}{m}
		\left( \slashed{p}_+ + m\right)P_{\iota\beta}(p_+)\right],\\
		\label{eq:F5TraceBeta}
		F_5^{\mu,\lambda\rho}(q)_{\rm GLW} =&  2 \, \ii \, \epsilon^{\mu\alpha\beta\tau} \, \frac{p_\tau}{m}
		\tr\left[ \left( \slashed{p}_- + m\right) P_\alpha^{\ \iota}(p_-)\frac{p^\lambda p^\rho}{m}
		\left( \slashed{p}_+ + m\right)P_{\iota\beta}(p_+)\right],
	\end{align}
\end{subequations}
and similarly for the linear response to $\SP$,
\begin{subequations}
\begin{align}
\label{eq:DeltaASeriesSP}
\Delta_\SP \mathcal{A}^{\mu,\alpha}_{\alpha,+}(x,\,p)_{\rm LE,GLW} =&
	\sum_{\stackrel{N=0}{\bar{y}(x,p)}}^\infty \frac{D^N_y(\bar{y})}{N!}
	\left.\left\{ -\frac{G(q)}{2} A^{\mu,\lambda,\rho\nu}(q)_{\rm GLW}
	\frac{n_\lambda(y)}{|p\cdot n(y)|}\SP_{\rho\nu}(y)\right\}\right|_{\stackrel{q=0}{y=\bar{y}}},\\
\label{eq:DeltaF5SeriesSP}
\Delta_\SP \mathcal{F}_{5,+}^{\mu}(x,\,p)_{\rm LE,GLW} =&
	\sum_{\stackrel{N=0}{\bar{y}(x,p)}}^\infty \frac{D^N_y(\bar{y})}{N!}
	\left.\left\{  -\frac{G(q)}{2} F_5^{\mu,\lambda,\rho\nu}(q)_{\rm GLW}
	\frac{n_\lambda(y)}{|p\cdot n(y)|}\SP_{\rho\nu}(y)\right\}\right|_{\stackrel{q=0}{y=\bar{y}}},
\end{align}
\begin{align}
\label{eq:ATraceSP}
A^{\mu,\lambda,\rho\nu}(q)_{\rm GLW} =&  \tr\left[ \gamma^\mu\gamma^5 g^{\alpha\beta} \left( \slashed{p}_- + m\right) P_{\alpha\iota}(p_-)\left(\mathcal{S}^{\lambda,\rho\nu}_{\rm GLW}(p)\right)^{\iota\zeta}
		\left( \slashed{p}_+ + m\right)P_{\zeta\beta}(p_+)\right],\\
\label{eq:F5TraceSP}
F_5^{\mu,\lambda,\rho\nu}(q)_{\rm GLW} =&  2\, \ii \, \epsilon^{\mu\alpha\beta\tau} \, \frac{p_\tau}{m}
		\tr\left[ \left( \slashed{p}_- + m\right) P_{\alpha\iota}(p_-)\left(\mathcal{S}^{\lambda,\rho\nu}_{\rm GLW}(p)\right)^{\iota\zeta}
		\left( \slashed{p}_+ + m\right)P_{\zeta\beta}(p_+)\right],
\end{align}
\end{subequations}
and for the response to $\zeta$,
\begin{subequations}
	\begin{align}
		\label{eq:DeltaASeriesZeta}
		\Delta_\zeta \mathcal{A}^{\mu,\alpha}_{\alpha,+}(x,\,p)_{\rm LE} =&
		\sum_{\stackrel{N=0}{\bar{y}(x,p)}}^\infty \frac{D^N_y(\bar{y})}{N!} 
		\left.\left\{ G(q) A^{\mu,\lambda}(q)
		\frac{n_\lambda(y)}{|p\cdot n(y)|}\Delta\zeta(y)\right\}\right|_{\stackrel{q=0}{y=\bar{y}}},\\
		\label{eq:DeltaF5SeriesZeta}
		\Delta_\zeta \mathcal{F}_{5,+}^{\mu}(x,\,p)_{\rm LE} =&
		\sum_{\stackrel{N=0}{\bar{y}(x,p)}}^\infty \frac{D^N_y(\bar{y})}{N!}
		\left.\left\{ G(q)F_5^{\mu,\lambda}(q)
		\frac{n_\lambda(y)}{|p\cdot n(y)|}\Delta\zeta(y)\right\}\right|_{\stackrel{q=0}{y=\bar{y}}},\\
		\label{eq:ATraceZeta}
		A^{\mu,\lambda}(q)=&  \tr\left[ \gamma^\mu\gamma^5 g^{\alpha\beta} \left( \slashed{p}_- + m\right) P_\alpha^{\ \iota}(p_-)\gamma^\lambda
		\left( \slashed{p}_+ + m\right)P_{\iota\beta}(p_+)\right],\\
		\label{eq:F5TraceZeta}
		F_5^{\mu,\lambda}(q) =&  2 \, \ii \, \epsilon^{\mu\alpha\beta\tau} \, \frac{p_\tau}{m}
		\tr\left[ \left( \slashed{p}_- + m\right) P_\alpha^{\ \iota}(p_-)\gamma^\lambda
		\left( \slashed{p}_+ + m\right)P_{\iota\beta}(p_+)\right].
	\end{align}
\end{subequations}
To perform these traces over Dirac indices, we employed the \textsc{FeynCalc} 10.1.0 \textsc{Mathematica} package~\cite{Mertig1991,Shtabovenko:2016sxi,Shtabovenko:2020gxv,Shtabovenko:2023idz}, obtaining
\begin{subequations}
	\begin{align}
		A^{\mu,\lambda\rho}(q)_{\rm GLW} =& 0,\quad
		F_5^{\mu,\lambda\rho}(q)_{\rm GLW} = 0,\\
        \label{eq:TraceAxialGLW}
		A^{\mu,\lambda,\rho\nu}(q)_{\rm GLW} =& \frac{40}{3}\epsilon^{\mu\rho\nu\tau}p_\tau p^\lambda
		+ \mathcal{O}\left(q\right),\quad
		F_5^{\mu,\lambda,\rho\nu}(q)_{\rm GLW} =  \frac{80}{3}\epsilon^{\mu\rho\nu\tau}p_\tau p^\lambda
		+ \mathcal{O}\left(q\right),\\
		A^{\mu,\lambda}(q) + F_5^{\mu,\lambda}(q) =& -\frac{\ii}{3}
		\left(40- \frac{40}{3}\frac{q^2}{m^2} + \frac{2}{3}\frac{q^4}{m^4}\right)
		\epsilon^{\mu\lambda\rho\tau}p_\rho q_\tau.
	\end{align}
\end{subequations}
Since the traces corresponding to the linear response to $\beta$ vanish identically, we conclude that
\begin{equation}
	\Delta_\beta \mathcal{A}^{\mu,\alpha}_{\alpha,+}(x,\,p)_{\rm LE,GLW} =0,\quad
	\Delta_\beta \mathcal{F}_{5,+}^{\mu}(x,\,p)_{\rm LE,GLW} = 0.
\end{equation}
Thus, as in the Dirac case~\cite{Buzzegoli:2021wlg}, there is no spin polarization induced by gradients of $\beta$ (including thermal vorticity and thermal shear) in the GLW pseudo-gauge for spin-$3/2$ fields. Instead, the spin polarization in the GLW pseudo-gauge is driven by the spin potential. 

Considering the leading terms of the series \eqref{eq:DeltaASeriesSP} and \eqref{eq:DeltaF5SeriesSP}, we obtain
\begin{equation*}
	\begin{split}
		\Delta_\SP \mathcal{A}^{\mu,\alpha}_{\alpha,+}(x,\,p)_{\rm LE,GLW} &+ \Delta_\SP \mathcal{F}_{5,+}^{\mu}(x,\,p)_{\rm LE,GLW} \simeq  -\sum_{\bar{y}(x,p)} 
		\frac{G(0)}{2} 40 \epsilon^{\mu\rho\nu\tau}p_\tau p^\lambda
		\frac{n_\lambda(\bar{y})}{|p\cdot n(\bar{y})|}\SP_{\rho\nu}(\bar{y})\\
		= & 40\frac{\delta(p^2-m^2)\theta(p^0)}{4(2\pi)^3} n_F(x,p)(1-n_F(x,p))
		\epsilon^{\mu\rho\nu\tau}p_\tau\\
		&\times \sum_{\bar{y}(x,p)}\text{sgn} [p\cdot n(\bar{y})]\SP_{\rho\nu}(\bar{y}).
	\end{split}
\end{equation*}
Similarly, noting that $A^{\mu,\lambda}(0) + F_5^{\mu,\lambda}(0)=0$,
the linear response to $\zeta$ eventually yields~\cite{Sheng:2025cjk}
\begin{align*}
\Delta_\zeta \mathcal{A}^{\mu,\alpha}_{\alpha,+}(x,\,p)_{\rm LE} + \Delta_\zeta \mathcal{F}_{5,+}^{\mu}(x,\,p)_{\rm LE} \simeq &
\sum_{\bar{y}(x,p)} \left(-\ii \Delta^{\kappa\sigma}  \right) G(0)
\left(-\ii\frac{40}{3} \epsilon^{\mu\lambda\rho\tau}p_\rho  \partial^q_\kappa q_\tau \right)\\
	&\times \partial^y_\sigma\left[\frac{n_\lambda(y)}{|p\cdot n(y)|}\Delta\zeta(y)\right]_{y=\bar{y}}\\
		= & -40\frac{\delta(p^2-m^2)\theta(p^0)}{4(2\pi)^3} n_F(x,p)(1-n_F(x,p))
		\epsilon^{\mu\rho\nu\tau}p_\tau \\
	&\times \sum_{\bar{y}(x,p)}\text{sgn}[p\cdot n(\bar{y})] \frac{1}{3}\frac{2 n_\rho(\bar{y})}{p\cdot n(\bar{y})}\partial_\nu^{\bar{y}}\zeta(\bar{y}).
\end{align*}
Inserting these expressions into Eq.~\eqref{eq:SpinPol3/2Short} and recalling Eq.~\eqref{eq:DenomSpin}, we obtain
\begin{equation}\label{eq:FinalLTEPolGLW}
\begin{split}
S^\mu(p)_{\rm GLW} =& -\frac{5}{8 m N_p}
	\int_{\Sigma_{\rm D}}\!\!\di\Sigma\cdot p \, n_F(x,p)(1-n_F(x,p))\epsilon^{\mu\rho\nu\tau}p_\tau\\
	&\times\sum_{\bar{y}(x,p)}\text{sgn} [p\cdot n(\bar{y})]\left\{\SP_{\rho\nu}(\bar{y})
	-\frac{1}{3}\frac{2 n_\rho(\bar{y})}{p\cdot n(\bar{y})}\partial_\nu^{\bar{y}}\zeta(\bar{y})\right\},    
\end{split}
\end{equation}
where
\begin{equation}
	N_p = \int_{\Sigma_{\rm D}}\di\Sigma\cdot p \, n_F(x,p).
\end{equation}
At global thermal equilibrium, the spin potential reduces to constant thermal vorticity~\cite{Becattini:2018duy,Buzzegoli:2021wlg}, i.e., $\SP_{\rho\nu}=\varpi_{\rho\nu}=\text{const.}$, $\zeta=\text{const.}$, and $n_F(x,p)$ becomes homogeneous, giving
\begin{equation}
		S^\mu(p)_{\rm GE} =-\frac{5}{8 m}\epsilon^{\mu\rho\nu\tau} p_\tau \varpi_{\rho\nu},
\end{equation}
in exact agreement with previous results~\cite{Becattini:2024uha} [see Eq.~\eqref{eq:SVectorAnySpin}].
The spin polarization induced by the gradient of the reduced chemical potential $\zeta=\beta\mu$ is known as the spin Hall effect~\cite{Liu:2020dxg,Buzzegoli:2022qrr,Fu:2022myl}. This direct calculation shows that the spin Hall effect does not scale with spin as the thermal vorticity and shear induced polarization, instead for the spin $3/2$ field result in an extra $1/3$ factor\footnote{As this factor is unexpected, to validate the result we checked that it is consistent with the Wigner equations \eqref{eq:WEqDFandS} and \eqref{eq:WEqpdotA}, where gradients of $\zeta$ are generated by gradients of the global equilibrium Wigner function in Eq.~\eqref{eq:WFGE}.} suggesting the following scaling:
\begin{equation}
	\Delta_\zeta S^\mu_{(S)}(p) \propto \frac{S(11-4S)}{18 m}\epsilon^{\mu\rho\nu\tau} p_\tau \frac{n_\rho}{p\cdot n}\partial_\nu\zeta.
\end{equation}
%

\subsection{Spin polarization in the Belinfante and canonical pseudo-gauges}
\label{subsec:SpinLTEBel}
Following the steps outlined above, one can readily evaluate the linear response of the Wigner function to gradients of $\beta$ and to the spin potential using a statistical operator constructed from different energy-momentum and spin tensors.
In the Belinfante pseudo-gauge, where the spin tensor vanishes, the local equilibrium statistical operator~\eqref{LE-density-operator} takes the form
\begin{equation}
	\widehat{\rho}_{\rm LTE, B} = \frac{1}{Z}\exp\left[
	-\int\di\Sigma_\lambda(y) \left(\widehat{T}^{\lambda\rho}_{\rm B}(y) \beta_\rho(y)
	-\zeta(y) \widehat{j}^\lambda(y)\right)\right].
\end{equation}
The contribution originating from the reduced chemical potential $\zeta$ remains identical to the one previously computed.
The contribution from $\beta$, on the other hand, can be derived from our earlier results by exploiting the pseudo-gauge transformation~\eqref{eq:pgt}, which relates the GLW and Belinfante representations. Substituting relation~\eqref{eq:pgt}, the Belinfante local thermal equilibrium statistical operator can be equivalently rewritten as
\begin{equation}
\begin{split}
\widehat{\rho}_{\rm LTE, B} = \frac{1}{Z}\exp&\left\{
	-\int\di\Sigma_\lambda(y) \left(\left[\widehat{T}^{\lambda\rho}_{\rm GLW}(y)\right.\right.\right.\\
	&\left.\left.\left.+ \frac{1}{2} \partial_\nu \left( \widehat{S}_{\rm GLW}^{\nu,\lambda\rho}(y)
	+ \widehat{S}_{\rm GLW}^{\rho,\lambda\nu}(y) +
	\widehat{S}_{\rm GLW}^{\lambda,\rho\nu}(y) \right)\right]\beta_\rho(y)
	-\zeta(y) \widehat{j}^\lambda(y)\right)\right\}.    
\end{split}
\end{equation}
Following the procedure established in Refs.~\cite{Becattini:2018duy,Buzzegoli:2021wlg}, the terms involving the spin tensor can be recast as
\begin{equation*}
\begin{split}
\int\di\Sigma_\lambda(y)  \partial_\nu \left(\widehat{S}_{\rm GLW}^{\nu,\lambda\rho}
	+ \widehat{S}_{\rm GLW}^{\rho,\lambda\nu} +
	\widehat{S}_{\rm GLW}^{\lambda,\rho\nu} \right)\beta_\rho=\int\di\Sigma_\lambda(y)  &
	\left[\partial_\nu\left( \beta_\rho\widehat{S}_{\rm GLW}^{\nu,\lambda\rho}
	+ \beta_\rho\widehat{S}_{\rm GLW}^{\rho,\lambda\nu} +
	\beta_\rho\widehat{S}_{\rm GLW}^{\lambda,\rho\nu} \right)\right.\\
&\left.-\left(\widehat{S}_{\rm GLW}^{\nu,\lambda\rho}
	+ \widehat{S}_{\rm GLW}^{\rho,\lambda\nu} +
	\widehat{S}_{\rm GLW}^{\lambda,\rho\nu} \right)\partial_\nu \beta_\rho\right],
\end{split}
\end{equation*}
where the total divergence term vanishes under suitable boundary conditions. Decomposing the gradient $\partial_\nu \beta_\rho$ into its symmetric and antisymmetric components—corresponding to thermal vorticity and thermal shear, respectively,
\begin{subequations}
	\begin{align}
		\varpi_{\mu\nu}(x)= & -\frac{1}{2}\left(\partial_\mu\beta_\nu(x) - \partial_\nu\beta_\mu(x)\right),\\
		\xi_{\mu\nu}(x)= & \frac{1}{2}\left(\partial_\mu\beta_\nu(x) + \partial_\nu\beta_\mu(x)\right),
	\end{align}
\end{subequations}
yields the statistical operator:
\begin{multline}
\widehat{\rho}_{\rm LTE, B} = \frac{1}{Z}\exp\left[
	-\int\di\Sigma_\lambda(y) \left(\widehat{T}^{\lambda\rho}_{\rm GLW}(y) \beta_\rho(y)
	- \frac{1}{2}\varpi_{\rho\nu}(y) \widehat{S}_{\rm GLW}^{\lambda,\rho\nu}(y)\right.\right.\\
	\left.\left.+ \xi_{\rho\nu}(y) \widehat{S}_{\rm GLW}^{\rho,\lambda\nu}(y)
	-\zeta(y) \widehat{j}^\lambda(y)\right)\right].    
\end{multline}
We observe that the linear response to the term
\begin{equation}
	\frac{1}{2}\varpi_{\rho\nu}(y) \widehat{S}_{\rm GLW}^{\lambda,\rho\nu}(y)
\end{equation}
matches the previous result for the spin potential upon replacing $\SP_{\rho\nu}$ with the thermal vorticity $\varpi_{\rho\nu}$.
Likewise, the linear response to the thermal shear term
\begin{equation}
	\xi_{\rho\nu}(y) \widehat{S}_{\rm GLW}^{\rho,\lambda\nu}(y)
\end{equation}
is obtained from the linear response to the GLW spin potential by replacing $\SP_{\rho\nu}$ with $-2\xi_{\rho\nu}$ and contracting the indices accordingly. With these observations, we directly arrive at the leading-order contributions to the spin polarization:
\begin{equation}\label{eq:FinalLTEPolBel}
\begin{split}
S^\mu(p)_{\rm B} =&-\frac{5}{8 m N_p}
	\int\di\Sigma\cdot p \, n_F(1-n_F)\epsilon^{\mu\rho\nu\tau}p_\tau\\
	&\times\sum_{\bar{y}(x,p)}\text{sgn} [p\cdot n(\bar{y})]
	\left\{\varpi_{\rho\nu}(\bar{y})
	+ \frac{2 n_\rho(\bar{y})}{p\cdot n(\bar{y})} \left[\xi_{\nu\alpha}(\bar{y})p^\alpha
	-\frac{1}{3}\partial_\nu^{\bar{y}}\zeta(\bar{y})\right]\right\}.    
\end{split}
\end{equation}

In the canonical pseudo-gauge the local equilibrium statistical operator~\eqref{LE-density-operator} reads
\begin{equation}
	\widehat{\rho}_{\rm LTE, C} = \frac{1}{Z}\exp\left[
	-\int\di\Sigma_\lambda(y) \left(\widehat{T}^{\lambda\rho}_C(y) \beta_\rho(y)
	\!-\!\frac{1}{2}{\wS_C^{\lambda,\rho\nu}(y)\SP_{\rho\nu}(y)}
	-\zeta(y) \widehat{j}^\lambda(y)\right)\right].
\end{equation}
Taking advantage of the transformation~\eqref{eq:pgt}, which relates the canonical and Belinfante  pseudo-gauges, the statistical operator can be equivalently written as~\cite{Buzzegoli:2021wlg}
\begin{multline}
\widehat{\rho}_{\rm LTE, C} = \frac{1}{Z}\exp\left[
	-\int\di\Sigma_\lambda(y) \left(\widehat{T}^{\lambda\rho}_B(y) \beta_\rho(y)
	\!+\!\frac{1}{2}\wS_C^{\lambda,\rho\nu}(y)(\varpi_{\rho\nu}(y)-\SP_{\rho\nu}(y))\right.\right.\\
	\left.\left.\!+\!\wS_C^{\rho,\lambda\nu}(y)\xi_{\rho\nu}(y)
	\!-\!\zeta(y) \widehat{j}^\lambda(y)\right)\right].    
\end{multline}
Note that, unlike in the Dirac case where the canonical spin tensor is totally antisymmetric, the Rarita-Schwinger canonical spin tensor contains a vector part $\wS_V^{\rho,\lambda\nu}(y)$ [see Eq.~\eqref{eq:SpinTenCanV}] which is not antisymmetric under the exchange of $\rho$ and $\nu$. Consequently, while for the Dirac field the canonical pseudo-gauge yields the same thermal-shear-induced polarization as the Belinfante one and differs only by the presence of the spin potential~\cite{Buzzegoli:2021wlg,Liu:2021nyg}, for the Rarita-Schwinger field the canonical and Belinfante pseudo-gauges yield different predictions for the thermal-shear-induced polarization. Explicitly, following the same procedure described above for the GLW pseudo-gauge, we obtain the canonical spin polarization
\begin{multline}\label{eq:FinalLTEPolCan}
S^\mu(p)_{\rm C} = -\frac{5}{8 m N_p}
	\int\di\Sigma\cdot p \, n_F(1-n_F)\epsilon^{\mu\rho\nu\tau}p_\tau
	\sum_{\bar{y}(x,p)}\text{sgn} [p\cdot n(\bar{y})]
	\left\{\SP_{\rho\nu}(\bar{y})\vphantom{\frac{2 n_\rho(\bar{y})}{p\cdot n(\bar{y})}}\right. \\
	\left.+ \frac{2}{3}\frac{n_\rho(\bar{y})}{p\cdot n(\bar{y})} \left[\left(\xi_{\nu\alpha}(\bar{y})
	+\SP_{\nu\alpha}(\bar{y}) - \varpi_{\nu\alpha}(\bar{y})\right)p^\alpha-\partial_\nu^{\bar{y}}\zeta(\bar{y})\right]\right\}.
\end{multline}
It is straightforward to verify that at global equilibrium, where $\SP=\varpi=\text{const.}$, $\xi=0$, and $\zeta=\text{const.}$,  the canonical spin polarization~\eqref{eq:FinalLTEPolCan} reduces to Eq.~\eqref{eq:SVectorAnySpin}. As mentioned earlier, it is interesting to note that if spin degrees of freedom equilibrate before reaching the decoupling hypersurface—that is, if the spin potential equals the thermal vorticity, $\SP(y)=\varpi(y)$ (not necessarily constant)—then, unlike for the Dirac field, the canonical and Belinfante pseudo-gauges yield different predictions for the contribution of thermal shear to the spin polarization.

This observation is crucial for using the local spin polarization of the $\Omega$ baryon to determine which of the two pseudo-gauges better describes heavy-ion collisions. The assumption of spin degree-of-freedom equilibration, i.e., $\SP(y)=\varpi(y)$, describes the spin polarization in high-energy heavy-ion collisions very well~\cite{Becattini:2024uha,Becattini:2021iol,Palermo:2024tza}, where including thermal-shear-induced polarization is essential to describing the local spin polarization of $\Lambda$ hyperons~\cite{Becattini:2024uha,Becattini:2021iol,Palermo:2024tza}. Even though the local spin polarization at LHC energies depends on bulk viscosity, because the canonical and Belinfante pseudo-gauges yield identical predictions for it, one can use the global and local spin polarization of $\Lambda$ to constrain the bulk viscosity. Once the $\Lambda$ spin polarization is accurately described and the transport parameters are constrained, the competing predictions for the $\Omega$ local spin polarization in the Belinfante and canonical pseudo-gauges can be compared against data to discriminate between the two frameworks.

\section{The axial vortical effect}
\label{sec:AVE}
The axial vortical effect (AVE) is a macroscopic quantum phenomenon describing the generation of an axial charge current along the rotation of a fluid~\cite{Vilenkin:1979ui,Landsteiner:2011iq,Gao:2012ix,Kharzeev:2015znc,Buzzegoli:2017cqy,Buzzegoli:2018wpy}. 
At global thermal equilibrium under rotation, described by the statistical operator
\begin{equation}\label{eq:GERho}
\wrho_{\rm GE} = \frac{1}{Z_{\rm GE}}\exp\left\{-\wP\cdot\beta(x)+\frac{1}{2}\varpi_{\rho\nu}\wJ_x^{\rho\nu}+\zeta\widehat{Q}\right\},\qquad
\beta_\mu(x)= b_\mu + \varpi_{\mu\nu}x^\nu,
\end{equation}
with $\wJ_x^{\rho\nu}$ the angular momentum-boost operator around the point $x$:
\begin{equation}\label{eq:JxOper}
\wJ_x^{\rho\nu} = \int_\Sigma \di\Sigma_\lambda(y) \left[\wT^{\lambda\nu}(y)\,(y-x)^\rho
    -\wT^{\lambda\rho}(y)\,(y-x)^\nu + \wS^{\lambda,\rho\nu}(y)\right],
\end{equation}
the expectation value of the axial current takes the form~\cite{Vilenkin:1979ui,Buzzegoli:2017cqy,Buzzegoli:2018wpy}
\begin{equation}
\langle \wj^\mu_A(x) \rangle_{\rm GE} = n_A u^\mu + W_A \frac{\omega^\mu}{T},
\end{equation}
where $n_A$ is the axial charge density, $u^\mu$ is the fluid four-velocity, $T$ is the temperature, and $\omega^\mu$ is the kinematic vorticity of the medium, related to the constant thermal vorticity tensor $\varpi_{\nu\rho}$ by
\begin{equation}\label{eq:RotThermalVort}
\frac{\omega^\mu}{T} = -\frac{1}{2}\epsilon^{\mu\nu\rho\sigma}\varpi_{\nu\rho}u_\sigma \, .
\end{equation}
For a free, massless Dirac field, the AVE conductivity is given by~\cite{Vilenkin:1979ui,Buzzegoli:2017cqy,Buzzegoli:2018wpy,Buzzegoli:2020ycf}
\begin{equation}
W_A^{(1/2)} = \left(\frac{T^2}{6} + \frac{\mu^2+\mu_A^2}{2\pi^2}\right)T,
\end{equation}
where $\mu=T \zeta$ is the vector chemical potential and $\mu_A$ is the axial chemical potential. 
While the AVE is generally understood as originating from spin-rotation coupling~\cite{Vilenkin:1979ui,Buzzegoli:2017cqy,Buzzegoli:2018wpy,Buzzegoli:2021jeh,Buzzegoli:2020ycf,Buzzegoli:2022qrr}—much like spin polarization—in the massless case, the thermal ($T^2$) and chemical potential ($\mu^2, \mu_A^2$) contributions to the conductivity are tied to the mixed-gravitational and chiral anomalies, respectively~\cite{Sadofyev:2010is,Landsteiner:2011cp,Avkhadiev:2017fxj,Buzzegoli:2020ycf,Prokhorov:2020npf,Prokhorov:2021bbv,Prokhorov:2022rna,Prokhorov:2022snx}.

The connection between the massless AVE conductivity and quantum anomalies for the spin-$3/2$ Rarita-Schwinger field was recently established in~\cite{Prokhorov:2021bbv,Prokhorov:2022rna,Prokhorov:2022snx}. Because the standard Rarita-Schwinger field exhibits Velo-Zwanziger pathologies~\cite{Velo:1969bt,Hortacsu:1974bm,Prokhorov:2021bbv,Prokhorov:2022rna,Prokhorov:2022snx} when coupled to background fields, the analyses in~\cite{Prokhorov:2021bbv,Prokhorov:2022rna,Prokhorov:2022snx} relied on the extended Rarita-Schwinger-Adler framework~\cite{Adler:2017shl}. In this work, we instead study a free, massive Rarita-Schwinger field. In this section, we compute its massive AVE conductivity, propose a resolution to the Velo-Zwanziger pathologies inspired by the spin polarization formula~\eqref{eq:SpinPol3/2}, and analyze its massless limit.

The axial current can be obtained via the Noether's theorem from the Lagrangian \eqref{rs_lagrangian} and the chiral transformation $\Psi_\mu\to\e^{\ii\theta\gamma_5}\Psi_\mu$:
\begin{equation}\label{eq:AxialCurrent}
\widehat{j}_A^\mu(x) = - \bar{\Psi}_\alpha(x) \gamma^\mu \gamma^5 \Psi^\alpha(x).
\end{equation}
Using the equations of motion~\eqref{eq:EoMAll}, we find
\begin{equation}
\partial_\mu \widehat{j}_A^\mu(x) = - 2\ii m\bar\Psi_\alpha(x) \gamma^5 \Psi^\alpha(x),
\end{equation}
which is conserved for a free and massless field, in exact analogy with the Dirac case.
Using Eqs.~\eqref{eq:WignBilin}, the mean value of the axial current is related to the direct axial  component of the Wigner function as
\begin{equation}
j^\mu_A(x) = \langle \wj^\mu_A(x) \rangle = -\int \di^4 p\; \mathcal{A}^{\mu\alpha}_\alpha(x,p),
\end{equation}
which, for a free field, decomposes into the sum of particle and antiparticle contributions:
\begin{equation}
j^\mu_A(x) =  -\int \di^4 p\; \left[\mathcal{A}^{\mu\alpha}_{\alpha,+}(x,p)
    + \mathcal{A}^{\mu\alpha}_{\alpha,-}(x,p)\right].
\end{equation}
The AVE conductivity is obtained by evaluating the expectation value of the axial current under the statistical operator~\eqref{eq:GERho} up to linear order in vorticity, namely:
\begin{equation}\label{eq:DeltaVortjA}
\Delta_\varpi j^\mu_A(x)_{\rm GE} =  -\int \di^4 p\; \left[\Delta_\varpi\mathcal{A}^{\mu\alpha}_{\alpha,+}(x,p)_{\rm GE}
    + \Delta_\varpi\mathcal{A}^{\mu\alpha}_{\alpha,-}(x,p)_{\rm GE}\right] = W_A^{(3/2)}\frac{\omega^\mu}{T}.
\end{equation}
Note that the global thermal equilibrium statistical operator~\eqref{eq:GERho} can be recovered from the GLW LTE statistical operator~\eqref{eq:StatOperGLW} by restricting the thermodynamic fields $\beta$ and $\SP$ to their global equilibrium form: $\beta_\mu(x)= b_\mu + \varpi_{\mu\nu}x^\nu$ and $\SP_{\rho\nu}=\varpi_{\rho\nu}$, with constant $b$ and $\varpi$.
Because our goal is to evaluate the AVE in a fluid rather than to extract a momentum spectrum in heavy-ion collisions, we evaluate the LTE statistical operator~\eqref{eq:StatOperGLW} on a generic spacelike 3D hypersurface with a single branch, see discussion in Section~\ref{subsec:hypersurface-int}. Furthermore, in global thermal equilibrium, the angular momentum-boost operator~\eqref{eq:JxOper} is a global conserved charge independent of the choice of hypersurface. For simplicity, we choose an hypersurface with normal vector $n^\mu=\delta^\mu_0$.

Consequently, the linear response of the Wigner function to thermal vorticity in global equilibrium~\eqref{eq:GERho} is obtained from the LTE linear response to $\Delta\beta$ and $\SP$ derived in Section~\ref{subsec:SpinLTEGLW} by setting  $\beta_\mu(x)= b_\mu + \varpi_{\mu\nu}x^\nu$, $\SP_{\rho\nu}=\varpi_{\rho\nu}$ and $n^\mu=\delta^\mu_0$.
Since the linear response to $\Delta\beta$ vanishes in the GLW pseudogauge, the direct axial Wigner function linear response to thermal vorticity in global equilibrium is obtained from Eqs.~\eqref{eq:DeltaASeriesSP} as
\begin{subequations}
	\begin{align}
	\Delta_\varpi \mathcal{A}^{\mu,\alpha}_{\alpha,+}(x,\,p)_{\rm GE} =&
		\sum_{N=0}^\infty \frac{1}{N!} D^N_y(\bar{y})
		\left.\left\{ -\frac{G(q)}{2} A^{\mu,\lambda,\rho\nu}(q)_{\rm GLW}
		\frac{\delta^0_\lambda}{|p_0|}\varpi_{\rho\nu}\right\}\right|_{q=0,y=x},\nonumber\\
	\label{eq:DeltaASeriesVort}
    =& -\frac{G(0)}{2} A^{\mu,\lambda,\rho\nu}(0)_{\rm GLW}
		\frac{\delta^0_\lambda}{|p_0|}\varpi_{\rho\nu}.
	\end{align}
\end{subequations}
Recalling Eq.~\eqref{eq:TraceAxialGLW}
\begin{equation}
A^{\mu,\lambda,\rho\nu}(0)_{\rm GLW} = \frac{40}{3} \epsilon^{\mu\rho\nu\tau} p_\tau p^\lambda,
\end{equation}
and the property
\begin{equation}
G(0) = -  \frac{1}{2(2\pi)^3}\frac{\delta(p_0-E_p)}{2E_p}n_F(x,p)(1-n_F(x,p)),
\end{equation}
we obtain for the particle component
\begin{equation}
\Delta_\varpi \mathcal{A}^{\mu,\alpha}_{\alpha,+}(x,\,p)_{\rm GE} = \frac{10}{3(2\pi)^3}\frac{\delta(p_0-E_p)}{2E_p}\epsilon^{\mu\rho\nu\tau} p_\tau
	\varpi_{\rho\nu}\,n_F(x,p)(1-n_F(x,p)),
\end{equation}
and similarly for the anti-particle component,
\begin{equation}
\Delta_\varpi \mathcal{A}^{\mu,\alpha}_{\alpha,-}(x,\,p)_{\rm GE} = \frac{10}{3(2\pi)^3}\frac{\delta(p_0-E_p)}{2E_p}\epsilon^{\mu\rho\nu\tau} p_\tau
	\varpi_{\rho\nu}\,\bar{n}_F(x,p)(1-\bar{n}_F(x,p)),
\end{equation}
where
\begin{equation}
n_F(x,p) = \frac{1}{\e^{\beta(x)\cdot p - \zeta(x)}+1},\quad
\bar{n}_F(x,p) = \frac{1}{\e^{\beta(x)\cdot p + \zeta(x)}+1}.
\end{equation}
Plugging these results into Eq.~\eqref{eq:DeltaVortjA}, noting that
\begin{equation}
\int\frac{\di^3 p}{2 E_p}\; p_\tau\left[n_F(1-n_F) + \bar{n}_F(1-\bar{n}_F)\right]=
u_\tau\int\frac{\di^3 p}{2 E_p}\; p\cdot u\left[n_F(1-n_F) + \bar{n}_F(1-\bar{n}_F)\right],
\end{equation}
and using the definition~\eqref{eq:RotThermalVort}, we arrive at the AVE current expression
\begin{equation}
\Delta_\varpi j^\mu_A(x)_{\rm GE} = W_A^{(3/2)}\frac{\omega^\mu}{T},
\end{equation}
with the conductivity coefficient
\begin{equation}
W_A^{(3/2)} = \frac{10}{3(2\pi)^3} \int\di^3 p\; \frac{p\cdot u}{E_p}
\left[n_F(x,p)(1-n_F(x,p)) + \bar{n}_F(x,p)(1-\bar{n}_F(x,p))\right].
\end{equation}
This quantity is a Lorentz scalar and can be conveniently evaluated in the local rest frame, where $u^\mu=\delta^\mu_0$ and
\begin{equation}
n_F(x,p) = \frac{1}{\e^{|\beta|(E_p-\mu)}+1},\quad
\bar{n}_F(x,p) = \frac{1}{\e^{|\beta|(E_p+\mu)}+1}.
\end{equation}
Noting that
\begin{equation}
n_F(x,p)(1-n_F(x,p)) = -\frac{1}{|\beta|}\frac{\partial}{\partial E_p}n_F(x,p),
\end{equation}
and similarly for $\bar{n}_F$, and integrating by parts yields
\begin{equation}\label{eq:WA_RS}
W^{(3/2)}_A(x) = \frac{10}{3}\frac{1}{2\pi^2|\beta|} \int_0^\infty\di |\bm{p}|\; \frac{E_p^2 +|\bm{p}|^2}{E_p}  \left[n_F(x,p) + \bar{n}_F(x,p)\right].
\end{equation}
The AVE conductivity of the spin-$3/2$ Rarita-Schwinger field thus matches that of a free (massive or massless) Dirac field~\cite{Buzzegoli:2017cqy,Buzzegoli:2021wlg}, up to an overall factor of $10/3$. The expression~\eqref{eq:WA_RS} is smooth in the massless limit, yielding
\begin{equation}\label{eq:WA_RS_massless}
\lim_{m\to 0}W^{(3/2)}_A(x) = \frac{10}{3}\left(\frac{T^2}{6} + \frac{\mu^2}{2\pi^2} \right)T.
\end{equation}
This outcome differs from the result in~\cite{Prokhorov:2021bbv,Prokhorov:2022rna,Prokhorov:2022snx} where the numerical prefactor is $5$ in accordance with the chiral anomaly, rather than $10/3$. As emphasized, the free field framework used to derive Eq.~\eqref{eq:WA_RS} cannot be directly extended to include minimal couplings to external gauge or gravitational fields due to the Velo-Zwanziger pathologies. Consequently, we cannot draw direct conclusions regarding its relationship with quantum anomalies. While this connection has been established for massless fields in~\cite{Prokhorov:2021bbv,Prokhorov:2022rna,Prokhorov:2022snx} within the Rarita-Schwinger-Adler formulation, non-minimal couplings to external fields have been explored for the massive field as an alternative approach; see~\cite{Deser:2000dz,Pascalutsa:2006up,Porrati:2009bs} and references therein.

To connect the axial current to the spin vector, we can use the spin polarization formula~\eqref{eq:SpinPol3/2} to define a modified axial current as
\begin{equation}\label{eq:jA1}
\widehat{j}_{A1}^\mu(x) = - \bar{\Psi}_\alpha(x) \gamma^\mu \gamma^5 \Psi^\alpha(x)
    +\epsilon^{\mu\alpha\beta\sigma}\bar{\Psi}_\beta(x)\frac{\codev_\sigma}{m} \Psi_\alpha(x).
\end{equation}
This current can be derived by adding the surface divergence term
\begin{equation}\label{eq:DeltaLag}
\Delta\mathcal{L} = \partial_\mu \left( \frac{\mathrm{i}}{m} \epsilon^{\mu\alpha\beta\sigma}
    \bar{\Psi}_\beta(x) \codev_\sigma \gamma^5 \Psi_\alpha(x) \right)
= \frac{\mathrm{i}}{m} \epsilon^{\mu\alpha\beta\sigma}
    \bar{\Psi}_\beta(x) \left( \codevR_\sigma\codevL_\mu - \codevR_\mu\codevL_\sigma\right)\gamma^5 \Psi_\alpha(x)
\end{equation}
to the Lagrangian~\eqref{rs_lagrangian} and applying Noether's theorem as before. Since a total divergence does not alter the classical equations of motion, and noting that $\epsilon^{\mu\alpha\beta\sigma} \partial_\mu \partial_\sigma = 0$, it straightforwardly follows that
\begin{equation}
\partial_\mu \widehat{j}^\mu_{A1} = \partial_\mu \widehat{j}^\mu_A
    = -2\mathrm{i} m \bar{\Psi}_\alpha(x)\gamma^5\Psi^\alpha(x).
\end{equation}
Interestingly, introducing an external gauge field via minimal substitution,
\begin{equation}
\codevR_\mu \to \codevR_\mu + \mathrm{i} q A_\mu,\qquad
\codevL_\mu \to \codevL_\mu - \mathrm{i} q A_\mu,
\end{equation}
yields
\begin{equation}\label{eq:DeltaLagD}
\Delta\mathcal{L} \to \frac{\mathrm{i}}{m} \epsilon^{\mu\alpha\beta\sigma}
    \bar{\Psi}_\beta(x) \left( \codevR_\sigma\codevL_\mu - \codevR_\mu\codevL_\sigma\right)\gamma^5 \Psi_\alpha(x)
    - \frac{2q}{m} \tilde{F}^{\alpha\beta}\bar{\Psi}_\beta(x)\gamma^5 \Psi_\alpha(x),
\end{equation}
where $\tilde{F}_{\alpha\beta} = \frac{1}{2}\epsilon_{\alpha\beta\mu\sigma}F^{\mu\sigma}$ is the dual electromagnetic field strength tensor. This term introduces a non-minimal gauge-invariant coupling, thereby eliminating the Velo-Zwanziger pathologies~\cite{Velo:1969bt}, restoring the proper constraints $\gamma^\mu \Psi_\mu=0$ and $D^\mu\Psi_\mu=0$, and preserving causal signal propagation.

However, the extra term in the modified axial current~\eqref{eq:jA1} contributes to the AVE. By construction, this term contains a contribution that can be evaluated using Eq.~\eqref{eq:DeltaF5SeriesSP}. Consequently, the resulting AVE conductivity for the current~\eqref{eq:jA1} is
\begin{equation}\label{eq:WA1_RS}
W^{(3/2)}_{A1}(x) = \frac{10}{2\pi^2|\beta|} \int_0^\infty\mathrm{d} |\bm{p}|\; \frac{E_p^2 +|\bm{p}|^2}{E_p}  \left[n_F(x,p) + \bar{n}_F(x,p)\right],
\end{equation}
yielding an overall numerical prefactor of $10$, rather than the factor of $10/3$ obtained previously or the factor of $5$ found in the massless Rarita-Schwinger-Adler theory. Despite the resolution of the Velo-Zwanziger pathologies and the existence of a continuous massless limit for expression~\eqref{eq:WA1_RS}, we cannot draw conclusions regarding a direct connection with quantum anomalies in the massless limit, as both the modified axial current and the extra Lagrangian term diverge as $m \to 0$.

It is essential to emphasize that the prefactors ($10/3$ and $10$) obtained here represent the purely kinematic spin-rotation coupling of the free massive field, rather than an anomaly-driven transport coefficient. Consequently, the discrepancy with the value of $5$ derived in the Rarita–Schwinger–Adler theory reflects the well-known non-commutativity of the massless limit when interaction constraints and quantum anomalies are taken into account. Because the connection with quantum anomalies is strictly valid only for massless fields, the general interpretation of the AVE as originating from spin-rotation coupling is sufficient to account for results~\eqref{eq:WA_RS} and \eqref{eq:WA1_RS}. A full resolution of the discrepancies and discontinuities between the massive and massless Rarita-Schwinger field theories remains beyond the scope of the present work. Promising avenues for future research include re-deriving the relation between the mean spin vector and the Wigner function directly for a massless field—in order to infer an optimal definition of the axial current linked to spin polarization—or analyzing global thermal equilibrium in the presence of external electromagnetic fields following the methods established for Dirac fields~\cite{Buzzegoli:2020ycf}. In this way, matching operator identities with expectation values makes it possible to determine which components of the conductivities are governed by the chiral anomaly, while extensions to curved spacetimes identifies manifestations of the chiral gravitational anomaly in the expectation value of the axial current~\cite{Prokhorov:2022udo}.

\section{Summary and outlook}
\label{sec:Summary}
In this paper, we have formulated the quantum field theory of the covariant Wigner function for a massive spin-$3/2$ Rarita-Schwinger field, defined in Eq.~\eqref{rswignerop}, and investigated its connection to the spin density matrix and mean spin polarization vector.

Our primary general result, given in Eq.~\eqref{eq:SpinPol3/2}, is the exact formula relating the mean spin polarization vector $S^\mu(p)$ to the covariant Wigner function components, see Eq.~\eqref{rs_clifford}:
\begin{equation}
	S^\mu(p) = \frac{1}{2}\frac{\int\di p^0\int\di\Sigma\cdot p\,\left[g^{\alpha\beta}\mathcal{A}^\mu_{\alpha\beta,+}(x,p)+2\ii\epsilon^{\mu\alpha\beta\sigma}\frac{p_\sigma}{m}\mathcal{F}_{\alpha\beta,+}(x,p)\right]}{\int\di p^0\int\di\Sigma\cdot p\, \mathcal{F}^\alpha_{\ \alpha,+}(x,p)}\;.
\end{equation}
Unlike the spin-$1/2$ Dirac case, the spin-$3/2$ polarization vector receives an additional contribution from the dual axial component $\mathcal{F}_{5,+}^\mu=2\ii\epsilon^{\mu\alpha\beta\sigma}\frac{p_\sigma}{m}\mathcal{F}_{\alpha\beta,+}$ associated with the tensor structure of the vector-spinor field.

For applications to heavy-ion collisions, we evaluated the Wigner function in local thermal equilibrium using linear response theory. In the de Groot--van Leeuwen--van Weert (GLW) pseudo-gauge, the spin polarization is driven solely by the spin potential $\SP_{\rho\nu}$ and chemical potential gradients, as shown in Eq.~\eqref{eq:FinalLTEPolGLW}:
\begin{multline}
S^\mu(p)_{\rm GLW} =-\frac{5}{8 m N_p}
	\int_{\Sigma_{\rm D}}\!\!\di\Sigma\cdot p \, n_F(x,p)(1-n_F(x,p))\epsilon^{\mu\rho\nu\tau}p_\tau\\
\times\sum_{\bar{y}(x,p)}\text{sgn} [p\cdot n(\bar{y})]\left\{\SP_{\rho\nu}(\bar{y})
	-\frac{1}{3}\frac{2 n_\rho(\bar{y})}{p\cdot n(\bar{y})}\partial_\nu^{\bar{y}}\zeta(\bar{y})\right\}\;.
\end{multline}
In contrast, in the Belinfante pseudo-gauge, where the spin potential is absent, the spin polarization is generated by the thermal vorticity $\varpi_{\rho\nu}$, thermal shear $\xi_{\rho\nu}$, and chemical potential gradients, as expressed in Eq.~\eqref{eq:FinalLTEPolBel}:
\begin{multline}
S^\mu(p)_{\rm B} =-\frac{5}{8 m N_p}
	\int\di\Sigma\cdot p \, n_F(1-n_F)\epsilon^{\mu\rho\nu\tau}p_\tau\\
	\times\sum_{\bar{y}(x,p)}\text{sgn} [p\cdot n(\bar{y})]\left\{\varpi_{\rho\nu}(\bar{y})
	+ \frac{2 n_\rho(\bar{y})}{p\cdot n(\bar{y})} \left[\xi_{\nu\alpha}(\bar{y})p^\alpha
	-\frac{1}{3}\partial_\nu^{\bar{y}}\zeta(\bar{y})\right]\right\}\;.    
\end{multline}
Similarly, for the canonical pseudo-gauge we obtained
\begin{multline}
S^\mu(p)_{\rm C} = -\frac{5}{8 m N_p}
	\int\di\Sigma\cdot p \, n_F(1-n_F)\epsilon^{\mu\rho\nu\tau}p_\tau
	\sum_{\bar{y}(x,p)}\text{sgn} [p\cdot n(\bar{y})]
	\left\{\SP_{\rho\nu}(\bar{y})\vphantom{\frac{2 n_\rho(\bar{y})}{p\cdot n(\bar{y})}}\right. \\
	\left.+ \frac{2}{3}\frac{n_\rho(\bar{y})}{p\cdot n(\bar{y})} \left[\left(\xi_{\nu\alpha}(\bar{y})
	+\SP_{\nu\alpha}(\bar{y}) - \varpi_{\nu\alpha}(\bar{y})\right)p^\alpha-\partial_\nu^{\bar{y}}\zeta(\bar{y})\right]\right\}.
\end{multline}
These results confirm the expectation that thermal-vorticity in every pseudo-gauge and thermal-shear in the Belinfante pseudo-gauge induced spin polarization at local equilibrium scales with the spin $S$ of the field as $S(S+1)/3$ as in global equilibrium. On the contrary, we found that the spin Hall effect, which is the spin polarization induced by the gradient of reduced chemical potential, and the thermal-shear contribution in the canonical pseudo-gauge scales differently with the spin of the field. It is important to note that the canonical and Belinfante spin polarization predictions differ also when the spin degrees of freedom already relaxed, that is when the spin potential equals the thermal vorticity $\SP=\varpi$. Specifically, in the canonical pseudo-gauge, the thermal-shear contribution is suppressed by an extra factor $1/3$ compared to the Belinfante one. This observation makes the measurements of local spin polarization for the $\Omega^-$ particularly important to determine which pseudo-gauge is more appropriate to describe heavy-ion collisions.

Future work will be dedicated to generating quantitative predictions for $\Omega^-$ hyperon spin polarization in heavy-ion collisions. Specifically, following the framework developed recently in Ref.~\cite{Buzzegoli:2026oxc}, ideal spin hydrodynamics in the GLW formalism will utilize Eq.~\eqref{eq:FinalLTEPolGLW}, while standard relativistic hydrodynamic simulations in the Belinfante pseudo-gauge will implement Eq.~\eqref{eq:FinalLTEPolBel}.

\begin{acknowledgments}
RS acknowledges the support of the Humboldt research fellowship for postdocs from the Alexander von Humboldt Foundation and COST Actions EXPAND CA24149 \& SHARP CA24159 supported by COST (European Cooperation in Science and Technology).
\end{acknowledgments}

\appendix

\section{Matrix elements of the Pauli-Lubanski operator in the spin-3/2 Rarita-Schwinger field representation}
\label{app:AppIntertwiner}
In this appendix, we prove identity~\eqref{eq:intertwinerLubansky}, which we restate here for convenience:
\begin{equation}\label{eq:intertwinerLubanskyApp}
\left[ D^{3/2}([p])^{-1} D^{3/2}(J_{\nu\rho}) D^{3/2}([p]) \right]_{sr} = -\frac{1}{2m} \bar{u}_s^\alpha(p) \mathcal{J}_{\nu\rho, \alpha \beta} u_r^\beta(p).
\end{equation}
Since a Lorentz transformation represents a symmetry, its action on a quantum field is generated by a unitary operator.
The unitary operator realizing the Lorentz transformation $\Lambda$ on a single-particle state is given by~\cite[Eq.~(2.5.23)]{Weinberg:1995mt}
\begin{equation}
U(\Lambda)\Psi_{p,r} = \sqrt{\frac{(\Lambda p)^0}{p^0}} \sum_{s'}
    D^S_{s',r}\left( W(\Lambda,p)\right) \Psi_{\Lambda p,s'},
\end{equation}
where $W(\Lambda, p) = [\Lambda p]^{-1} \Lambda [p]$ is the Wigner little group element.
It follows that for the Rarita-Schwinger field transforming under the finite-dimensional representation $\mathcal{D}^{16}(\Lambda)$, the single-particle wavefunctions $u_r(p)$ satisfy the general intertwiner relation:
\begin{equation}
\mathcal{D}^{16}(\Lambda)^\mu_{\ \nu} u_r^\nu(p) = \sum_{s'} u_{s'}^\mu(\Lambda p) D^{(j)}_{s'r}(W(\Lambda, p)).
\end{equation}
Evaluating this relation in the rest frame, $p = k$ (where $k^\mu=(m,\,\bm{0})$), and noting that $[k] = \mathbb{1}$, the Wigner element simplifies to $W(\Lambda, k) = [\Lambda k]^{-1} \Lambda$:
\begin{equation}\label{eq:MasterIntertwiner}
\mathcal{D}^{16}(\Lambda)^\mu_{\ \nu} u_r^\nu(k) = \sum_{s'} u_{s'}^\mu(\Lambda k) D^{3/2}_{s'r}\left(  [\Lambda k]^{-1} \Lambda \right).
\end{equation}
Evaluating this identity for infinitesimal rotations and boosts yields the master intertwiner relation
\begin{equation}\label{eq:MasterIntertwinerJ}
\mathcal{J}_{\alpha\beta, \ \nu}^{\ \ \mu}\, u_r^\nu(k) = \sum_{s'} u_{s'}^\mu(k) \left[ D^{3/2}(J_{\alpha\beta}) \right]_{s'r},
\end{equation}
relating the irreducible representation of the Lorentz generators $D^{3/2}(J_{\alpha\beta})$ and its 16-dimensional field representation
\begin{equation}\label{eq:16DLorentzJApp}
\mathcal{J}_{\nu\rho, \ \ \beta}^{\ \ \alpha} = \frac{1}{2}\sigma_{\nu\rho} \delta^\alpha_\beta
    + \ii\left(\delta^\alpha_\nu g_{\rho\beta} - \delta^\alpha_\rho g_{\nu\beta}\right).
\end{equation}
Consider first an infinitesimal rotation $R = \mathbb{1} + \frac{1}{2}\omega^{ij}\mathcal{J}_{ij}$.
A pure rotation leaves the rest-frame momentum invariant, $R k = k$, yielding $[R k] = [k] = \mathbb{1}$ and the Wigner rotation $W(R, k) = \mathbb{1}^{-1} R\, \mathbb{1} = R$. Inserting these relations into the master equation~\eqref{eq:MasterIntertwiner}, the left-hand side gives
\begin{equation}
\mathcal{D}^{16}(R)^\mu_{\ \nu} u_r^\nu(k) = \left( \delta^\mu_\nu + \frac{1}{2}\omega^{ij} \mathcal{J}_{ij, \ \ \nu}^{\ \ \mu} \right) u_r^\nu(k) = u_r^\mu(k) + \frac{1}{2}\omega^{ij} \mathcal{J}_{ij, \ \ \nu}^{\ \ \mu} u_r^\nu(k),
\end{equation}
while, using $u_{s'}^\mu(R k) = u_{s'}^\mu(k)$, the right-hand side becomes
\begin{equation}
\begin{split}
\sum_{s'} u_{s'}^\mu(k) D^{3/2}_{s'r}(R) =& \sum_{s'} u_{s'}^\mu(k) \left( \delta_{s'r} + \frac{1}{2}\omega^{ij} \left[ D^{3/2}(J_{ij}) \right]_{s'r} \right)\\
=& u_r^\mu(k) + \frac{1}{2}\omega^{ij} \sum_{s'} u_{s'}^\mu(k) \left[ D^{3/2}(J_{ij}) \right]_{s'r}.    
\end{split}
\end{equation}
Equating terms of order $\mathcal{O}(\omega)$ yields the exact unprojected identity for rotations:
\begin{equation}
\mathcal{J}_{ij, \ \ \nu}^{\ \ \mu} u_r^\nu(k) = \sum_{s'} u_{s'}^\mu(k) \left[ D^{3/2}(J_{ij}) \right]_{s'r}.
\end{equation}
Similarly, for an infinitesimal boost $B = \mathbb{1} + \frac{1}{2}\omega^{0i}\mathcal{J}_{0i}$, we have $B k = k + \delta k \neq k$. Since the boost taking $k$ to $B k$ is $B$ itself, $[B k] = B$, and the Wigner rotation vanishes: $W(B, k) = B^{-1} B = \mathbb{1}$ (pure boosts at rest produce no little group rotation).
Inserting these properties into the master equation~\eqref{eq:MasterIntertwiner} yields
\begin{equation}
\mathcal{D}^{16}(B)^\mu_{\ \nu} u_r^\nu(k)
    = u_r^\mu(k) + \frac{1}{2}\omega^{0i} \mathcal{J}_{0i, \ \ \nu}^{\ \ \mu} u_r^\nu(k) + \frac{1}{2}\omega^{0i} \sum_{s'} u_{s'}^\mu(k) \left[ D^{3/2}(J_{0i}) \right]_{s'r}.
\end{equation}
This completes the proof of Eq.~\eqref{eq:MasterIntertwinerJ}.

To isolate the matrix elements $\left[ D^{3/2}(J_{\alpha\beta}) \right]_{sr}$, we contract both sides from the left with the Dirac adjoint wavefunction $\bar{u}_{s \mu}(k)$:
\begin{equation}
\bar{u}_{s \mu}(k) \mathcal{J}_{\alpha\beta, \ \ \nu}^{~\ \ \mu} u_r^\nu(k) = \sum_{s'} \bar{u}_{s \mu}(k) u_{s'}^\mu(k) \left[ D^{3/2}(J_{\alpha\beta}) \right]_{s'r}.
\end{equation}
Substituting the normalization condition $\bar{u}_{s \mu}(k) u_{s'}^\mu(k) = -2m \delta_{ss'}$ into the right-hand side gives
\begin{equation}
\bar{u}_{s \mu}(k) \mathcal{J}_{\alpha\beta, \ \ \nu}^{~\ \ \mu} u_r^\nu(k) = \sum_{s'} \left( -2m \delta_{ss'} \right) \left[ D^{3/2}(J_{\alpha\beta}) \right]_{s'r}.
\end{equation}
The Kronecker delta collapses the sum over $s'$, leaving
\begin{equation}
\bar{u}_{s \mu}(k) \mathcal{J}_{\alpha\beta, \ \ \nu}^{~\ \ \mu} u_r^\nu(k) = -2m \left[ D^{3/2}(J_{\alpha\beta}) \right]_{sr}.
\end{equation}
Dividing by $-2m$ leads directly to the rest-frame identity:
\begin{equation}\label{eq:intertwinerRest}
\left[ D^{3/2}(J_{\alpha\beta}) \right]_{sr} = -\frac{1}{2m} \bar{u}_{s \mu}(k) \mathcal{J}_{\alpha\beta, \ \ \nu}^{~\ \ \mu} u_r^\nu(k).
\end{equation}
Note that relation~\eqref{eq:intertwinerRest} involves spinors evaluated at the rest-frame momentum $k$, whereas the matrix elements in Eq.~\eqref{eq:intertwinerLubanskyApp} also involve the standard Lorentz transformation. Combining the transformation properties of the spinors under standard Lorentz boosts with Eq.~\eqref{eq:intertwinerRest} yields Eq.~\eqref{eq:intertwinerLubanskyApp}.
Indeed, the vector-spinor wavefunction $u_s^\mu(p)$ transforms under a standard Lorentz transformation $[p]$ as a direct tensor product of vector and spinorial transformations:
\begin{equation}
u_s^\mu(p) = \mathcal{D}^{16}([p])^\mu_{\ \nu} u_s^\nu(k) = [p]^\mu_{\ \nu} D^{1/2}([p]) u_s^\nu(k),
\end{equation}
where $D^{1/2}([p])$ is the standard Lorentz transformation in the spin-$1/2$ representation. This relation follows from expressing the wavefunction as a matrix element of the field operator (see, e.g., Eq.~\eqref{rsfieldexp}):
\begin{equation}
u^\mu_s(p) = \langle 0 | \psi^\mu(0) | p, s \rangle.
\end{equation}
By definition, the standard Lorentz transformation maps $k$ to $p$, and by Wigner's theorem, it acts via the unitary boost operator $U([p])$:
\begin{equation}
\vert{}p, s\rangle = U([p]) \vert{} k, s \rangle.
\end{equation}
Using the invariance of the vacuum, $U([p])\vert{}0\rangle = \vert{}0\rangle$ (which implies $\langle 0 \vert{} = \langle 0 \vert{} U([p])^{-1}$), we write
\begin{equation}
u_s^\mu(p) = \langle 0 \vert{} U([p])^{-1} \psi^\mu(0) U([p]) \vert{} k, s \rangle.
\end{equation}
The Rarita-Schwinger field operator transforms under a Lorentz transformation $\Lambda$ as
\begin{equation}
U(\Lambda)^{-1} \psi^\mu(x) U(\Lambda) = \Lambda^\mu_{\ \nu} D^{1/2}(\Lambda) \psi^\nu(\Lambda^{-1} x),
\end{equation}
that is, via the 16D finite-dimensional representation $\Lambda^\mu_{\ \nu} \otimes D^{1/2}(\Lambda)$.
Setting $x=0$ and choosing $\Lambda=[p]$ gives
\begin{equation}
U([p])^{-1} \psi^\mu(0) U([p]) = [p]^\mu_{\ \nu} D^{1/2}([p]) \psi^\nu(0).
\end{equation}
Substituting this identity into the matrix element yields
\begin{equation}
\begin{split}
u_s^\mu(p) =& \langle 0 \vert{} U([p])^{-1} \psi^\mu(0) U([p]) \vert{} k, s \rangle
 = \langle 0 \vert{} \left( [p]^\mu_{\ \nu} D^{1/2}([p]) \psi^\nu(0) \right) \vert{} k, s \rangle\\
 =& [p]^\mu_{\ \nu} D^{1/2}([p]) \langle 0 \vert{} \psi^\nu(0) \vert{} k, s \rangle
 = [p]^\mu_{\ \nu} D^{1/2}([p]) u_s^\nu(k).
\end{split}
\end{equation}
Inverting this relation gives
\begin{equation}
\label{eq:BoostedSpinor}
u(k) = \mathcal{D}^{16}([p])^{-1} u(p),\quad
     \bar{u}(k) = \bar{u}(p) \mathcal{D}^{16}([p]).
\end{equation}
Finally, we can boost the identity~\eqref{eq:intertwinerRest}.
Under the standard Lorentz transformation $[p]$, the Lie algebra generators transform as rank-2 tensors:
\begin{align}
\label{eq:tensor_transform_4d}
D^{3/2}([p])^{-1} D^{3/2}(J_{\nu\rho}) D^{3/2}([p]) =& ([p]^{-1})^\alpha_{\ \nu} ([p]^{-1})^\beta_{\ \rho} D^{3/2}(J_{\alpha\beta}),\\
\label{eq:tensor_transform_16d}
\mathcal{D}^{16}([p])^{-1} \mathcal{J}_{\nu\rho} \mathcal{D}^{16}([p]) =& ([p]^{-1})^\alpha_{\ \nu} ([p]^{-1})^\beta_{\ \rho} \mathcal{J}_{\alpha\beta}.
\end{align}
Expressing the rest-frame spinors in terms of the boosted wavefunctions $u(p)$ and using the relations~\eqref{eq:BoostedSpinor}, we find:
\begin{align}
-\frac{1}{2m} \bar{u}_{s \alpha}(p) \mathcal{J}_{\nu\rho, \ \ \beta}^{~\ \ \alpha} u_r^\beta(p) 
    &= -\frac{1}{2m} \bar{u}_{s \alpha}(k) \left( \mathcal{D}^{16}([p])^{-1} \mathcal{J}_{\nu\rho} \mathcal{D}^{16}([p]) \right)_{\ \ \beta}^{\alpha} u_r^\beta(k) \nonumber \\
&= ([p]^{-1})^\alpha_{\ \nu} ([p]^{-1})^\beta_{\ \rho} \left( -\frac{1}{2m} \bar{u}_{s\mu}(k) \mathcal{J}_{\alpha\beta, \ \ \delta}^{~\ \ \mu}\, u_r^\delta(k) \right) \nonumber \\
&= ([p]^{-1})^\alpha_{\ \nu} ([p]^{-1})^\beta_{\ \rho} \left[ D^{3/2}(J_{\alpha\beta}) \right]_{sr},
\end{align}
where identity~\eqref{eq:intertwinerRest} has been applied.
Comparing this result directly with Eq.~\eqref{eq:tensor_transform_4d} completes the proof of the identity:
\begin{equation}
    \left[ D^{3/2}([p])^{-1} D^{3/2}(J_{\nu\rho}) D^{3/2}([p]) \right]_{sr} = -\frac{1}{2m} \bar{u}_s^\alpha(p) \mathcal{J}_{\nu\rho, \alpha \beta} u_r^\beta(p).
\end{equation}

\providecommand{\href}[2]{#2}\begingroup\raggedright\endgroup


\begin{thebibliography}{10}
	
	\bibitem{STAR:2017ckg}
	{\scshape STAR} collaboration, \emph{{Global $\Lambda$ hyperon polarization in nuclear collisions: evidence for the most vortical fluid}}, \href{https://doi.org/10.1038/nature23004}{\emph{Nature} {\bfseries 548} (2017) 62} [\href{https://arxiv.org/abs/1701.06657}{{\ttfamily 1701.06657}}].
	
	\bibitem{STAR:2018gyt}
	{\scshape STAR} collaboration, \emph{{Global polarization of $\Lambda$ hyperons in Au+Au collisions at $\sqrt{s_{_{NN}}}$ = 200 GeV}}, \href{https://doi.org/10.1103/PhysRevC.98.014910}{\emph{Phys. Rev. C} {\bfseries 98} (2018) 014910} [\href{https://arxiv.org/abs/1805.04400}{{\ttfamily 1805.04400}}].
	
	\bibitem{STAR:2019erd}
	{\scshape STAR} collaboration, \emph{{Polarization of $\Lambda$ ($\bar{\Lambda}$) hyperons along the beam direction in Au+Au collisions at $\sqrt{s_{_{NN}}}$ = 200 GeV}}, \href{https://doi.org/10.1103/PhysRevLett.123.132301}{\emph{Phys. Rev. Lett.} {\bfseries 123} (2019) 132301} [\href{https://arxiv.org/abs/1905.11917}{{\ttfamily 1905.11917}}].
	
	\bibitem{ALICE:2019onw}
	{\scshape ALICE} collaboration, \emph{{Global polarization of $\Lambda \bar \Lambda$ hyperons in Pb-Pb collisions at $\sqrt {s_{NN}}$ = 2.76 and 5.02 TeV}}, \href{https://doi.org/10.1103/PhysRevC.101.044611}{\emph{Phys. Rev. C} {\bfseries 101} (2020) 044611} [\href{https://arxiv.org/abs/1909.01281}{{\ttfamily 1909.01281}}].
	
	\bibitem{STAR:2022fan}
	{\scshape STAR} collaboration, \emph{{Pattern of global spin alignment of \ensuremath{\phi} and K$^{*0}$ mesons in heavy-ion collisions}}, \href{https://doi.org/10.1038/s41586-022-05557-5}{\emph{Nature} {\bfseries 614} (2023) 244} [\href{https://arxiv.org/abs/2204.02302}{{\ttfamily 2204.02302}}].
	
	\bibitem{Liang:2004ph}
	Z.-T.~Liang and X.-N.~Wang, \emph{{Globally polarized quark-gluon plasma in non-central A+A collisions}}, \href{https://doi.org/10.1103/PhysRevLett.94.102301}{\emph{Phys. Rev. Lett.} {\bfseries 94} (2005) 102301} [\href{https://arxiv.org/abs/nucl-th/0410079}{{\ttfamily nucl-th/0410079}}].
	
	\bibitem{Liang:2004xn}
	Z.-T.~Liang and X.-N.~Wang, \emph{{Spin alignment of vector mesons in non-central A+A collisions}}, \href{https://doi.org/10.1016/j.physletb.2005.09.060}{\emph{Phys. Lett. B} {\bfseries 629} (2005) 20} [\href{https://arxiv.org/abs/nucl-th/0411101}{{\ttfamily nucl-th/0411101}}].
	
	\bibitem{Becattini:2024uha}
	F.~Becattini, M.~Buzzegoli, T.~Niida, S.~Pu, A.-H.~Tang and Q.~Wang, \emph{{Spin polarization in relativistic heavy-ion collisions}}, \href{https://doi.org/10.1142/S0218301324300066}{\emph{Int. J. Mod. Phys. E} {\bfseries 33} (2024) 2430006} [\href{https://arxiv.org/abs/2402.04540}{{\ttfamily 2402.04540}}].
	
	\bibitem{STAR:2020xbm}
	{\scshape STAR} collaboration, \emph{{Global Polarization of $\Xi$ and $\Omega$ Hyperons in Au+Au Collisions at $\sqrt {s_{NN}}$ = 200 GeV}}, \href{https://doi.org/10.1103/PhysRevLett.126.162301}{\emph{Phys. Rev. Lett.} {\bfseries 126} (2021) 162301} [\href{https://arxiv.org/abs/2012.13601}{{\ttfamily 2012.13601}}].
	
	\bibitem{Alice:2026Moriond}
	C.D.~Martin, ``{Measurement of hyperon polarization in light- and heavy-ion collisions with ALICE}.'' \href{https://indico.cern.ch/event/1648766/contributions/6964252/}{Talk presented} at the \textit{60th Rencontres de Moriond: QCD \& High Energy Interactions}, La Thuile, Italy, Mar., 2026.
	
	\bibitem{Becattini:2020sww}
	F.~Becattini, \emph{{Polarization in Relativistic Fluids: A Quantum Field Theoretical Derivation}}, \href{https://doi.org/10.1007/978-3-030-71427-7_2}{\emph{Lect. Notes Phys.} {\bfseries 987} (2021) 15} [\href{https://arxiv.org/abs/2004.04050}{{\ttfamily 2004.04050}}].
	
	\bibitem{Xie:2017upb}
	Y.~Xie, D.~Wang and L.P.~Csernai, \emph{{Global \ensuremath{\Lambda} polarization in high energy collisions}}, \href{https://doi.org/10.1103/PhysRevC.95.031901}{\emph{Phys. Rev. C} {\bfseries 95} (2017) 031901} [\href{https://arxiv.org/abs/1703.03770}{{\ttfamily 1703.03770}}].
	
	\bibitem{Fu:2020oxj}
	B.~Fu, K.~Xu, X.-G.~Huang and H.~Song, \emph{{Hydrodynamic study of hyperon spin polarization in relativistic heavy ion collisions}}, \href{https://doi.org/10.1103/PhysRevC.103.024903}{\emph{Phys. Rev. C} {\bfseries 103} (2021) 024903} [\href{https://arxiv.org/abs/2011.03740}{{\ttfamily 2011.03740}}].
	
	\bibitem{Li:2026udm}
	G.-H.~Li, C.~Yi, X.-Y.~Wu, S.~Pu and G.-Y.~Qin, \emph{{Global polarization of $\Lambda$, $Xi^{-}$, and $\Omega^{-}$ hyperons in Au+Au collisions at RHIC BES-II energies}},  6, 2026.
	
	\bibitem{Becattini:2021suc}
	F.~Becattini, M.~Buzzegoli and A.~Palermo, \emph{{Spin-thermal shear coupling in a relativistic fluid}}, \href{https://doi.org/10.1016/j.physletb.2021.136519}{\emph{Phys. Lett. B} {\bfseries 820} (2021) 136519} [\href{https://arxiv.org/abs/2103.10917}{{\ttfamily 2103.10917}}].
	
	\bibitem{Liu:2021uhn}
	S.Y.F.~Liu and Y.~Yin, \emph{{Spin polarization induced by the hydrodynamic gradients}}, \href{https://doi.org/10.1007/JHEP07(2021)188}{\emph{JHEP} {\bfseries 07} (2021) 188} [\href{https://arxiv.org/abs/2103.09200}{{\ttfamily 2103.09200}}].
	
	\bibitem{Becattini:2021iol}
	F.~Becattini, M.~Buzzegoli, G.~Inghirami, I.~Karpenko and A.~Palermo, \emph{{Local Polarization and Isothermal Local Equilibrium in Relativistic Heavy Ion Collisions}}, \href{https://doi.org/10.1103/PhysRevLett.127.272302}{\emph{Phys. Rev. Lett.} {\bfseries 127} (2021) 272302} [\href{https://arxiv.org/abs/2103.14621}{{\ttfamily 2103.14621}}].
	
	\bibitem{Fu:2021pok}
	B.~Fu, S.Y.F.~Liu, L.~Pang, H.~Song and Y.~Yin, \emph{{Shear-Induced Spin Polarization in Heavy-Ion Collisions}}, \href{https://doi.org/10.1103/PhysRevLett.127.142301}{\emph{Phys. Rev. Lett.} {\bfseries 127} (2021) 142301} [\href{https://arxiv.org/abs/2103.10403}{{\ttfamily 2103.10403}}].
	
	\bibitem{Liu:2020dxg}
	S.Y.F.~Liu and Y.~Yin, \emph{{Spin Hall effect in heavy-ion collisions}}, \href{https://doi.org/10.1103/PhysRevD.104.054043}{\emph{Phys. Rev. D} {\bfseries 104} (2021) 054043} [\href{https://arxiv.org/abs/2006.12421}{{\ttfamily 2006.12421}}].
	
	\bibitem{Buzzegoli:2022qrr}
	M.~Buzzegoli, \emph{{Spin polarization induced by magnetic field and the relativistic Barnett effect}}, \href{https://doi.org/10.1016/j.nuclphysa.2023.122674}{\emph{Nucl. Phys. A} {\bfseries 1036} (2023) 122674} [\href{https://arxiv.org/abs/2211.04549}{{\ttfamily 2211.04549}}].
	
	\bibitem{Fu:2022myl}
	B.~Fu, L.-G.~Pang, H.~Song and Y.~Yin, \emph{{Signatures of a baryonic spin-Hall effect in dense QCD matter}}, \href{https://doi.org/10.1103/m5rj-rp7n}{\emph{Phys. Rev. C} {\bfseries 113} (2026) L041901} [\href{https://arxiv.org/abs/2201.12970}{{\ttfamily 2201.12970}}].
	
	\bibitem{Rarita:1941mf}
	W.~Rarita and J.~Schwinger, \emph{{On a theory of particles with half integral spin}}, \href{https://doi.org/10.1103/PhysRev.60.61}{\emph{Phys. Rev.} {\bfseries 60} (1941) 61}.
	
	\bibitem{Fierz:1939ix}
	M.~Fierz and W.~Pauli, \emph{{On relativistic wave equations for particles of arbitrary spin in an electromagnetic field}}, \href{https://doi.org/10.1098/rspa.1939.0140}{\emph{Proc. Roy. Soc. Lond. A} {\bfseries 173} (1939) 211}.
	
	\bibitem{Lurie:1968}
	D.~Lurie, \emph{Particles and Fields}, John Wiley \& Sons, Nashville, TN (Aug., 1968).
	
	\bibitem{Weinberg:1995mt}
	S.~Weinberg, \emph{{The Quantum theory of fields. Vol. 1: Foundations}}, Cambridge University Press (6, 2005), \href{https://doi.org/10.1017/CBO9781139644167}{10.1017/CBO9781139644167}.
	
	\bibitem{Pascalutsa:2006up}
	V.~Pascalutsa, M.~Vanderhaeghen and S.N.~Yang, \emph{{Electromagnetic excitation of the Delta(1232)-resonance}}, \href{https://doi.org/10.1016/j.physrep.2006.09.006}{\emph{Phys. Rept.} {\bfseries 437} (2007) 125} [\href{https://arxiv.org/abs/hep-ph/0609004}{{\ttfamily hep-ph/0609004}}].
	
	\bibitem{Sheng:2025cjk}
	X.-L.~Sheng, F.~Becattini and D.~Roselli, \emph{{Improved formula for Wigner function and spin polarization in a decoupling relativistic fluid at local thermodynamic equilibrium}}, \href{https://doi.org/10.1103/5rpc-hg99}{\emph{Phys. Rev. C} {\bfseries 114} (2026) 034903} [\href{https://arxiv.org/abs/2509.14301}{{\ttfamily 2509.14301}}].
	
	\bibitem{Hehl:1976vr}
	F.W.~Hehl, \emph{{On the Energy Tensor of Spinning Massive Matter in Classical Field Theory and General Relativity}}, \href{https://doi.org/10.1016/0034-4877(76)90016-1}{\emph{Rept. Math. Phys.} {\bfseries 9} (1976) 55}.
	
	\bibitem{Buzzegoli:2021wlg}
	M.~Buzzegoli, \emph{{Pseudogauge dependence of the spin polarization and of the axial vortical effect}}, \href{https://doi.org/10.1103/PhysRevC.105.044907}{\emph{Phys. Rev. C} {\bfseries 105} (2022) 044907} [\href{https://arxiv.org/abs/2109.12084}{{\ttfamily 2109.12084}}].
	
	\bibitem{Singh:2024qvg}
	R.~Singh, \emph{{On the nonuniqueness of the energy{\textendash}momentum and spin currents}}, \href{https://doi.org/10.1142/S0217751X26500211}{\emph{Int. J. Mod. Phys. A} {\bfseries 41} (2026) 2650021} [\href{https://arxiv.org/abs/2406.02127}{{\ttfamily 2406.02127}}].
	
	\bibitem{DeGroot:1980dk}
	S.R.~De~Groot, W.A.~Van~Leeuwen and C.G.~Van~Weert, \emph{{Relativistic Kinetic Theory. Principles and Applications}}, {Elsevier Science Ltd} (1980).
	
	\bibitem{Speranza:2020ilk}
	E.~Speranza and N.~Weickgenannt, \emph{{Spin tensor and pseudo-gauges: from nuclear collisions to gravitational physics}}, \href{https://doi.org/10.1140/epja/s10050-021-00455-2}{\emph{Eur. Phys. J. A} {\bfseries 57} (2021) 155} [\href{https://arxiv.org/abs/2007.00138}{{\ttfamily 2007.00138}}].
	
	\bibitem{Wigner:1932eb}
	E.P.~Wigner, \emph{{On the quantum correction for thermodynamic equilibrium}}, \href{https://doi.org/10.1103/PhysRev.40.749}{\emph{Phys. Rev.} {\bfseries 40} (1932) 749}.
	
	\bibitem{Das:2022azr}
	A.~Das, W.~Florkowski, A.~Kumar, R.~Ryblewski and R.~Singh, \emph{{Semi-classical Kinetic Theory for Massive Spin-half Fermions with Leading-order Spin Effects}}, \href{https://doi.org/10.5506/APhysPolB.54.8-A4}{\emph{Acta Phys. Polon. B} {\bfseries 54} (2023) 8} [\href{https://arxiv.org/abs/2203.15562}{{\ttfamily 2203.15562}}].
	
	\bibitem{Collins:1974ky}
	J.C.~Collins and M.J.~Perry, \emph{{Superdense Matter: Neutrons Or Asymptotically Free Quarks?}}, \href{https://doi.org/10.1103/PhysRevLett.34.1353}{\emph{Phys. Rev. Lett.} {\bfseries 34} (1975) 1353}.
	
	\bibitem{Shuryak:1977ut}
	E.V.~Shuryak, \emph{{Theory of Hadronic Plasma}}, {\emph{Sov. Phys. JETP} {\bfseries 47} (1978) 212}.
	
	\bibitem{Busza:2018rrf}
	W.~Busza, K.~Rajagopal and W.~van~der Schee, \emph{{Heavy Ion Collisions: The Big Picture, and the Big Questions}}, \href{https://doi.org/10.1146/annurev-nucl-101917-020852}{\emph{Ann. Rev. Nucl. Part. Sci.} {\bfseries 68} (2018) 339} [\href{https://arxiv.org/abs/1802.04801}{{\ttfamily 1802.04801}}].
	
	\bibitem{Stoecker:1986ci}
	H.~Stoecker and W.~Greiner, \emph{{High-Energy Heavy Ion Collisions: Probing the Equation of State of Highly Excited Hadronic Matter}}, \href{https://doi.org/10.1016/0370-1573(86)90131-6}{\emph{Phys. Rept.} {\bfseries 137} (1986) 277}.
	
	\bibitem{Rischke:1995ir}
	D.H.~Rischke, S.~Bernard and J.A.~Maruhn, \emph{{Relativistic hydrodynamics for heavy ion collisions. 1. General aspects and expansion into vacuum}}, \href{https://doi.org/10.1016/0375-9474(95)00355-1}{\emph{Nucl. Phys. A} {\bfseries 595} (1995) 346} [\href{https://arxiv.org/abs/nucl-th/9504018}{{\ttfamily nucl-th/9504018}}].
	
	\bibitem{Rischke:1995mt}
	D.H.~Rischke, Y.~Pursun and J.A.~Maruhn, \emph{{Relativistic hydrodynamics for heavy ion collisions. 2. Compression of nuclear matter and the phase transition to the quark - gluon plasma}}, \href{https://doi.org/10.1016/0375-9474(95)00356-3}{\emph{Nucl. Phys. A} {\bfseries 595} (1995) 383} [\href{https://arxiv.org/abs/nucl-th/9504021}{{\ttfamily nucl-th/9504021}}].
	
	\bibitem{Jaiswal:2020hvk}
	A.~Jaiswal et~al., \emph{{Dynamics of QCD matter {\textemdash} current status}}, \href{https://doi.org/10.1142/S0218301321300010}{\emph{Int. J. Mod. Phys. E} {\bfseries 30} (2021) 2130001} [\href{https://arxiv.org/abs/2007.14959}{{\ttfamily 2007.14959}}].
	
	\bibitem{Arslandok:2023utm}
	M.~Arslandok et~al., \emph{{Hot QCD White Paper}}, \href{https://doi.org/10.48550/arXiv.2303.17254}{\emph{arXiv eprint} (2023) } [\href{https://arxiv.org/abs/2303.17254}{{\ttfamily 2303.17254}}].
	
	\bibitem{Zubarev:1966}
	D.N.~Zubarev, \emph{A statistical operator for non-stationary processes}, {\emph{Soviet Physics Doklady} {\bfseries 10} (1966) 850}.
	
	\bibitem{Zubarev:1979afm}
	D.N.~Zubarev, A.V.~Prozorkevich and S.A.~Smolyanskii, \emph{{Derivation of nonlinear generalized equations of quantum relativistic hydrodynamics}}, \href{https://doi.org/10.1007/BF01032069}{\emph{Theor. Math. Phys.} {\bfseries 40} (1979) 821}.
	
	\bibitem{Zubarev:1989su}
	D.N.~Zubarev and M.V.~Tokarchuk, \emph{{Nonequilibrium Thermo Field Dynamics and the Method of the Nonequilibrium Statistical Operator}}, \href{https://doi.org/10.1007/BF01019114}{\emph{Teor. Mat. Fiz.} {\bfseries 88N2} (1991) 286}.
	
	\bibitem{vanWeert1982}
	C.G.~van Weert, \emph{Maximum entropy principle and relativistic hydrodynamics}, \href{https://doi.org/10.1016/0003-4916(82)90338-4}{\emph{Annals of Physics} {\bfseries 140} (1982) 133}.
	
	\bibitem{Becattini:2014yxa}
	F.~Becattini, L.~Bucciantini, E.~Grossi and L.~Tinti, \emph{{Local thermodynamical equilibrium and the beta frame for a quantum relativistic fluid}}, \href{https://doi.org/10.1140/epjc/s10052-015-3384-y}{\emph{Eur. Phys. J. C} {\bfseries 75} (2015) 191} [\href{https://arxiv.org/abs/1403.6265}{{\ttfamily 1403.6265}}].
	
	\bibitem{Becattini:2019dxo}
	F.~Becattini, M.~Buzzegoli and E.~Grossi, \emph{{Reworking the Zubarev's approach to non-equilibrium quantum statistical mechanics}}, \href{https://doi.org/10.3390/particles2020014}{\emph{Particles} {\bfseries 2} (2019) 197} [\href{https://arxiv.org/abs/1902.01089}{{\ttfamily 1902.01089}}].
	
	\bibitem{Buzzegoli:2020ycf}
	M.~Buzzegoli, \emph{{Thermodynamic Equilibrium of Massless Fermions with Vorticity, Chirality and Electromagnetic Field}}, \href{https://doi.org/10.1007/978-3-030-71427-7_3}{\emph{Lect. Notes Phys.} {\bfseries 987} (2021) 59} [\href{https://arxiv.org/abs/2011.09974}{{\ttfamily 2011.09974}}].
	
	\bibitem{Becattini:2013fla}
	F.~Becattini, V.~Chandra, L.~Del~Zanna and E.~Grossi, \emph{{Relativistic distribution function for particles with spin at local thermodynamical equilibrium}}, \href{https://doi.org/10.1016/j.aop.2013.07.004}{\emph{Annals Phys.} {\bfseries 338} (2013) 32} [\href{https://arxiv.org/abs/1303.3431}{{\ttfamily 1303.3431}}].
	
	\bibitem{Buzzegoli:2025zud}
	M.~Buzzegoli, \emph{{Kubo formulas for spin polarization in dissipative relativistic spin hydrodynamics: a first-order gradient expansion approach}}, \href{https://doi.org/10.1007/JHEP07(2025)255}{\emph{JHEP} {\bfseries 07} (2025) 255} [\href{https://arxiv.org/abs/2502.15520}{{\ttfamily 2502.15520}}].
	
	\bibitem{Becattini:2012tc}
	F.~Becattini, \emph{{Covariant statistical mechanics and the stress-energy tensor}}, \href{https://doi.org/10.1103/PhysRevLett.108.244502}{\emph{Phys. Rev. Lett.} {\bfseries 108} (2012) 244502} [\href{https://arxiv.org/abs/1201.5278}{{\ttfamily 1201.5278}}].
	
	\bibitem{Becattini:2018duy}
	F.~Becattini, W.~Florkowski and E.~Speranza, \emph{{Spin tensor and its role in non-equilibrium thermodynamics}}, \href{https://doi.org/10.1016/j.physletb.2018.12.016}{\emph{Phys. Lett. B} {\bfseries 789} (2019) 419} [\href{https://arxiv.org/abs/1807.10994}{{\ttfamily 1807.10994}}].
	
	\bibitem{Pang:2018zzo}
	L.-G.~Pang, H.~Petersen and X.-N.~Wang, \emph{{Pseudorapidity distribution and decorrelation of anisotropic flow within the open-computing-language implementation CLVisc hydrodynamics}}, \href{https://doi.org/10.1103/PhysRevC.97.064918}{\emph{Phys. Rev. C} {\bfseries 97} (2018) 064918} [\href{https://arxiv.org/abs/1802.04449}{{\ttfamily 1802.04449}}].
	
	\bibitem{Wu:2021fjf}
	X.-Y.~Wu, G.-Y.~Qin, L.-G.~Pang and X.-N.~Wang, \emph{{(3+1)-D viscous hydrodynamics at finite net baryon density: Identified particle spectra, anisotropic flows, and flow fluctuations across energies relevant to the beam-energy scan at RHIC}}, \href{https://doi.org/10.1103/PhysRevC.105.034909}{\emph{Phys. Rev. C} {\bfseries 105} (2022) 034909} [\href{https://arxiv.org/abs/2107.04949}{{\ttfamily 2107.04949}}].
	
	\bibitem{Wu:2022mkr}
	X.-Y.~Wu, C.~Yi, G.-Y.~Qin and S.~Pu, \emph{{Local and global polarization of \ensuremath{\Lambda} hyperons across RHIC-BES energies: The roles of spin hall effect, initial condition, and baryon diffusion}}, \href{https://doi.org/10.1103/PhysRevC.105.064909}{\emph{Phys. Rev. C} {\bfseries 105} (2022) 064909} [\href{https://arxiv.org/abs/2204.02218}{{\ttfamily 2204.02218}}].
	
	\bibitem{Sheng:2026Data}
	X.-L.~Sheng, F.~Becattini and D.~Roselli, ``{Hydrodynamic freeze-out hypersurface data for heavy-ion collisions}.'' \href{https://doi.org/10.5281/zenodo.18336680}{Zenodo}, 2026.
	\newblock 10.5281/zenodo.18336680.
	
	\bibitem{Becattini:2025oyi}
	F.~Becattini and R.~Singh, \emph{{On the local thermodynamic relations in relativistic spin hydrodynamics}}, \href{https://doi.org/10.1140/epjc/s10052-025-15071-3}{\emph{Eur. Phys. J. C} {\bfseries 85} (2025) 1338} [\href{https://arxiv.org/abs/2506.20681}{{\ttfamily 2506.20681}}].
	
	\bibitem{Mertig1991}
	R.~Mertig, M.~B\"{o}hm and A.~Denner, \emph{Feyn calc - computer-algebraic calculation of feynman amplitudes}, \href{https://doi.org/10.1016/0010-4655(91)90130-d}{\emph{Computer Physics Communications} {\bfseries 64} (1991) 345–359}.
	
	\bibitem{Shtabovenko:2016sxi}
	V.~Shtabovenko, R.~Mertig and F.~Orellana, \emph{{New Developments in FeynCalc 9.0}}, \href{https://doi.org/10.1016/j.cpc.2016.06.008}{\emph{Comput. Phys. Commun.} {\bfseries 207} (2016) 432} [\href{https://arxiv.org/abs/1601.01167}{{\ttfamily 1601.01167}}].
	
	\bibitem{Shtabovenko:2020gxv}
	V.~Shtabovenko, R.~Mertig and F.~Orellana, \emph{{FeynCalc 9.3: New features and improvements}}, \href{https://doi.org/10.1016/j.cpc.2020.107478}{\emph{Comput. Phys. Commun.} {\bfseries 256} (2020) 107478} [\href{https://arxiv.org/abs/2001.04407}{{\ttfamily 2001.04407}}].
	
	\bibitem{Shtabovenko:2023idz}
	V.~Shtabovenko, R.~Mertig and F.~Orellana, \emph{{FeynCalc 10: Do multiloop integrals dream of computer codes?}}, \href{https://doi.org/10.1016/j.cpc.2024.109357}{\emph{Comput. Phys. Commun.} {\bfseries 306} (2025) 109357} [\href{https://arxiv.org/abs/2312.14089}{{\ttfamily 2312.14089}}].
	
	\bibitem{Liu:2021nyg}
	Y.-C.~Liu and X.-G.~Huang, \emph{{Spin polarization formula for Dirac fermions at local equilibrium}}, \href{https://doi.org/10.1007/s11433-022-1903-8}{\emph{Sci. China Phys. Mech. Astron.} {\bfseries 65} (2022) 272011} [\href{https://arxiv.org/abs/2109.15301}{{\ttfamily 2109.15301}}].
	
	\bibitem{Palermo:2024tza}
	A.~Palermo, E.~Grossi, I.~Karpenko and F.~Becattini, \emph{{$\Lambda $ polarization in very high energy heavy ion collisions as a probe of the quark\textendash{}gluon plasma formation and properties}}, \href{https://doi.org/10.1140/epjc/s10052-024-13229-z}{\emph{Eur. Phys. J. C} {\bfseries 84} (2024) 920} [\href{https://arxiv.org/abs/2404.14295}{{\ttfamily 2404.14295}}].
	
	\bibitem{Vilenkin:1979ui}
	A.~Vilenkin, \emph{{Macroscopic parity violating effects: neutrino fluxes from rotating black holes and in rotating thermal radiation}}, \href{https://doi.org/10.1103/PhysRevD.20.1807}{\emph{Phys. Rev. D} {\bfseries 20} (1979) 1807}.
	
	\bibitem{Landsteiner:2011iq}
	K.~Landsteiner, E.~Megias, L.~Melgar and F.~Pena-Benitez, \emph{{Holographic Gravitational Anomaly and Chiral Vortical Effect}}, \href{https://doi.org/10.1007/JHEP09(2011)121}{\emph{JHEP} {\bfseries 09} (2011) 121} [\href{https://arxiv.org/abs/1107.0368}{{\ttfamily 1107.0368}}].
	
	\bibitem{Gao:2012ix}
	J.-H.~Gao, Z.-T.~Liang, S.~Pu, Q.~Wang and X.-N.~Wang, \emph{{Chiral Anomaly and Local Polarization Effect from Quantum Kinetic Approach}}, \href{https://doi.org/10.1103/PhysRevLett.109.232301}{\emph{Phys. Rev. Lett.} {\bfseries 109} (2012) 232301} [\href{https://arxiv.org/abs/1203.0725}{{\ttfamily 1203.0725}}].
	
	\bibitem{Kharzeev:2015znc}
	D.E.~Kharzeev, J.~Liao, S.A.~Voloshin and G.~Wang, \emph{{Chiral magnetic and vortical effects in high-energy nuclear collisions{\textemdash}A status report}}, \href{https://doi.org/10.1016/j.ppnp.2016.01.001}{\emph{Prog. Part. Nucl. Phys.} {\bfseries 88} (2016) 1} [\href{https://arxiv.org/abs/1511.04050}{{\ttfamily 1511.04050}}].
	
	\bibitem{Buzzegoli:2017cqy}
	M.~Buzzegoli, E.~Grossi and F.~Becattini, \emph{{General equilibrium second-order hydrodynamic coefficients for free quantum fields}}, \href{https://doi.org/10.1007/JHEP10(2017)091}{\emph{JHEP} {\bfseries 10} (2017) 091} [\href{https://arxiv.org/abs/1704.02808}{{\ttfamily 1704.02808}}].
	
	\bibitem{Buzzegoli:2018wpy}
	M.~Buzzegoli and F.~Becattini, \emph{{General thermodynamic equilibrium with axial chemical potential for the free Dirac field}}, \href{https://doi.org/10.1007/JHEP12(2018)002}{\emph{JHEP} {\bfseries 12} (2018) 002} [\href{https://arxiv.org/abs/1807.02071}{{\ttfamily 1807.02071}}].
	
	\bibitem{Buzzegoli:2021jeh}
	M.~Buzzegoli and D.E.~Kharzeev, \emph{{Anomalous gravitomagnetic moment and nonuniversality of the axial vortical effect at finite temperature}}, \href{https://doi.org/10.1103/PhysRevD.103.116005}{\emph{Phys. Rev. D} {\bfseries 103} (2021) 116005} [\href{https://arxiv.org/abs/2102.01676}{{\ttfamily 2102.01676}}].
	
	\bibitem{Sadofyev:2010is}
	A.V.~Sadofyev, V.I.~Shevchenko and V.I.~Zakharov, \emph{{Notes on chiral hydrodynamics within effective theory approach}}, \href{https://doi.org/10.1103/PhysRevD.83.105025}{\emph{Phys. Rev. D} {\bfseries 83} (2011) 105025} [\href{https://arxiv.org/abs/1012.1958}{{\ttfamily 1012.1958}}].
	
	\bibitem{Landsteiner:2011cp}
	K.~Landsteiner, E.~Megias and F.~Pena-Benitez, \emph{{Gravitational Anomaly and Transport}}, \href{https://doi.org/10.1103/PhysRevLett.107.021601}{\emph{Phys. Rev. Lett.} {\bfseries 107} (2011) 021601} [\href{https://arxiv.org/abs/1103.5006}{{\ttfamily 1103.5006}}].
	
	\bibitem{Avkhadiev:2017fxj}
	A.~Avkhadiev and A.V.~Sadofyev, \emph{{Chiral Vortical Effect for Bosons}}, \href{https://doi.org/10.1103/PhysRevD.96.045015}{\emph{Phys. Rev. D} {\bfseries 96} (2017) 045015} [\href{https://arxiv.org/abs/1702.07340}{{\ttfamily 1702.07340}}].
	
	\bibitem{Prokhorov:2020npf}
	G.Y.~Prokhorov, O.V.~Teryaev and V.I.~Zakharov, \emph{{Chiral vortical effect for vector fields}}, \href{https://doi.org/10.1103/PhysRevD.103.085003}{\emph{Phys. Rev. D} {\bfseries 103} (2021) 085003} [\href{https://arxiv.org/abs/2009.11402}{{\ttfamily 2009.11402}}].
	
	\bibitem{Prokhorov:2021bbv}
	G.Y.~Prokhorov, O.V.~Teryaev and V.I.~Zakharov, \emph{{Chiral vortical effect in extended Rarita-Schwinger field theory and chiral anomaly}}, \href{https://doi.org/10.1103/PhysRevD.105.L041701}{\emph{Phys. Rev. D} {\bfseries 105} (2022) L041701} [\href{https://arxiv.org/abs/2109.06048}{{\ttfamily 2109.06048}}].
	
	\bibitem{Prokhorov:2022rna}
	G.Y.~Prokhorov, O.V.~Teryaev and V.I.~Zakharov, \emph{{Gravitational chiral anomaly for spin 3/2 field interacting with spin 1/2 field}}, \href{https://doi.org/10.1103/PhysRevD.106.025022}{\emph{Phys. Rev. D} {\bfseries 106} (2022) 025022} [\href{https://arxiv.org/abs/2202.02168}{{\ttfamily 2202.02168}}].
	
	\bibitem{Prokhorov:2022snx}
	G.Y.~Prokhorov, O.V.~Teryaev and V.I.~Zakharov, \emph{{Gravitational chiral anomaly and anomalous transport for fields with spin 3/2}}, \href{https://doi.org/10.1016/j.physletb.2023.137839}{\emph{Phys. Lett. B} {\bfseries 840} (2023) 137839} [\href{https://arxiv.org/abs/2211.03865}{{\ttfamily 2211.03865}}].
	
	\bibitem{Velo:1969bt}
	G.~Velo and D.~Zwanziger, \emph{{Propagation and quantization of Rarita-Schwinger waves in an external electromagnetic potential}}, \href{https://doi.org/10.1103/PhysRev.186.1337}{\emph{Phys. Rev.} {\bfseries 186} (1969) 1337}.
	
	\bibitem{Hortacsu:1974bm}
	M.~Hortacsu, \emph{{Demonstration of noncausality for the rarita-schwinger equation}}, \href{https://doi.org/10.1103/PhysRevD.9.928}{\emph{Phys. Rev. D} {\bfseries 9} (1974) 928}.
	
	\bibitem{Adler:2017shl}
	S.L.~Adler, \emph{{Analysis of a gauged model with a spin-$\frac{1}{2}$ field directly coupled to a Rarita-Schwinger spin-$\frac{3}{2}$ field}}, \href{https://doi.org/10.1103/PhysRevD.97.045014}{\emph{Phys. Rev. D} {\bfseries 97} (2018) 045014} [\href{https://arxiv.org/abs/1711.00907}{{\ttfamily 1711.00907}}].
	
	\bibitem{Deser:2000dz}
	S.~Deser, V.~Pascalutsa and A.~Waldron, \emph{{Massive spin 3/2 electrodynamics}}, \href{https://doi.org/10.1103/PhysRevD.62.105031}{\emph{Phys. Rev. D} {\bfseries 62} (2000) 105031} [\href{https://arxiv.org/abs/hep-th/0003011}{{\ttfamily hep-th/0003011}}].
	
	\bibitem{Porrati:2009bs}
	M.~Porrati and R.~Rahman, \emph{{Causal Propagation of a Charged Spin 3/2 Field in an External Electromagnetic Background}}, \href{https://doi.org/10.1103/PhysRevD.80.025009}{\emph{Phys. Rev. D} {\bfseries 80} (2009) 025009} [\href{https://arxiv.org/abs/0906.1432}{{\ttfamily 0906.1432}}].
	
	\bibitem{Prokhorov:2022udo}
	G.Y.~Prokhorov, O.V.~Teryaev and V.I.~Zakharov, \emph{{Hydrodynamic Manifestations of Gravitational Chiral Anomaly}}, \href{https://doi.org/10.1103/PhysRevLett.129.151601}{\emph{Phys. Rev. Lett.} {\bfseries 129} (2022) 151601} [\href{https://arxiv.org/abs/2207.04449}{{\ttfamily 2207.04449}}].
	
	\bibitem{Buzzegoli:2026oxc}
	M.~Buzzegoli, A.~Gecic and R.~Singh, \emph{{Modeling {\ensuremath{\Lambda}} polarization in Au+Au collisions at sNN=200{\,}{\,}GeV using relativistic spin hydrodynamics}}, \href{https://doi.org/10.1103/6mm9-hrj7}{\emph{Phys. Rev. D} {\bfseries 114} (2026) 074017} [\href{https://arxiv.org/abs/2605.08219}{{\ttfamily 2605.08219}}].
	
\end{thebibliography}
\end{document}